\documentclass[a4paper,11pt]{article}
\pdfoutput=1
\usepackage{jheppub}
\usepackage{amssymb}
\usepackage{amsmath}
\usepackage{latexsym}
\usepackage{fancybox}
\usepackage{float}
\usepackage{simplewick}
\usepackage{comment}
\usepackage{bm}
\usepackage{framed}
\definecolor{shadecolor}{rgb}{0.9,0.9,0.95}
\usepackage{setspace}
\usepackage[normalem]{ulem}
\definecolor{darkgreen}{rgb}{0,0.5,0}
\definecolor{darkblue}{cmyk}{0.9,0.9,0,0}
\definecolor{darkred}{rgb}{0.6,0,0.3}

\usepackage{graphicx}
\usepackage{subcaption}
\usepackage{adjustbox}
\usepackage{placeins}
\usepackage[curve,matrix,arrow,color]{xy}
\usepackage{tikz}
\usetikzlibrary{intersections}
\usetikzlibrary{er,positioning}
\usetikzlibrary{arrows,shapes}
\usepackage{hyperref}

\def\eqref#1{(\ref{#1})}

\def\beq{\begin{equation}}
\def\eeq{\end{equation}}

\renewcommand{\thefigure}{\thesection.\arabic{figure}}

\newcommand{\ri}{\mathrm{i}}

\numberwithin{equation}{section}

\title{Quantum Entanglement of Bethe States}
\author[a]{Yu Hao}
\author[a]{Yunfeng Jiang}
\author[a]{Bi-Quan Yang}
\author[b,c]{and De-liang Zhong}

\affiliation[a]{School of physics \& Shing-Tung Yau Center, Southeast University, Nanjing 211189, P. R. China}
\affiliation[b]{Institute of Theoretical Physics, Chinese Academy of Sciences, Beijing 100190, China}
\affiliation[c]{School of Physical Sciences, University of Chinese Academy of Sciences,
Beijing 100049, China}

\abstract{
We investigate the quantum entanglement of Bethe states across a family of integrable spin chains, including the XXX$_{\frac{1}{2}}$ model, its higher-spin generalizations (XXX$_s$), and the non-compact $SL(2,\mathbb{R})$ chain. For on-shell eigenstates, we perform a comprehensive scan of the bipartite entanglement entropy across the entire spectrum of finite chains with periodic boundary conditions, and identify the Bethe solutions that minimize and maximize the entanglement. These extremal solutions follow systematic, spin-dependent patterns in the Bethe quantum numbers. In the XXX$_{\frac{1}{2}}$ spin chain, for the antiferromagnetic chain, the state with minimal entropy always coincides with the lowest-energy state (the ground state) within a given fixed-magnon sector. For the higher-spin XXX$_s$ model, however, the lowest-entropy state is not always identical to the ground state, and can even be the state of highest energy. By contrast, the Bethe roots that maximize entropy exhibit considerably more intricate structure. Our analysis further reveals how special Bethe root configurations, such as singular and strange solutions, affect entanglement, and it uncovers characteristic entanglement features in the non-compact $SL(2,\mathbb{R})$ chain that are absent from compact spin chains. For off-shell Bethe states, we develop an optimization algorithm that extremizes the entanglement entropy over rapidity distributions, enabling us to explore the maximum entanglement achievable by a Bethe state without imposing the Bethe ansatz equations.
}

\begin{document}

\maketitle

\section{Introduction}
\label{sec:intro}
Bethe states occupy a central place in quantum integrable systems. Since Bethe's original solution of the Heisenberg spin chain \cite{Bethe:1931hc}, the Bethe ansatz has developed into a family of powerful methods for constructing exact eigenstates and computing spectra. In the coordinate Bethe ansatz, wavefunctions are expressed as superpositions of plane waves with coefficients determined by factorized scattering. In the algebraic Bethe ansatz \cite{Sklyanin:1979pfu,Takhtajan:1979iv,Faddeev:1996iy}, Bethe states are generated by non-local creation operators built from an $R$-matrix satisfying the Yang--Baxter equation. In both formulations, a Bethe state is characterized by a set of rapidities $\{u_1,\ldots,u_M\}$. When these rapidities satisfy the Bethe ansatz equations, the state is an on-shell eigenstate; otherwise it is an off-shell Bethe state. Although off-shell states are not eigenstates, they are indispensable for computing important physical quantities such as form factors and correlation functions. Moreover, they can serve as useful variational or approximate states in settings that go beyond strictly integrable models~\cite{Zhao:2026vyp,Wang:2026fhk,krauth1991bethe,kiwata1994bethe,okunishi1999magnetization}.\par

The rapidities of an on-shell Bethe state encode its energy, momentum, and higher conserved charges. In this sense, Bethe roots provide a microscopic set of quasiparticle coordinates for the state. However, spectral information is not the only structure of physical interest. A quantum many-body wavefunction also possesses a quantum-information structure, captured by quantities such as entanglement entropy, mutual information, and symmetry-resolved entanglement. Entanglement has become one of the primary diagnostics of quantum many-body physics, both in condensed-matter systems and in high-energy physics; see, \emph{e.g.}, \cite{Amico:2007ag,Eisert:2008ur,Laflorencie:2015eck,Latorre:2009zz,Nishioka:2018khk} for reviews. Several exact and integrability-based approaches to entanglement entropy have been developed, which we will briefly review in Section~\ref{sec:integrabilityApproach}. The present work complements existing results in the literature: rather than focusing on the ground state or specific excited states of \emph{particular models}, we are interested in the quantum entanglement of a \emph{special class of states} — namely, the \emph{Bethe states}.\par

The central question of this work is how the entanglement structure of a Bethe state is encoded in its Bethe roots. More concretely, we study the map
\begin{equation}
\{u_j\}\ \text{or}\ \{I_j\}
\quad \longrightarrow \quad
\mathcal{S}_{\ell},
\end{equation}
where $\{u_j\}$ are rapidities, $\{I_j\}$ are Bethe quantum numbers for on-shell states, and $\mathcal{S}_{\ell}$ is the bipartite entanglement entropy with subsystem size $\ell$. 
This question is natural: ground states and excited states exhibit very different entanglement behavior---ground states typically obey an area law, whereas highly excited states often display volume-law entanglement---yet in an integrable spin chain both are constructed in the same manner, their differences lying primarily in the Bethe roots. This leads us to ask: How much entanglement can a Bethe state generate? For given length $L$ and magnon number $M$, which Bethe roots yield low entanglement and which yield high entanglement? And can different rapidity configurations give rise to the same entanglement entropy? \par

We approach these questions in two complementary ways. First, for on-shell Bethe states, we perform full-spectrum scans of Bethe eigenstates. For each state, we compute the bipartite entanglement entropy for all possible subsystem sizes $\ell$, yielding numerical data for $\mathcal{S}(L,M;\ell)$. These complete data allow us to identify the generic features of the entanglement entropy for finite-size Bethe states. Furthermore, we extract the entanglement extrema in each sector labeled by chain length \(L\), magnon number \(M\), and subsystem size \(\ell\),
\begin{equation}
\mathcal{S}_{\min}(L,M;\ell),
\qquad
\mathcal{S}_{\max}(L,M;\ell).
\end{equation}
We identify the Bethe roots (or Bethe quantum numbers) corresponding to these extrema and investigate the patterns exhibited by such Bethe quantum numbers. Second, for off-shell Bethe states, we relax the Bethe equations, treat the rapidities as free parameters, and optimize \(\mathcal{S}_\ell\) directly over the rapidities. This approach probes how much entanglement can be generated by a generic Bethe state, before the Bethe equations select physical eigenstates.
\par

\subsection{Main results}
\label{sec:main-results}
We summarize the main results before presenting the details.

\begin{enumerate}
\item \textbf{A new approach for computing entanglement entropy.} We develop a novel method to compute the entanglement entropy for Bethe states, applicable to both on-shell and off-shell cases. This approach involves cutting the Bethe state into two Bethe states of smaller length and employing the Schmidt decomposition. The key ingredient is the overlap of off-shell Bethe states, whose evaluation cost does not grow with the chain length. The method works for a wide range of models, including both compact and non-compact spin chains, in a unified fashion, and is efficient for finite magnon numbers in both short and long chains.

\item \textbf{On-shell Bethe states of compact chains.}
For the compact XXX$_{\frac{1}{2}}$ chain, we find that the minima of the entanglement entropy are always associated with compact, symmetric Bethe quantum number distributions: the quantum numbers are symmetric about zero and packed as tightly as possible. These configurations correspond precisely to the lowest-energy configurations in the corresponding fixed-\(M\) sector. For the higher-spin XXX$_s$ chain, we investigate the cases $s=1$ and $s=\frac{3}{2}$ as examples. In both, the lowest-entropy states do not always coincide with the ground states; they can sometimes correspond to the highest-energy state, and at other times to neither. This reveals a qualitative difference from the spin-$\frac{1}{2}$ case and indicates that the dimension of the local Hilbert space affects the entanglement properties of quantum states in a nontrivial way. The Bethe quantum number patterns that yield the maxima are less universal: they can involve spread-out mode numbers, singular or strange solutions, repeated roots, or complex-root configurations (Section~\ref{subsec:onshell_xxx} and Section~\ref{subsec:onshell_xxxs}).

\item \textbf{On-shell Bethe states of non-compact chains.}
For the non-compact \(SL(2,\mathbb{R})\) chain, we investigate the entanglement entropy for larger magnon numbers $M$ and uncover a universal logarithmic dependence on $M$. For \(L=2\), the coefficient is very stable and close to \(0.48\). For larger \(L\), the data are well described by logarithmic fits over the sampled range, but the fitted slopes show finite-size and fitting-window dependence. We therefore regard the logarithmic growth as robust numerical evidence, while the universality of the coefficient in front of $\log M$ remains a conjecture.

\item \textbf{Off-shell Bethe states for compact chains.} For a Bethe state with $M$ magnons, the entanglement entropy for any length and subsystem size is bounded by $M$. This follows directly from \eqref{eq:cutBS}: cutting a Bethe state yields at most $2^M$ terms; via the Schmidt decomposition, the theoretical maximum is simply $M$.
For the XXX$_{\frac{1}{2}}$ chain at the middle cut \(\ell_*=\lfloor L/2\rfloor\), Section~\ref{subsec:offshell_xxx} reveals two regimes of maximum entropy. At fixed magnon number \(M\), with $L\to\infty$, we find
\begin{align*}
\mathcal{S}_{\max}^{\rm off}(L,M;\ell_*)\longrightarrow M ,
\end{align*}
which saturates the upper bound. Conversely, when the magnon number scales with $L$, for example at half filling, the asymptotic behavior exhibits a volume law, \emph{e.g.}
\begin{align*}
\mathcal{S}_{\max}^{\rm off}(L,L/2;\ell_*)\simeq 0.35\,L\,,
\end{align*}
which lies below the upper bound. {The higher-spin compact chains are consistent with the same off-shell scaling picture: fixed-magnon maxima approach the partition bound.} {As for the minimum, compact off-shell chains can reach zero through product-state rapidity loci: singular rapidities \(u_j=\pm\ri/2\) for XXX$_{1/2}$ and boundary-string analogues for higher spin.}

\item \textbf{Off-shell Bethe states for non-compact chains.}
For the non-compact \(SL(2,\mathbb{R})\) chain, the theoretical upper bound for an $M$-magnon state depends on the subsystem size. We find that while the maximum nearly saturates the bound \(\log(M+1)\) for $\ell=1$, for larger chains with half subsystem size $\ell=\lfloor L/2\rfloor$, the optima over finite roots remain below the upper bound (Section~\ref{subsec:offshell_sl2}). {The off-shell maxima admit the same large-$M$ fit as on-shell, \(\mathcal{S}_{\max}^{\rm off}\simeq \alpha \log M+\beta\): for \(L=2\) (\(\ell=1\)) we find a stable slope \(\alpha=1\), about twice the on-shell value, so the maximum saturates \(\log(M+1)\); for larger \(L\) (\(\ell\ge2\)) the growth remains logarithmic, but our current data are not yet sufficient for us to make a convincing statement about the universality of the slope \(\alpha\) for these cases.}

\item {\textbf{The map from rapidities to entropy is many-to-one.}
The entanglement entropy is not an injective coordinate on Bethe-root space. For on-shell states, Bethe roots that are related by parity transformation, \emph{i.e.} \(\{u_j\}\) and \(\{-u_j\}\), have the same Schmidt spectrum, and additional accidental equalities can occur in finite scans (Section~\ref{subsec:onshell_xxx}). For off-shell states, this many-to-one structure becomes continuous: the real-rapidity plots for XXX$_{\frac{1}{2}}$ show iso-entropy contours for \(M=2\) and iso-entropy surfaces for \(M=3\) (Figures~\ref{fig:offshell_xxx_heatmap} and~\ref{fig:offshell_xxx_iso}). For \(L=6\), \(M=3\), and \(\ell=3\), the real slice contains several visible components at the same entropy level, including the \(S_3\) permutation orbits near the maximum. A bounded complex-path search did not find bridges between the tested real components, so these data give numerical evidence that equal-entropy rapidity sets can have non-trivial geometry, although they do not constitute a full classification of the complex level sets.}

\item \textbf{On-shell versus off-shell.}
Comparing on-shell and off-shell results (Sections~\ref{subsec:offshell_xxx}, \ref{subsec:offshell_xxxs}, \ref{subsec:offshell_sl2}) yields the following observations. For compact chains, going off shell does not dramatically raise the maximum entropy: {the on-shell maxima lie close to the off-shell optima, and both are at or below the upper bound, with saturation only in special low-rank sectors.} The larger change occurs at the minimum, where off-shell rapidities can access product-state loci that are excluded by the regular on-shell construction. {In the non-compact chain, the \(\ell=1\) cut shows a stronger on/off-shell difference: the off-shell maximum nearly reaches the upper bound, whereas the Bethe equations keep the on-shell value much lower. For larger cuts with \(\ell>1\), the off-shell maxima also remain below the upper bound, as seen in the \(L=4,\ell=2\) and \(L=6,\ell=3\) data.}
\end{enumerate}

The structure of the paper is as follows. Section~\ref{sec:EEint} reviews bipartite entanglement entropy and introduces the algebraic Bethe ansatz method used to compute the entanglement entropy of Bethe states. Section~\ref{sec:offshell} describes the numerical optimization of off-shell entanglement entropy. Sections~\ref{sec:HXXX}, \ref{sec:highSpin}, and \ref{sec:nonComp} present the results for the XXX$_{\frac{1}{2}}$ chain, the higher-spin XXX$_s$ chains, and the non-compact $SL(2,\mathbb{R})$ spin chain, respectively. We conclude in Section~\ref{sec:conclude}. The appendices collect scalar-product formulas, the Gram--Schmidt construction, symmetry properties, implementation details, analytic product-state and saturation limits, and supplementary numerical data.

\section{Entanglement entropy of Bethe States}
\label{sec:EEint}
We begin by reviewing fundamental concepts of entanglement entropy, and then introduce an integrability-based method for computing the entanglement entropy of Bethe states.

\subsection{Entanglement entropy}
\paragraph{Entanglement entropy.} Consider a spin chain of length $L$, partitioned into two subsystems labeled $A$ and $B$, of lengths $\ell_A=\ell$ and $\ell_B=L-\ell$, respectively. For a given quantum state $|\psi\rangle$, the reduced density matrices are defined as
\begin{align}
\hat{\rho}_A=\text{Tr}_B|\psi\rangle\langle\psi|\,\qquad \hat{\rho}_B=\text{Tr}_A|\psi\rangle\langle\psi|\,.
\end{align} 
The von Neumann entanglement entropy between subsystems $A$ and $B$ is then given by
\begin{align}
\mathcal{S}_{\ell}=-\text{Tr}(\hat{\rho}_A\log\hat{\rho}_A)=-\text{Tr}(\hat{\rho}_B\log\hat{\rho}_B)\,.
\end{align}
Throughout this work, we take the logarithm to base 2 and adopt the following notation
\begin{align}
\log x\equiv\log_2 x,\qquad \ln x\equiv\log_e x,\qquad \log x=\frac{\ln x}{\ln 2}\,.
\end{align}
For $\alpha>0$, the R\'enyi entropy is defined as
\begin{align}
\mathcal{S}_{\ell}^{(\alpha)}=\frac{1}{1-\alpha}\log \text{Tr}(\rho_A^{\alpha})\,,
\end{align}
which is related to the von Neumann entropy by
\begin{align}
\mathcal{S}_{\ell}=\lim_{\alpha\to 1}\mathcal{S}_{\ell}^{(\alpha)}\,.
\end{align}

\paragraph{Schmidt decomposition.} A practical way of computing entanglement entropy is via the Schmidt decomposition, which we employ in this work. We choose \emph{orthonormal bases} $\{|i\rangle_A\}$ and $\{|j\rangle_B\}$ for the two subsystems, with dimensions $N_A$ and $N_B$, respectively. Any pure state $|\psi\rangle$ can be expressed as
\begin{align}
\label{eq:decomPsi}
|\psi\rangle=\sum_{i,j}\Psi_{ij}|i\rangle_A\otimes|j\rangle_B\,.
\end{align}
The coefficients $\Psi_{ij}$ form an $N_A\times N_B$ matrix. Performing a singular value decomposition (SVD) on this matrix yields 
\begin{align}
\label{eq:SVD}
\Psi_{ij}=U_{ia}S_{aa}V_{ja}^*, 
\end{align}
{or in matrix form} $\Psi=USV^{\dagger}$ where $U$ is a matrix of dimension $N_A\times\min(N_A,N_B)$ satisfying $U^{\dagger}U=1$, $V^{\dagger}$ is a matrix of dimension $\min(N_A,N_B)\times N_B$ satisfying $V^{\dagger}V=1$, and $S$ is a diagonal matrix of dimension $\min(N_A,N_B)\times\min(N_A,N_B)$ with \emph{non-negative} entries $S_{aa}=s_a$. The values $s_a$ are the \emph{singular values}. The number $r$ of non-zero singular values is the Schmidt rank of $\Psi$. Using the SVD, the state can be rewritten as
\begin{align}
|\psi\rangle=\sum_{a=1}^{\text{min}(N_A,N_B)}\left(\sum_{i=1}^{N_A}U_{ia}|i\rangle_A\right) s_a\left(\sum_{j=1}^{N_B}V_{ja}^*|j\rangle_B \right)
=\sum_{a=1}^{r}s_a|a\rangle_A|a\rangle_B\,,
\end{align}
where $\{|a\rangle_A\}$ and $\{|a\rangle_B\}$ are orthonormal sets. In this representation, the reduced density matrices take the following simple forms
\begin{align}
\hat{\rho}_A=\sum_{a=1}^r s_a^2|a\rangle_A{_A\langle} a|,\qquad \hat{\rho}_B=\sum_{a=1}^r s_a^2|a\rangle_B{_B\langle} a|\,,
\end{align}
from which the von Neumann and R\'enyi entropies can be expressed compactly as
\begin{align}
\mathcal{S}_{\ell}=-\sum_{a=1}^r s_a^2\log s_a^2,\qquad \mathcal{S}_{\ell}^{(\alpha)}=\frac{1}{1-\alpha}\log\sum_{a=1}^r (s_a^{2})^{\alpha}\,.
\end{align}

\subsection{Integrability approach}
\label{sec:integrabilityApproach}
For short spin chains with relatively small subsystem dimensions $N_A$ and $N_B$, the entanglement entropy can be computed directly by constructing the wavefunction and performing a Schmidt decomposition. In integrable models, however, several more specialized methods are available. These methods exploit free-particle structures, exact reduced density matrices, transfer-matrix techniques, form-factor expansions, or quasiparticle data. We briefly summarize the main approaches most relevant to integrable spin chains.\footnote{The literature on entanglement entropy is vast. The following discussion is not meant as a complete review, but as a guide to integrability-based methods that are related to the present work.}

\begin{itemize}
\item Spin chains that can be mapped to free lattice fermions through the Jordan--Wigner transformation have been studied extensively. The entanglement entropy of the XX chain was first computed numerically in \cite{Vidal:2002rm}, where logarithmic growth with subsystem size was observed in the critical ground state. This scaling was confirmed analytically using Toeplitz determinants and the Fisher--Hartwig conjecture in \cite{Jin:2003lzd}, in agreement with the universal conformal field theory result of \cite{Korepin:2004zz}. The analysis was later generalized to the XY spin chain in \cite{Its2004,Peschel2004}. More generally, for quadratic fermionic Hamiltonians the reduced density matrix can be reconstructed from the two-point correlation matrix \cite{Peschel:2003rdm}. For more details, see \cite{Peschel:2009free} for a review of reduced density matrices and entanglement entropy in free lattice models.
\item For infinite bipartitions, Baxter's corner transfer matrix method provides another exact route to the reduced density matrix and the entanglement spectrum. This perspective is closely related to the study of density matrix spectra in integrable models \cite{Peschel:1999dms}. It has been applied to several solvable chains, including the transverse-field Ising model and the XXZ chain \cite{Calabrese:2004eu,Weston:2006bg}, as well as the XYZ chain and its sine-Gordon scaling limit \cite{Ercolessi:2009kc}.
\item For finite interacting chains, one can compute the reduced density matrix more directly using integrability. In the spin-$\frac{1}{2}$ XXZ spin chain, density matrix elements can be obtained by solving the quantum Knizhnik--Zamolodchikov equation, and exact formulas are obtained in \cite{Boos:2005dc,Sato:2006pk,Sato:2007qs}. A complementary approach uses the algebraic Bethe ansatz together with the solution of the inverse problem \cite{Kitanine1998}. This approach was used in \cite{Alba:2009th} to study the entanglement entropy of excited states in finite spin chains.
\item The replica trick and the related branch-point twist field method give another integrability-based route to R\'enyi and von Neumann entropies. In quantum field theory, branch-point twist fields allow one to compute the exact ground state entanglement entropy for conformal field theories \cite{Calabrese:2004eu} and express entanglement entropies in terms of form factor expansions for massive integrable quantum field theories \cite{Cardy:2007mb}. Motivated by this construction, a lattice analogue of the branch-point twist field was introduced for the XXZ spin chain in \cite{Castro-Alvaredo:2010scy}.
\item A closely related line of work studies the entanglement of excited Bethe eigenstates and magnon excitations more directly. Bound states in Bethe spectra were shown to leave clear signatures in the entanglement of excited spin-chain states \cite{Molter:2014hna}. Bethe-ansatz methods have also been applied to entanglement in rank-one sectors of the $\mathcal{N}=4$ SYM spin chain \cite{Georgiou:2016wff}. More recently, universal formulas have been developed for the entanglement of magnon and quasiparticle excitations in spin chains \cite{Castro-Alvaredo:2018dja,Zhang:2020vtc,Zhang:2020dtd,Zhang:2021mte} within certain regimes.

\item Finally, symmetry-resolved entanglement provides a refined diagnostic when the model has a conserved charge. Although this is not restricted to integrable models, integrability supplies exact tools for studying the charge-resolved sectors; see, for example, form-factor/bootstrap approaches in integrable field theory \cite{Horvath:2020vzs}. This viewpoint is naturally connected to the block decomposition by magnon number used below.
\end{itemize}

{The present work is closest in spirit to the finite-chain Bethe state approaches \cite{Alba:2009th} above, in the sense that we consider both ground states and excited states. However, there are also important differences. In \cite{Alba:2009th}, one first constructs the reduced density matrix by explicitly computing each matrix element using the quantum inverse scattering method. This method is technically more complicated for higher-spin models and does not work for non-compact spin chains. Instead, in our method, we first cut the Bethe state into two Bethe states of smaller length, and then perform a Schmidt decomposition. The Schmidt matrix can be computed analytically while its singular values are computed numerically. Our method works for higher-spin and non-compact chains with only minor modifications.}

\paragraph{Algebraic Bethe ansatz.} The models studied in the current work include the integrable XXX$_s$ chain\footnote{The standard Heisenberg XXX chain is the XXX$_{\frac{1}{2}}$ model in this notation.} and the non-compact $SL(2,\mathbb{R})$ spin chain. All of them are solved by the algebraic Bethe ansatz, which we review briefly.\par 

We start with the Lax operator, which is defined at each site $n$ as
\begin{align}
\label{eq:defLax}
\mathbf{L}_{an}(u)=\left(\begin{array}{cc} u+\ri\,\mathbf{S}_n^z & \ri\,\mathbf{S}_n^- \\ \ri\,\mathbf{S}_n^+ & u-\ri\,\mathbf{S}_n^z  \end{array}\right)_a\,,
\end{align}
where ``a'' denotes the auxiliary space, chosen to be $\mathbb{C}^2$ throughout this paper. The local spin operators $\mathbf{S}_n^{\alpha}$ satisfy the algebra
\begin{align}
[\mathbf{S}^+,\mathbf{S}^-]=2\mathbf{S}^z,\qquad [\mathbf{S}^z,\mathbf{S}^{\pm}]=\pm\mathbf{S}^{\pm}\,.
\end{align}
The Lax operator constructed in \eqref{eq:defLax} satisfies the $RLL$-relation
\begin{align}
R_{ab}(u-v)\mathbf{L}_{an}(u)\mathbf{L}_{bn}(v)=\mathbf{L}_{bn}(v)\mathbf{L}_{an}(u)R_{ab}(u-v)
\end{align}
where $R_{ab}(u)$ is the $R$-matrix, which is a solution of the Yang--Baxter equation. Explicitly, it reads
\begin{align}
R_{ab}(u)=\left(\begin{array}{cccc} u+\ri & 0 & 0 & 0\\ 0 & u & \ri & 0\\ 0 &\ri & u & 0\\ 0 & 0& 0& u+\ri \end{array}\right)_{ab}\,.
\end{align}
Multiplying the Lax operators at each site in order gives the monodromy matrix 
\begin{align}
\mathbf{M}_a(u)=\mathbf{L}_{a1}(u)\ldots\mathbf{L}_{aL}(u)=\left(\begin{array}{cc} \mathbf{A}(u) & \mathbf{B}(u)\\ \mathbf{C}(u) &\mathbf{D}(u)     \end{array}\right)_a\,.
\end{align}
The transfer matrix is defined by $\mathbf{T}(u)=\text{Tr}_a\mathbf{M}_a(u)=\mathbf{A}(u)+\mathbf{D}(u)$. From the $RLL$-relation, it follows that the monodromy matrix satisfies the following Yangian algebra
\begin{align}
\label{eq:RMM}
R_{ab}(u-v)\mathbf{M}_{a}(u)\mathbf{M}_{b}(v)=\mathbf{M}_{b}(v)\mathbf{M}_{a}(u)R_{ab}(u-v)\,.
\end{align}
The Yangian algebra encodes the algebraic relations among the operators $\mathbf{A}(u)$, $\mathbf{B}(u)$, $\mathbf{C}(u)$, and $\mathbf{D}(u)$.\par

\paragraph{Construction of eigenstates.} The models under consideration in this paper share the same Yangian algebra~\eqref{eq:RMM}, but are in different representations which will be specified in the relevant sections. The eigenstates are constructed as follows.
\begin{itemize}
\item For the compact XXX$_s$ spin chain, an $M$-magnon Bethe state takes the form
\begin{align}
|\{u_j\}\rangle^{(s)}\equiv \mathbf{B}^{(s)}(u_1)\ldots \mathbf{B}^{(s)}(u_M)|0\rangle^{(s)}
\end{align}
where $|0_s\rangle$ is the highest weight state satisfying
\begin{align}
\mathbf{A}^{(s)}(u)|0\rangle^{(s)}=a_s(u)|0\rangle^{(s)},\quad \mathbf{D}^{(s)}(u)|0\rangle^{(s)}=d_s(u)|0\rangle^{(s)},\quad \mathbf{C}^{(s)}(u)|0\rangle^{(s)}=0\,.
\end{align}
The functions $a_s(u)$ and $d_s(u)$ are given by
\begin{align}
\label{eq:defad}
a_s(u)=(u+\ri s)^L,\qquad d_s(u)=(u-\ri s)^L\,.
\end{align}
\item For the non-compact $SL(2,\mathbb{R})$ spin chain, an $M$-magnon Bethe state takes the form
\begin{align}
|\{u_j\}\rangle\equiv \mathbf{C}(u_1)\ldots \mathbf{C}(u_M)|\bar{0}\rangle
\end{align}
where $|\bar{0}\rangle$ is the lowest weight state satisfying
\begin{align}
\mathbf{A}(u)|\bar{0}\rangle=a_{-1/2}(u)|\bar{0}\rangle,\qquad \mathbf{D}(u)|\bar{0}\rangle=d_{-1/2}(u)|\bar{0}\rangle,\qquad \mathbf{B}(u)|\bar{0}\rangle=0\,.
\end{align}
Here $a_s(u)$ and $d_s(u)$ are defined in \eqref{eq:defad}. Thus the $SL(2,\mathbb{R})$ spin chain corresponds to the $s=-1/2$ representation of the Yangian algebra \eqref{eq:RMM}.
\end{itemize}
For a Bethe state to be an eigenstate of the spin chain Hamiltonian, or the transfer matrix, the rapidities $\{u_j\}$ must satisfy the BAE
\begin{align}
\frac{a_s(u_j)}{d_s(u_j)}=\prod_{k=1\atop k\ne j}^M\frac{u_j-u_k+\ri}{u_j-u_k-\ri}\,,\qquad j=1,\ldots,M\,.
\end{align}

\paragraph{Cutting Bethe states.} Consider a spin chain of length $L$, decomposed into two parts $A$ and $B$ with lengths $\ell_A=\ell$ and $\ell_B=L-\ell$. The monodromy matrix factorizes as
\begin{align}
\mathbf{M}_a(u)=\left[\mathbf{L}_{a1}(u)\ldots\mathbf{L}_{a\ell}(u)\right]\left[\mathbf{L}_{a,\ell+1}(u)\ldots\mathbf{L}_{aL}(u)\right]\equiv\mathbf{M}_{a,1}(u)\mathbf{M}_{a,2}(u)\,,
\end{align}
where $\mathbf{M}_{a,1}$ and $\mathbf{M}_{a,2}$ act on subsystems $A$ and $B$ respectively. Writing
\begin{align}
\mathbf{M}_{a,j}(u)=\left(\begin{array}{cc} \mathbf{A}_j(u) & \mathbf{B}_j(u)\\ \mathbf{C}_j(u) & \mathbf{D}_j(u) \end{array}\right)_a,\qquad j=1,2\,,
\end{align}
we obtain the decomposition
\begin{align}
\left(\begin{array}{cc} \mathbf{A}(u) & \mathbf{B}(u)\\ \mathbf{C}(u) & \mathbf{D}(u) \end{array}\right)_a
=\left(\begin{array}{cc} \mathbf{A}_1(u) & \mathbf{B}_1(u)\\ \mathbf{C}_1(u) & \mathbf{D}_1(u) \end{array}\right)_a
\left(\begin{array}{cc} \mathbf{A}_2(u) & \mathbf{B}_2(u)\\ \mathbf{C}_2(u) & \mathbf{D}_2(u) \end{array}\right)_a\,.
\end{align}
Thus the matrix elements of the full monodromy matrix can be expressed in terms of those of the subchain monodromy matrices. In particular,
\begin{align} \label{eqn-Bparts}
\mathbf{B}(u)=&\,\mathbf{A}_1(u)\mathbf{B}_2(u)+\mathbf{B}_1(u)\mathbf{D}_2(u)\,,\\\nonumber
\mathbf{C}(u)=&\,\mathbf{C}_1(u)\mathbf{A}_2(u)+\mathbf{D}_1(u)\mathbf{C}_2(u)\,.
\end{align}
Acting with $\mathbf{B}(u)$ on the pseudovacuum $|0\rangle=|0\rangle_A\otimes|0\rangle_B$, which factorizes naturally, gives
\begin{align}
\mathbf{B}(u)|0\rangle=&\,\left(\mathbf{A}_1(u)\mathbf{B}_2(u)+\mathbf{B}_1(u)\mathbf{D}_2(u)\right)|0\rangle_A\otimes|0\rangle_B\\\nonumber
=&\,\left(\mathbf{A}_1(u)|0\rangle_A\right)\otimes\left(\mathbf{B}_2(u)|0\rangle_B\right)+\left(\mathbf{B}_1(u)|0\rangle_A\right)\otimes\left(\mathbf{D}_2(u)|0\rangle_B\right)\\\nonumber
=&\,a_A(u)|0\rangle_A\otimes |\{u\}\rangle_B+d_B(u)|\{u\}\rangle_A\otimes|0\rangle_B
\end{align}
with
\begin{align}
\label{eq:subchainad}
a_A(u)=(u+\tfrac{\ri}{2})^{\ell},\qquad d_B(u)=(u-\tfrac{\ri}{2})^{L-\ell}\,.
\end{align}
As can be seen, cutting a Bethe state leads to a sum of tensor products of Bethe states. When the Bethe state is cut, each magnon is distributed into one of the two subchains, with the two possibilities summed over. This procedure generalizes to the $M$-magnon case. Using the Yangian algebra, one obtains the following formula \cite{Escobedo:2010xs}
\begin{align}
\label{eq:cutBS}
|\{u_j\}\rangle=\sum_{\alpha\cup\bar{\alpha}=\{u_i\}}H(\alpha,\bar{\alpha})|\alpha\rangle_A\otimes|\bar{\alpha}\rangle_B
\end{align}
where the sum runs over all partitions of the rapidity set $\{u_i\}$. Similar decomposition formulas for Bethe states like \eqref{eq:subchainad} can be derived straightforwardly for higher-spin and non-compact chains. The states $|\alpha\rangle_A$ and $|\bar{\alpha}\rangle_B$ are \emph{Bethe states} of the subchain $A$ and $B$ respectively. It is important to note that even if the original state is on-shell (\emph{i.e.}, $\{u_j\}$ satisfy BAE), the resulting subchain Bethe states are generally off-shell.\par

The explicit form of the factor $H(\alpha,\bar{\alpha})$ in \eqref{eq:cutBS} reads
\begin{align}
\label{eq:Hfactor}
H(\alpha,\bar{\alpha})=f^{\alpha\bar{\alpha}}d_B^{\alpha}a_A^{\bar{\alpha}}
\end{align}
where we use the shorthand notation introduced in \cite{Escobedo:2010xs},
\begin{align}
\label{eq:shorthandNotation}
F^{\alpha}\equiv\prod_{u_j\in\alpha}F(u_j),\qquad F^{\alpha\bar{\alpha}}=\prod_{u_i\in\alpha\atop v_j\in\bar{\alpha}}F(u_i,v_j)\,,
\end{align}
with the functions $a_A(u)$ and $d_B(u)$ defined in \eqref{eq:subchainad} and
\begin{align}
f(u,v)=\frac{u-v+\ri}{u-v}\,.
\end{align}

\paragraph{Block diagonal form.} The models under consideration possess a U(1) symmetry, which implies that states with different magnon numbers are orthogonal. Consequently, an $M$-magnon state decomposes further as 
\begin{align}
\label{eq:detailedSum}
|\{u_i\}\rangle=\sum_{n}\sum_{\alpha\cup\bar{\alpha}=\{u_i\}\atop |\alpha|=n}H^{(n)}(\alpha,\bar{\alpha})|\alpha\rangle_A\otimes|\bar{\alpha}\rangle_B
\end{align}
where $\alpha$ and $\bar{\alpha}$ are subsets of $\{u_j\}$ and $|\alpha|$ denotes the cardinality of $\alpha$. Note that for the non-compact chain, we have $n=0,1,\ldots,M$, while for the compact chain the allowed range is
\begin{align}
\max(0,M-2s(L-\ell))\le n\le \min(M,2s\ell),
\end{align}
as each site can host at most $2s$ magnons for the XXX$_s$ spin chain.\par 

For a fixed $n$ there are ${M\choose n}$ partitions. We label the corresponding subchain Bethe states as $|a_j^{(n)}\rangle\!\rangle_A$ and $|\bar{a}_j^{(n)}\rangle\!\rangle_B$ with $j=1,\ldots, {M\choose n}$, ordering them so that the rapidities in $|a_j^{(n)}\rangle\!\rangle$ and $|\bar{a}_j^{(n)}\rangle\!\rangle$ are complements of each other. We denote the corresponding factor by $H^{(n)}(\alpha,\bar{\alpha})\equiv H_{jj}^{(n)}$. In this notation, \eqref{eq:detailedSum} becomes
\begin{align}
\label{eq:decomposeState}
|\{u_i\}\rangle=\sum_{n}\sum_{j=1}^{{M\choose n}}H_{jj}^{(n)}|a_j^{(n)}\rangle\!\rangle_A\otimes|\bar{a}_j^{(n)}\rangle\!\rangle_B\,.
\end{align}
Orthogonality between different magnon sectors renders the wave function matrix block-diagonal. Singular values can therefore be computed independently within each sector. The entanglement entropy within each sector corresponds precisely to the symmetry-resolved entanglement entropy \cite{Goldstein:2017bua,2018PhRvB..98d1106X,Castro-Alvaredo:2024azg} with respect to the U(1) charge.\par

Although \eqref{eq:decomposeState} resembles the Schmidt form \eqref{eq:decomPsi}, there is a crucial difference: the sets $|a_j^{(n)}\rangle\!\rangle_A$ and $|\bar{a}_j^{(n)}\rangle\!\rangle_B$ are \emph{not orthonormal}. To bring the decomposition into the standard Schmidt form, we must transform to orthonormal bases via, for example, a Gram--Schmidt process. Denoting the resulting orthonormal states by $|e_j^{(n)}\rangle\!\rangle_A$ and $|\bar{e}_j^{(n)}\rangle\!\rangle_B$, we can relate the original set of states and the new set by
\begin{align}
|a_j^{(n)}\rangle\!\rangle_A=L_{jl}^{(n)}|e_l^{(n)}\rangle\!\rangle_A\,,\qquad
|\bar{a}_j^{(n)}\rangle\!\rangle_B=R_{jm}^{(n)}|\bar{e}_m^{(n)}\rangle\!\rangle_B\,.
\end{align}
For the XXX$_{\frac{1}{2}}$ chain, the dimension of the $n$-magnon Hilbert space of subsystem $A$ is ${\ell \choose n}$, while $\{|\alpha\rangle_A\}$ with $|\alpha|=n$ contains ${M\choose n}$ states. This implies that for $M\ge \ell$, the states $\{|\alpha\rangle_A\}$ are not linearly independent. For a compact spin-$s$ chain, the corresponding dimension is the coefficient of $x^n$ in $(1+x+\cdots+x^{2s})^\ell$, while for the non-compact chain it is the number of weak compositions of $n$ among $\ell$ sites. Therefore, to construct an orthonormal set, we can pick a maximal linearly independent subset of $\{|\alpha\rangle_A\}$ and then perform the Gram--Schmidt process. More details are given in Appendix~\ref{app:GSprocess}. \par

After choosing an orthonormal basis for each subchain, the decomposition of a Bethe state can be written as
\begin{align}
\label{eq:HLRdecomp}
|\{u_i\}\rangle=\sum_{n} H_{jj}^{(n)}L_{jl}^{(n)}R_{jm}^{(n)}|e_l^{(n)}\rangle\!\rangle_A\otimes|\bar{e}_m^{(n)}\rangle\!\rangle_B\,.
\end{align}
Here $H_{jk}^{(n)}$, $L_{jk}^{(n)}$ and $R_{jk}^{(n)}$ form matrices, which we denote by $\mathbf{H}^{(n)}$, $\mathbf{L}^{(n)}$ and $\mathbf{R}^{(n)}$. Notice that $\mathbf{H}^{(n)}$ is diagonal. To compute the entanglement entropy within the $n$-magnon sector in the sum \eqref{eq:HLRdecomp}, we perform the SVD for the matrix
\begin{align}
\mathbf{\Psi}^{(n)}=\frac{(\mathbf{L}^{(n)})^{\text{T}}\cdot\mathbf{H}^{(n)}\cdot\mathbf{R}^{(n)}}{\sqrt{\langle\{u_i\}|\{u_i\}\rangle}}\,.
\end{align}
To illustrate the method, we now consider two simple examples.

\paragraph{Example 1. One magnon Bethe states.}
We begin by considering the simplest one-magnon Bethe state, which admits the decomposition
\begin{align}
|\{u\}\rangle=d_B(u)|\{u\}\rangle_A\otimes|0\rangle_B+a_A(u)|0\rangle_A\otimes|\{u\}\rangle_B\,.
\end{align}
Each term corresponds to a distinct sector. Hence we may rewrite the state as
\begin{align}
|\{u\}\rangle=d_B(u)\sqrt{\langle\{u\}|\{u\}\rangle_A}|e_1^{(1)}\rangle\otimes|\bar{e}_1^{(0)}\rangle_B
+a_A(u)\sqrt{\langle\{u\}|\{u\}\rangle_B}|e_1^{(0)}\rangle_A\otimes|\bar{e}_1^{(1)}\rangle_B\,.
\end{align}
Consequently, we must determine the singular values of the coefficients
\begin{align}
\Psi^{(0)}=\frac{a_A(u)\sqrt{\langle\{u\}|\{u\}\rangle_B}}{\sqrt{\langle \{u\}|\{u\}\rangle}},\qquad 
\Psi^{(1)}=\frac{d_B(u)\sqrt{\langle\{u\}|\{u\}\rangle_A}}{\sqrt{\langle \{u\}|\{u\}\rangle}}\,.
\end{align}
Treating $\Psi^{(0)}$ and $\Psi^{(1)}$ as $1\times 1$ matrices, their singular values are
\begin{align}
s_0=\frac{|a_A(u)|\sqrt{\langle\{u\}|\{u\}\rangle_B}}{\sqrt{\langle \{u\}|\{u\}\rangle}},\qquad
s_1=\frac{|d_B(u)|\sqrt{\langle\{u\}|\{u\}\rangle_A}}{\sqrt{\langle \{u\}|\{u\}\rangle}}
\end{align}
and the corresponding eigenvalues of the reduced density matrix read
\begin{align}
\lambda_0=\frac{|a_A(u)|^2\langle\{u\}|\{u\}\rangle_B}{{\langle \{u\}|\{u\}\rangle}},\qquad 
\lambda_1=\frac{|d_B(u)|^2{\langle\{u\}|\{u\}\rangle_A}}{{\langle \{u\}|\{u\}\rangle}}\,.
\end{align}
The functions $|a(u)|^2$ and $|d(u)|^2$ are given by
\begin{equation}
\begin{aligned}
&|a(u)|^2=\left[(u+\tfrac{\ri}{2})(u^*-\tfrac{\ri}{2})\right]^L=\left(|u|^2+\text{Im}(u)+\tfrac{1}{4}\right)^L\,,\\
&|d(u)|^2=\left[(u-\tfrac{\ri}{2})(u^*+\tfrac{\ri}{2})\right]^L=\left(|u|^2-\text{Im}(u)+\tfrac{1}{4}\right)^L\,.
\end{aligned}
\end{equation}
For an off-shell one-magnon Bethe state of length $L$, the scalar product takes the form
\begin{align}
S_1(\{v\}|\{u\})=&\langle 0|\mathbf{C}(v)\mathbf{B}(u)|0\rangle=\frac{\ri(a(u)d(v)-a(v)d(u))}{u-v}\,.
\end{align}
The norm of the off-shell state follows as
\begin{equation}
\langle\{u\}|\{u\}\rangle=\lim_{v\to u}S_1(\{u^*\}|\{v\})
=\frac{1}{2\text{Im}(u)}\left(|a(u)|^2-|d(u)|^2 \right)
\end{equation}
Therefore,
\begin{align}
\lambda_0=\frac{|a_A(u)|^2\left(|a_B(u)|^2-|d_B(u)|^2\right)}{|a(u)|^2-|d(u)|^2},\qquad
\lambda_1=\frac{|d_B(u)|^2\left(|a_A(u)|^2-|d_A(u)|^2\right)}{|a(u)|^2-|d(u)|^2}\,,
\end{align}
and the entanglement entropy is given by
\begin{align}
\mathcal{S}(L,1;\ell)=-\lambda_0\log\lambda_0-\lambda_1\log\lambda_1
\end{align}
For real $u$ where $\text{Im}(u)=0$, the expressions simplify considerably. Applying l'Hospital's rule yields
\begin{align}
\langle\{u\}|\{u\}\rangle=L\,\left(|u|^2+\tfrac{1}{4}\right)^{L-1}\,.
\end{align}
In this case, $\lambda_0$ and $\lambda_1$ become independent of $u$
\begin{align}
\lambda_0=\frac{\ell_B}{L}=\frac{L-\ell}{L},\qquad \lambda_1=\frac{\ell_A}{L}=\frac{\ell}{L}\,.
\end{align}
Consequently, the entanglement entropy reduces to
\begin{align}
\label{eq:universalone}
\mathcal{S}(L,1;\ell)=-x\log x-(1-x)\log(1-x),\qquad x=\frac{\ell}{L}\,,
\end{align}
in agreement with \cite{Alba:2009th}. The fact that for $u\in\mathbb{R}$, the entanglement entropy for the one-magnon state \eqref{eq:universalone} is independent of $u$ also holds for XXX$_s$ and the non-compact spin chain.

\paragraph{Example 2. Two-magnon Bethe states.}
For two magnons the decomposition consists of four terms, organized into three magnon sectors $n=0,1,2$:
\begin{align}
\label{eq:decompose2Mag}
|\{u_1,u_2\}\rangle=&\,\phantom{+}H(\emptyset,\{u_1,u_2\})\,|0\rangle_A\otimes|\{u_1,u_2\}\rangle_B\\\nonumber
&+H(\{u_1\},\{u_2\})|\{u_1\}\rangle_A\otimes|\{u_2\}\rangle_B\\\nonumber
&+H(\{u_2\},\{u_1\})|\{u_2\}\rangle_A\otimes|\{u_1\}\rangle_B\\\nonumber
&+H(\{u_1,u_2\},\emptyset)|\{u_1,u_2\}\rangle_A\otimes|0\rangle_B\,.
\end{align}
Relabelling the four coefficients as $H_{11}^{(0)},H_{11}^{(1)},H_{22}^{(1)},H_{11}^{(2)}$ in the notation of \eqref{eq:decomposeState}, the $n=0$ and $n=2$ sectors each contain a single partition, so their coefficient ``matrices'' are the numbers
\begin{align}
{\Psi}^{(0)}=\frac{a_A(u_1)a_A(u_2)\sqrt{{_B}\langle \{u_1,u_2\}|\{u_1,u_2\}\rangle_B}}{\sqrt{\langle\{u_1,u_2\}|\{u_1,u_2\}\rangle}}\,,\quad
{\Psi}^{(2)}=\frac{d_B(u_1)d_B(u_2)\sqrt{{_A}\langle \{u_1,u_2\}|\{u_1,u_2\}\rangle_A}}{\sqrt{\langle\{u_1,u_2\}|\{u_1,u_2\}\rangle}}\,,
\end{align}
with $a_A(u)=(u+\tfrac{\ri}{2})^{\ell_A}$, $d_B(u)=(u-\tfrac{\ri}{2})^{\ell_B}$. The off-shell subchain norms ${_A}\langle\cdot\rangle_A$, ${_B}\langle\cdot\rangle_B$ can be evaluated using the Izergin--Korepin formula. Their singular values are simply $|\Psi^{(0)}|$ and $|\Psi^{(2)}|$. The $n=1$ sector is genuinely two-dimensional: orthonormalizing the pairs $\{|u_1\rangle_A,|u_2\rangle_A\}$ and $\{|u_1\rangle_B,|u_2\rangle_B\}$ by the Gram--Schmidt process produces lower-triangular matrices $\mathbf{L}^{(1)}$ and $\mathbf{R}^{(1)}$ whose entries are fixed by the off-shell overlaps ${_A}\langle\{u_i\}|\{u_j\}\rangle_A$ and ${_B}\langle\{u_i\}|\{u_j\}\rangle_B$. The explicit $2\times2$ expressions are the $m=1$ case of the general construction recorded in Appendix~\ref{app:GSprocess}. The singular values of $\mathbf{\Psi}^{(1)}=(\mathbf{L}^{(1)})^{\mathrm{T}}\,\mathbf{H}^{(1)}\,\mathbf{R}^{(1)}/\sqrt{\langle\{u_i\}|\{u_i\}\rangle}$ then follow directly.

Several remarks are in order. First, despite being fully explicit, the singular values, and hence the entanglement entropy, depend on the Bethe roots in a highly nontrivial way, even for the simple two-magnon case. Second, our method is numerically efficient. We can generalize the procedure for the two-magnon case systematically, with the details delegated to Appendices~\ref{app:IZformula} and~\ref{app:GSprocess}. A key advantage is that the computational complexity depends neither on the total length of the spin chain nor on the subsystem sizes $\ell_A$ and $\ell_B$, because these quantities enter only through the functions $a_A(u)$, $a_B(u)$, $d_A(u)$, and $d_B(u)$ and can be modified easily. Such functions appear explicitly in \eqref{eq:Hfactor} and in the scalar products of Bethe states. Finally, the approach applies universally to all rank-1 models solvable by ABA. For the XXX$_s$ chain we take $a_A(u)=(u+ \ri s)^{\ell_A}$ and $d_B(u)=(u- \ri s)^{\ell_B}$, while for the $SL(2,\mathbb{R})$ chain we take $a_A(u)=(u-\tfrac{\ri}{2})^{\ell_A}$ and $d_B(u)=(u+\tfrac{\ri}{2})^{\ell_B}$.

\section{Numerical optimization of entanglement entropy}
\label{sec:offshell}
We now consider the entanglement entropy of off-shell Bethe states, where the rapidities $\{u_j\}$ are regarded as free parameters to be optimized. We will explain the optimization procedure and some details of the implementation.

\paragraph{Variational formulation.}
As introduced in Section~\ref{sec:EEint}, an off-shell $M$-magnon Bethe state $|\{u_j\}\rangle$ is parameterized by complex rapidities $\{u_j\}\in\mathbb{C}^M$ and defines a map to a normalized many-body wavefunction. The bipartite entanglement entropy
\begin{equation}
\mathcal{S}_\ell(\{u_j\}) \equiv -\mathrm{Tr}_A\!\left(\hat{\rho}_A(\{u_j\})\log \hat{\rho}_A(\{u_j\})\right),\quad
\hat{\rho}_A(\{u_j\})=\mathrm{Tr}_B\!\left(\frac{|\{u_j\}\rangle\langle\{u_j\}|}{\langle\{u_j\}|\{u_j\}\rangle}\right),
\end{equation}
is therefore a real-valued function on the continuous parameter space $\mathbb{C}^M\simeq \mathbb{R}^{2M}$, and finding the extremal Bethe states reduces to the non-linear optimization problem
\begin{equation}
\min_{\{u_j\}\in\mathbb{C}^M} \mathcal{S}_\ell(\{u_j\}),\qquad
\max_{\{u_j\}\in\mathbb{C}^M} \mathcal{S}_\ell(\{u_j\})\, .
\end{equation}
The entropy function has many local extrema and is invariant under permutations of the $u_j$. We therefore repeat the search from many starting rapidity sets, so that extrema reached from different regions of rapidity space can be found.

\paragraph{Numerical strategy.}
\label{subsec:algorithm_summary_physics}
{
We optimize \(\mathcal{S}_\ell\) by writing each rapidity as \(u_j=x_j+\ri y_j\) and treating the \(2M\) real numbers \((x_1,\ldots,x_M,y_1,\ldots,y_M)\) as variational parameters. The basic idea is standard gradient descent or ascent: move the rapidities in the direction that decreases or increases the entropy most rapidly. A schematic update is 
\begin{equation}
q^{(t+1)}=q^{(t)}-\eta\,\nabla_q\mathcal{S}_\ell(q^{(t)})\quad\text{(min)},\qquad
q^{(t+1)}=q^{(t)}+\eta\,\nabla_q\mathcal{S}_\ell(q^{(t)})\quad\text{(max)},
\label{eq:gd_update}
\end{equation}
where \(q=(x_1,\ldots,x_M,y_1,\ldots,y_M)\), \(t\) labels the iteration, \(\nabla_q\mathcal{S}_\ell\) is the gradient with respect to these real variables, and \(\eta\) is the learning rate. In the actual numerical search we use Adam~\cite{kingma2014adam}, which implements the same descent/ascent idea but adapts the effective step size separately for each variable. The gradient is obtained by automatic differentiation in \texttt{PyTorch}~\cite{paszke2019pytorch}, because \(\mathcal{S}_\ell\) depends on the rapidities through Bethe-state construction, normalization, partial trace, and entropy evaluation. Since the landscape is non-convex, each reported extremum is the best value found from \(R=30\) independently chosen initial rapidity sets unless stated otherwise. Chain-specific numerical procedures are recorded in Appendix~\ref{sec:TechDetailOffShell}, especially Appendix~\ref{subsec:offshell_chain_specific}.
}

\paragraph{Renormalization and the singular-state locus.}

A subtle issue is the growth or decay of the unnormalized Bethe state during the sequential application of $\mathbf{B}(u_j)$: as $\{u_j\}$ changes, $|\{u_j\}\rangle$ can vary by many orders of magnitude. In most optimization problems periodic renormalization would be a routine numerical safeguard. Here it is more than that: the off-shell \emph{minimum-entropy} configurations studied in Section~\ref{sec:HXXX} lie on singular loci of the rapidity map. Along some regulating paths the unnormalized vector becomes small before normalization, while the normalized vector has a finite limit. The optimizer is therefore driven toward a boundary of the normalized Bethe-state map, and the gradient of $\mathcal{S}_\ell$ must remain well-defined on the approach. We periodically renormalize,
\begin{equation}
|\{u_j\}\rangle \ \leftarrow\ \frac{|\{u_j\}\rangle}{\sqrt{\langle\{u_j\}|\{u_j\}\rangle+\varepsilon}}\, ,
\label{eq:renorm}
\end{equation}
which leaves all physical observables invariant but keeps the differentiated numerical evaluation stable and, crucially, keeps the gradient of $\mathcal{S}_\ell$ well-defined as $\|\,|\{u_j\}\rangle\,\|$ approaches zero. The small $\varepsilon=\mathcal{O}(10^{-30})$ avoids division by zero at pathological points; its precise value is unimportant provided it lies far below any physical norm we encounter and well above the underflow scale.

The output of this procedure is a set of trial extremal rapidity configurations together with their entanglement entropies. In the main text we use those outputs only to support the physics claims; the reproducibility-level numerical details are given in Appendix~\ref{sec:TechDetailOffShell}.

\paragraph{The product-state mechanism.}
Before turning to the model-by-model results, it is worth isolating the mechanism behind one robust feature of the entropy function: for every chain studied below, the off-shell \emph{minimum} entropy vanishes on a model-dependent rapidity locus (the optimizer reaches the numerical floor). The reason is already visible in the cutting formula~\eqref{eq:cutBS}, which expands the bipartite state as a sum over the ways the $M$ magnons can be shared between the two subchains. For generic rapidities every term contributes and the Schmidt rank is large. Along suitable singular regulating paths, however, the coefficients select a single leading tensor-product component across the cut. The normalized state then factorizes and the entanglement entropy vanishes. The minimum-entropy optimization therefore does not converge to a generic nearby local minimum; it is driven onto this product-state locus. This is also why the renormalization~\eqref{eq:renorm} is essential rather than cosmetic near mixed singular approaches, where the unnormalized Bethe norm can become small. The locus is model-dependent. It is \(u_j\to\pm\ri/2\) for spin-$1/2$, compact boundary strings for XXX$_{s\ge1}$, and half-integer imaginary towers for the non-compact chain. The detailed product-state limits are collected in Appendix~\ref{app:offshell_product_state_proofs}.

\section{The XXX$_{\frac{1}{2}}$ chain}
\label{sec:HXXX}
In this section, we present results for the entanglement entropy of the XXX$_{\frac{1}{2}}$ chain. Although this model is a special case of the general XXX$_s$ chain, its importance warrants a separate discussion. We therefore present the XXX$_{\frac{1}{2}}$ results on their own here, deferring to Section~\ref{sec:highSpin} the additional higher-spin features, such as repeated roots, that do not occur for XXX$_{\frac{1}{2}}$.\par

\subsection{Model and Bethe states}
For given length $L$ and magnon number $M$, we solve the rational $Q$-system \cite{Marboe:2016yyn} to obtain all physical solutions of the BAE. Each solution is then inserted into the Bethe ansatz construction, yielding the corresponding Bethe states. These states represent primary (highest-weight) states under the $\mathfrak{su}(2)$ algebra. They can be classified into two types: \emph{regular states} and \emph{singular states}, which originate from regular and singular solutions of the BAE, respectively. The singular solutions of $M$ magnons are the solutions of the following form
\begin{align}
\{-\tfrac{\ri}{2},\tfrac{\ri}{2},u_3,\ldots,u_M\}\,.
\end{align}
Singular states require a separate treatment for two reasons. Mathematically, substituting singular solutions directly into the ABA leads to null states and divergent eigenvalues, necessitating regularization. Physically, singular solutions correspond to length-2 bound states, namely composite particles formed by two magnons.\par

\begin{figure}[h!]
\centering
\includegraphics[scale=0.5]{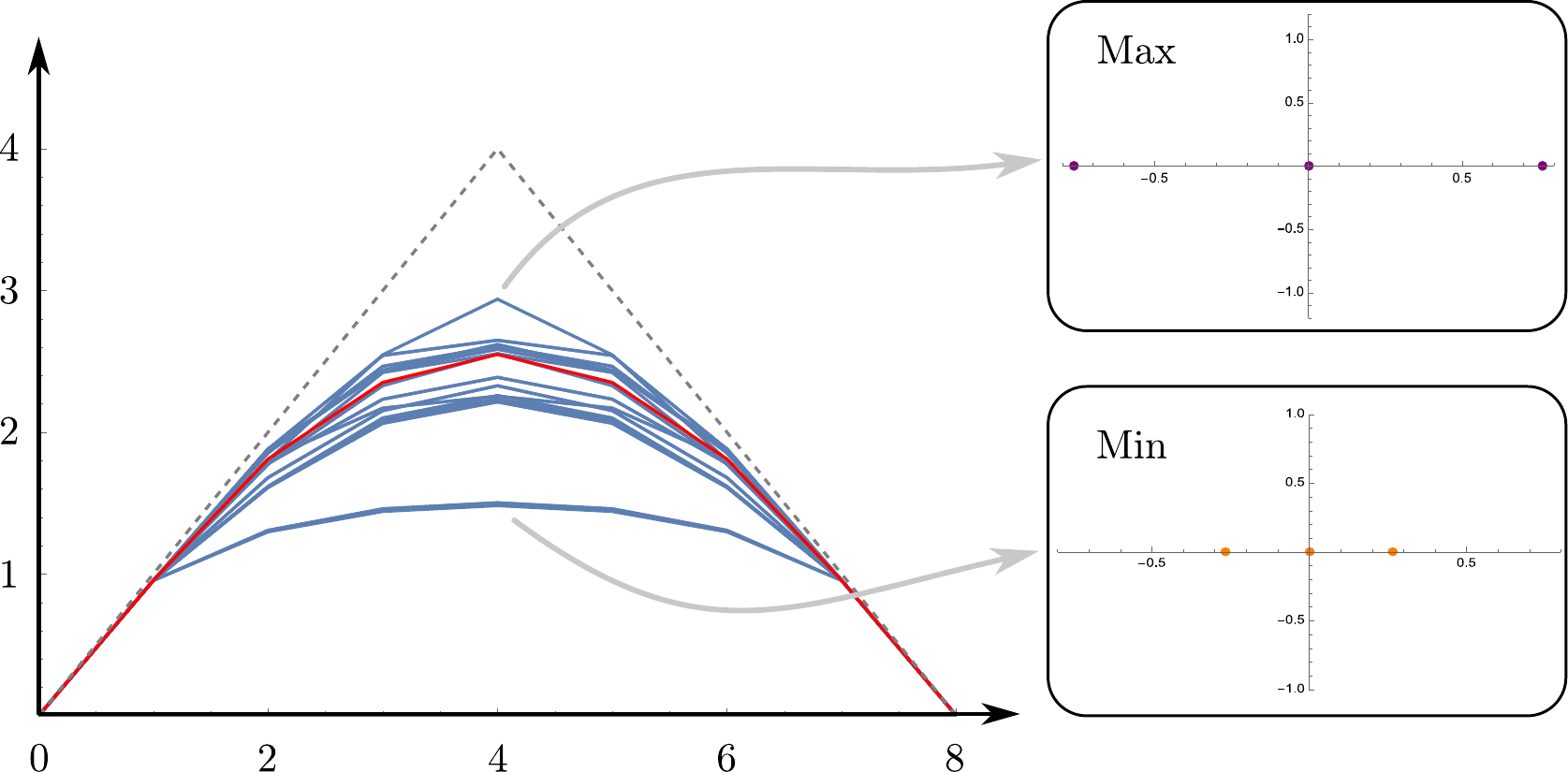}
\caption{Entanglement entropy of primary states for $L=8$, $M=3$. Blue lines correspond to Bethe states obtained from regular solutions, while the red line represents the singular solution. The dashed gray line indicates the crosscap state, which has maximal entanglement.}
\label{fig:EEL8M3}
\end{figure}

Regularizing Bethe states is a subtle procedure. As noted in prior work \cite{Nepomechie:2013mua}, a naive regularization does not yield the correct eigenstate. To obtain the proper state, we adopt the following regularization for $\pm \ri/2$
\begin{align}
u_1^{\epsilon}=\frac{\ri}{2}+\epsilon+c\,\epsilon^L,\qquad u_2^{\epsilon}=-\frac{\ri}{2}+\epsilon
\end{align}
with the parameter $c$ given by
\begin{align}
c=-\frac{2}{\ri^{L+1}}\prod_{j=3}^M\frac{u_j-\frac{3\ri}{2}}{u_j+\frac{\ri}{2}}\,.
\end{align}
Using this scheme, the Bethe state corresponding to a singular solution is expressed as
\begin{align}
|\{-\ri/2,\ri/2,u_3,\ldots,u_M\}\rangle=\lim_{\epsilon\to 0}\frac{1}{\epsilon^L}\mathbf{B}(u_1^{\epsilon})\mathbf{B}(u_2^{\epsilon})\mathbf{B}(u_3)\ldots \mathbf{B}(u_M)|0\rangle\,.
\end{align}
Applying this regularization, we can construct finite Bethe states for all physical solutions of the BAE. Then we compute the entanglement entropy for all Bethe states. Across different quantum numbers, the results exhibit similar qualitative features. A representative example is shown in Figure~\ref{fig:EEL8M3}. Each curve corresponds to the entanglement entropy of a Bethe state as a function of subsystem size. Blue curves denote regular solutions, while the red curve corresponds to the singular solution. For comparison, we also display the entanglement entropy of the maximally entangled crosscap state\footnote{The crosscap state pairs each site $j$ with its antipode $j+L/2$ into a maximally entangled Bell pair; across any contiguous bipartition its reduced density matrix is maximally mixed, so it attains the maximal entropy $\mathcal{S}=\min(\ell,L-\ell)$.} to assess how closely the entanglement growth approaches the theoretical maximum.

\subsection{Results for on-shell Bethe states}
\label{subsec:onshell_xxx}
We now present the on-shell primary Bethe-state data for the XXX$_{\frac{1}{2}}$ chain. The scan covers $L=6,\ldots,13$ and all admissible magnon sectors. Instead of displaying a separate plot and table for every chain length, we keep the main text to one representative entropy plot, Figure~\ref{fig:EEL8M3}, and one representative summary table, Table~\ref{tab:L8}. The remaining length-by-length summary tables are collected in Appendix~\ref{app:extraEE}.

To expose the main qualitative pattern in these data, we first define the Bethe quantum number for each solution $\{u_j\}$ as
\begin{align}
I_j=-\frac{L}{\pi}\arctan(2u_j)+\frac{1}{\pi}\sum_{k=1}^M\arctan(u_j-u_k),\qquad j=1,\ldots,M\,.
\end{align}
The Bethe quantum numbers $\{I_j\}$ are integers or half-integers. Note that for singular solutions, the Bethe quantum numbers are not well-defined, and in those cases we will simply give the Bethe roots themselves. 
Table~\ref{tab:L8} then displays the resulting pattern: in each magnon sector the lowest entanglement state corresponds to the solution of the BAE whose Bethe quantum numbers form a symmetric and compact distribution around the origin, while the largest entanglement is produced by more spread-out Bethe quantum numbers, and in some sectors by singular or complex-root solutions. In this table, $\mathcal{N}_R$ and $\mathcal{N}_S$ count regular and singular states, while $\mathcal{S}_{\text{min}/\text{max}}$ and $\{I_j\}_{\text{min}/\text{max}}$ give the extremal entropies and the corresponding Bethe quantum numbers. 
\begin{table}[h!]
\centering
\begin{tabular}{c|c|c|c|c|c|c}
  \hline\hline
  $M$ & $\mathcal{N}_R$ & $\mathcal{N}_S$ & $\mathcal{S}_{\text{min}}$ & $\{I_j\}_{\text{min}}$ & $\mathcal{S}_{\text{max}}$ & $\{I_j\}_{\text{max}}$\\
  \hline
  1 & 7 & 0 & 1.0  & / & 1.0 & / \\
  \hline
 2 & 19 & 1 & 1.34017  & $\{-\frac{1}{2},\frac{1}{2}\}$ & 1.99144 & $\{-\frac{3}{2},\frac{5}{2}\}_p$\\
 \hline
 3 & 27 & 1 & 1.48246  & $\{-1,0,1\}$ & 2.93892 & $\{-2, 0,2\}$ \\
 \hline
 4 & 11 & 3 & 1.51651  & $\{-\frac{3}{2},-\frac{1}{2},\frac{1}{2},\frac{3}{2}\}$ & 3.37924 & $\{-\frac{3}{2},\frac{3}{2},\frac{5}{2},\frac{7}{2}\}$ \\
   \hline\hline
\end{tabular}
\caption{Extrema of entanglement entropy and the corresponding Bethe quantum numbers for on-shell Bethe states of the XXX$_{\frac{1}{2}}$ chain with length $L=8$. Here $\{-\frac{3}{2},\frac{5}{2}\}_p$ indicates that there is another set of quantum numbers obtained by performing a parity transformation, \emph{i.e.} $\{\frac{3}{2},-\frac{5}{2}\}$.}
\label{tab:L8}
\end{table}

\paragraph{Minimum and maximum entanglement entropy.}
The minimum entanglement entropy follows a particularly simple rule. For fixed $(L,M,\ell)$ it is attained by the most compact set of allowed Bethe quantum numbers centered around the origin,
\begin{align}
I_j=-\frac{M-1}{2}+j-1,\qquad j=1,\ldots,M,
\end{align}
up to the usual integer or half-integer convention. Thus the minimizing configurations are, for example,
\begin{align}
\left\{-\frac12,\frac12\right\},\qquad
\{-1,0,1\},\qquad
\left\{-\frac32,-\frac12,\frac12,\frac32\right\}.
\end{align}
The corresponding Bethe states are also the lowest-energy states in those sectors\footnote{Here we consider the antiferromagnetic spin chain.}. The minimizing pattern is stable as the subsystem size $\ell$ is varied.

This behavior has a natural wavefunction interpretation. In the coordinate Bethe ansatz one may write schematically
\begin{align}
\Psi(x_1,\ldots,x_M)
=
\sum_{\sigma\in S_M} A_\sigma
\exp\left(\ri\sum_{j=1}^M p_{\sigma_j}x_j\right),
\end{align}
where the momenta $p_j$ are determined by the rapidities, or equivalently by the Bethe quantum numbers. Compact mode numbers correspond to nearby quasiparticle momenta. The resulting wavefunction is comparatively smooth and coherent, so after a spatial cut many restricted wavefunction components remain similar to one another. The Schmidt spectrum is then concentrated, giving a smaller entropy.

The maximum behaves differently. The maximizing Bethe quantum numbers are usually not compactly filled around the origin. They tend to occupy a wider interval, often with holes near the center and with some modes pushed toward the edge of the allowed range. For two magnons this often reduces to a widely separated pair of mode numbers. For larger $M$, however, the pattern is less universal: the maximum may involve a mixture of edge and interior modes, and in high magnon sectors it may be associated with singular or complex root solutions rather than with a simple real root pattern.

Thus the maximal state should not be interpreted simply as the state with the largest possible mode numbers. What matters is how the quasiparticle wavelengths and Bethe scattering phases decompose across the bipartition. Spread-out mode numbers generate more distinct oscillatory components and more varied scattering phases, making the left and right Schmidt vectors more distinguishable and flattening the Schmidt spectrum. 

\paragraph{Degeneracies and multiplets.}
A systematic source of degeneracy is parity. If two Bethe states are related by
\begin{align}
\{u_j\}\longmapsto \{-u_j\},
\end{align}
or equivalently by the parity transformation of their Bethe quantum numbers, then their bipartite entanglement entropies are equal. This follows from the invariance of the Schmidt spectrum under unitary transformations acting separately on the two subsystems, as reviewed in Appendix~\ref{app:EEinvariance}. In the tables such parity-related extrema are denoted by the subscript $_p$. Self-paired configurations satisfying $\{u_j\}=\{-u_j\}$ fit naturally into the same symmetry structure, although they do not generate a distinct partner. Additional accidental degeneracies can occur, but they are not fixed by parity alone.

The finite Bethe solutions are highest-weight states of the $\mathfrak{su}(2)$ algebra,
\begin{align}
S^+|\{u_j\}\rangle=0,
\end{align}
and each belongs to a spin multiplet of total spin
\begin{align}
s=\frac{L}{2}-M.
\end{align}
The descendant states are generated by
\begin{align}
(S^-)^n|\{u_j\}\rangle,\qquad n=0,1,\ldots,L-2M.
\end{align}
They share the same finite Bethe roots, supplemented by additional roots at infinity, and therefore the same energy and higher conserved charges. Their entanglement entropies need not be equal, however, because $S^-$ is a global operator rather than a product of independent local unitaries on the two subsystems.

As $n$ moves away from the highest-weight state, spin flips can be distributed across the two subsystems in more ways, so more Schmidt sectors may contribute. The two ends of the multiplet are related by the global spin-flip operator
\begin{align}
(S^-)^{L-2M}|\{u_j\}\rangle\propto
\Sigma|\{u_j\}\rangle,
\qquad
\Sigma=\prod_{j=1}^L\sigma^x_j,
\end{align}
which leaves the Schmidt spectrum invariant. Consequently the entropy along a multiplet is symmetric under
\begin{align}
n\longmapsto L-2M-n.
\end{align}
In the finite chains studied here it typically increases from the highest-weight state toward the middle of the multiplet and then decreases symmetrically toward the spin-flipped state.

\subsection{Results for off-shell Bethe states}
\label{subsec:offshell_xxx}

We now turn to the off-shell sector of the XXX$_{1/2}$ chain. Throughout this subsection the cut is the middle cut, \(\ell_*=\lfloor L/2\rfloor\). We use the abbreviation
\begin{align}
\mathcal{S}_{\max}^{\rm off}(L,M)
\equiv
\mathcal{S}_{\max}^{\rm off}(L,M;\ell_*)
\label{eq:middle_cut_smax_abbrev}
\end{align}
within this subsection. In \eqref{eq:cutBS}, each rapidity is assigned, after the spatial cut, to either the left subchain or the right subchain. For fixed \(M\), this gives at most \(2^M\) product terms. Therefore
\begin{align}
\mathrm{rank}_{\rm Schmidt}\le 2^M,\qquad
\mathcal{S}_\ell\le M .
\label{eq:xxx_half_fixed_M_partition_bound}
\end{align}
We will refer to \eqref{eq:xxx_half_fixed_M_partition_bound} as the fixed-\(M\) partition upper bound. Relaxing the Bethe equations turns each fixed \((L,M)\) sector into a continuous rapidity manifold, and the optimization of Section~\ref{sec:offshell} reveals three simple features:
\begin{enumerate}
\item at fixed \(M\), \(\mathcal{S}_{\max}^{\rm off}(L,M)\) approaches the fixed-\(M\) partition upper bound \(M\) with growing $L$;
\item at half filling, \(\mathcal{S}_{\max}^{\rm off}(L,L/2)\) grows linearly with $L$, with slope \(\simeq0.35\);
\item for \(M\ge2\), \(\mathcal{S}_{\min}^{\rm off}\to0\) as the rapidities approach the singular values \(u=\pm\ri/2\).
\end{enumerate}
The rest of the subsection presents these three statements and then compares them with the on-shell spectrum.

The middle-cut convention is not essential but gives the largest entanglement for the finite examples we scanned. At fixed \((L,M)\), \(\mathcal{S}_{\max}^{\rm off}(L,M;\ell)\) forms a symmetric arc under \(\ell\leftrightarrow L-\ell\), with the maximum at or near the middle cut and the edge cuts limited by the small number of ways to assign magnons to the two subchains. The optimizing rapidities can depend on \(\ell\), so the middle-cut entries below should be read as a fixed-cut optimization problem rather than as a single rapidity configuration evaluated at all cuts. The diagnostic \(\ell\)-scan is shown in Appendix~\ref{app:offshell_xxx_diagnostics}.

\subsubsection{Maximum off-shell entropy}

Table~\ref{tab:offshell_xxx} collects $\mathcal{S}_{\max}^{\rm off}$ as a function of $L$ and $M$.

\begin{table}[h!]
\centering
\footnotesize
\begin{tabular}{|c|c|c|c|c|c|c|c|c|c|c|c|}
  \hline
  $L \backslash M$ & 1 & 2 & 3 & 4 & 5 & 6 & 7 & 8 & 9 & 10 & 11 \\
  \hline
  3  & 1.00 & & & & & & & & & & \\
  4  & 1.00 & 2.00 & & & & & & & & & \\
  5  & 1.00 & 1.96 & & & & & & & & & \\
  6  & 1.00 & 1.99 & 2.87 & & & & & & & & \\
  7  & 1.00 & 1.98 & 2.83 & & & & & & & & \\
  8  & 1.00 & 1.99 & 2.95 & 3.47 & & & & & & & \\
  9  & 1.00 & 1.99 & 2.92 & 3.60 & & & & & & & \\
  10 & 1.00 & 2.00 & 2.96 & 3.78 & 4.11 & & & & & & \\
  11 & 1.00 & 1.99 & 2.95 & 3.82 & 4.25 & & & & & & \\
  12 & 1.00 & 2.00 & 2.96 & 3.86 & 4.61 & 4.70 & & & & & \\
  13 & 1.00 & 2.00 & 2.96 & 3.88 & 4.54 & 5.07 & & & & & \\
  14 & 1.00 & 2.00 & 2.98 & 3.88 & 4.85 & 5.39 & 5.49 & & & & \\
  15 & 1.00 & 2.00 & 2.97 & 3.90 & 4.75 & 5.47 & 5.76 & & & & \\
  16 & 1.00 & 2.00 & 2.99 & 3.96 & 4.81 & 5.57 & 6.09 & 6.28 & & & \\
  \hline
  17 & 1.00 & 2.00 & 2.98 & 3.93 & 4.81 & 5.70 & 6.29 & 6.45 & & & \\
  18 & 1.00 & 2.00 & 2.99 & 3.96 & 4.92 & 5.73 & 6.49 & 6.87 & 6.98 & & \\
  19 & 1.00 & 2.00 & 2.98 & 3.94 & 4.81 & 5.78 & 6.23 & 6.98 & 7.19 & & \\
  20 & 1.00 & 2.00 & 2.99 & 3.96 & 4.90 & 5.77 & 6.66 & 7.17 & 7.38 & 7.63 & \\
  21 & 1.00 & 2.00 & 2.98 & 3.95 & 4.83 & 5.84 & 6.64 & 7.04 & 7.58 & 7.85 & \\
  22 & 1.00 & 2.00 & 2.99 & 3.97 & 4.93 & 5.88 & 6.72 & 7.45 & 7.69 & 8.09 & 8.26 \\
  \hline
\end{tabular}
\caption{Fixed-$M$ XXX$_{\frac{1}{2}}$ off-shell maxima approach the partition upper bound \(M\), while the half-filled diagonal grows with \(L\). Entries use the middle cut \(\ell=\lfloor L/2\rfloor\).}
\label{tab:offshell_xxx}
\end{table}

Two complementary trends are visible in the table. First, for \emph{fixed} magnon number $M$, the maximum entropy saturates rapidly as $L$ grows: each column approaches a finite limit. Second, along the \emph{diagonal} $M=\lfloor L/2\rfloor$, the entropy grows linearly with $L$. We now analyze each regime, starting with the simplest.

\paragraph{One-magnon sector.}
For $M=1$, equation~\eqref{eq:cutBS} has exactly two partitions, so the reduced density matrix is a $2\times 2$ block-diagonal matrix with at most two non-vanishing eigenvalues. The entropy is therefore bounded above by \(1\), attained when both Schmidt coefficients are equal. In the one-magnon expression, changing the complex rapidity changes the relative norm of the two partition terms, so the equal-weight point is accessible. Hence $\mathcal{S}_{\max}^{\rm off}(L,M\!=\!1;\ell_*)=1$ for every $L$, matching the numerics to the displayed precision.

\begin{figure}[h!]
\centering
\includegraphics[width=0.7\textwidth]{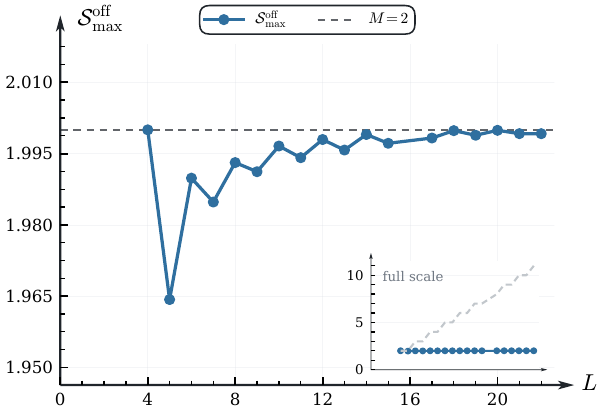}
\caption{The fixed-$M=2$ off-shell maximum approaches the partition upper bound \(2\) as $L$ grows. Data are for the XXX$_{\frac{1}{2}}$ chain at the middle cut \(\ell=\lfloor L/2\rfloor\).}
\label{fig:offshell_xxx_smax_M2}
\end{figure}

\paragraph{Fixed-$M$ saturation.}
For $M\ge 2$ and fixed, the off-shell maximum saturates to an $L$-independent limit equal to the upper bound \(M\): $\mathcal{S}_{\max}^{\rm off}\to 2$ for $M=2$ (reached already at $L=4$ and recovered to within $10^{-2}$ for all larger $L$; see Figure~\ref{fig:offshell_xxx_smax_M2}), $\approx 3$ for $M=3$, $\approx 4$ for $M=4$, and so on through the table. The upper bound would be saturated when the \(2^M\) partition terms become effectively orthogonal and carry equal weights. The off-shell data realize this condition to within the \(10^{-2}\) residual already noted. The approach to the upper bound is modulated by a mild even/odd-$L$ alternation, with even-$L$ values (middle cut $\ell=L/2$) systematically closer to saturation than neighboring odd-$L$ values (middle cut $\ell=(L-1)/2$).

\paragraph{Half-filling: volume-law growth.}
The picture changes qualitatively along the even-$L$ half-filled diagonal $M=L/2$. Here the number of magnons scales with the chain and the fixed-\(M\) partition upper bound itself grows with $L$. A linear fit to the even-$L$ diagonal data from $L=4$ to $L=22$ gives\footnote{{Here \(R^2\) denotes the coefficient of determination of the linear fit.}}
\begin{align}
\label{eq:xxx_linear_fit}
\mathcal{S}_{\max}^{\rm off}\!\left(L,\,M\!=\!L/2;\ell_*\right) \approx 0.347\, L + 0.67\,,\qquad R^2=0.999\,.
\end{align}
This is a volume law: the off-shell maximum scales with subsystem size. The slope $0.347$ is below the local Hilbert space upper-bound slope \(1/2\), set by $\ell=L/2$ spin-$1/2$ sites, so the data show extensive but non-maximal entanglement. This submaximal slope reflects the algebraic constraints of the Bethe vector parametrization: even without imposing the BAE constraints, the state is not a generic vector in the Hilbert space. The same scaling is visible in the odd chain lengths: for odd \(L\), \(M=(L-1)/2\), and a fit over \(3\le L\le21\) gives $\mathcal{S}_{\max}^{\rm off}(L,(L-1)/2;\ell_*)\approx 0.374\,L+0.10$ with $R^2=0.998$. A combined fit over \(3\le L\le22\) gives slope $0.362$ and intercept $0.36$ with $R^2=0.994$. The even and odd sequences share comparable slopes, so the linear scaling is not an artifact of restricting to even $L$. The small offset between the two lines traces back to the even/odd Schmidt-dimension mismatch between $\ell=L/2$ and $\ell=(L-1)/2$ bipartitions.

\begin{figure}[h!]
\centering
\includegraphics[width=0.7\textwidth]{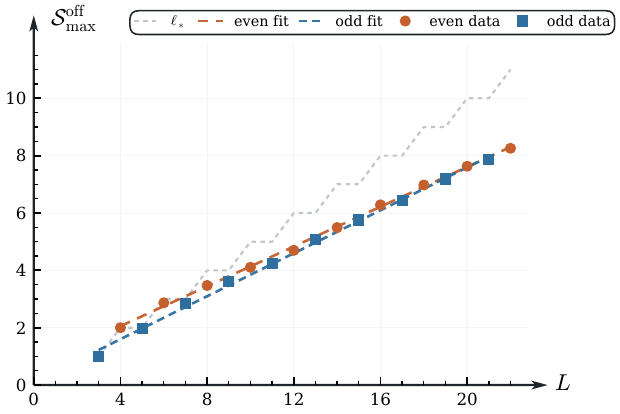}
\caption{Half-filled XXX$_{1/2}$ off-shell maxima grow linearly with $L$. The dashed lines are separate linear fits to the even-\(L\) data for \(4\le L\le22\) and the odd-\(L\) data for \(3\le L\le21\), with \(\ell=\lfloor L/2\rfloor\).}
\label{fig:offshell_xxx_smax_half}
\end{figure}

The optimizing rapidities for the off-shell maxima appear less universal than the entropy extrema themselves, much like the maximizing Bethe quantum numbers in the on-shell scan. We therefore focus here on the entropy values and their scaling with \(L\) and \(M\). The aggregate root-cloud plot in Appendix~\ref{app:offshell_xxx_diagnostics} records the sampled maximizing rapidities for reference.

\subsubsection{Minimum off-shell entropy}

The minimum sector is more tightly constrained. For every $L$ and every $M\ge 2$ that we have surveyed, the optimizer drives $\mathcal{S}_\ell$ to numerical zero: the best residual values are $\mathcal{S}_{\min}^{\rm off}\lesssim 10^{-4}$, consistent with a true vanishing entropy modulo finite-precision arithmetic. This is not a new lower bound to be optimized against: entropy is non-negative, and there are explicit off-shell rapidities whose normalized Bethe vector is a product state. For example, taking all rapidities to the one-sided singular value \(u_j=\ri/2\) gives
\begin{align}
\mathbf{B}(\ri/2)^M|0\rangle
\propto
\mathbf{S}_{L-M+1}^-\cdots \mathbf{S}_{L}^-|0\rangle,
\qquad 1\le M\le L,
\end{align}
and the singular value \(u_j=-\ri/2\) gives the analogous product vector localized at the left edge. These vectors have Schmidt rank one across every bipartition, hence \(\mathcal{S}_{\ell}=0\). The optimizer also finds mixed singular configurations, with some rapidities near \(+\ri/2\) and others near \(-\ri/2\). Their limiting state depends on how the rapidities approach \(u=\pm\ri/2\). This point is discussed in detail in Appendix~\ref{subsec:xxx_half_product_approaches}.

\subsubsection{Equal-entropy sets and real-slice visualization}

For fixed \(L\), \(M\), \(\ell\), and a fixed entropy value, the equal-entropy set is the complex off-shell level set \(\mathcal{S}_\ell(u_1,\ldots,u_M)=\mathrm{const}\) with \(u_j\in\mathbb{C}\). This is the geometric object we want to probe, but it cannot be visualized directly. We therefore first restrict to the \emph{real-rapidity slice} \(u_j\in\mathbb{R}\). This is a visualization slice of the off-shell entropy map, not an on-shell real-root condition. In the discussion below, components mean connected components of the plotted real level sets. Since the real slice can miss connections that pass through complex rapidities, we also test whether selected real-slice components can be connected in the complex off-shell space; this test is presented below, following Figure~\ref{fig:offshell_xxx_iso}.

\begin{figure}[h!]
\centering
\begin{subfigure}[t]{0.310\textwidth}
\centering
\includegraphics[width=\linewidth]{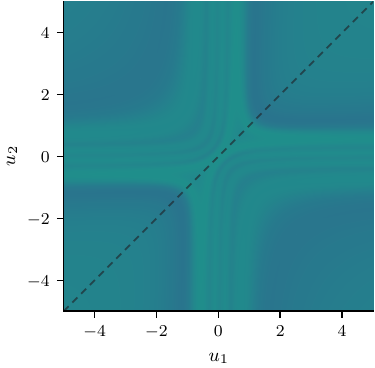}
\caption{\(\ell=1\)}
\end{subfigure}
\hspace{0.012\textwidth}
\begin{subfigure}[t]{0.265\textwidth}
\centering
\includegraphics[width=\linewidth]{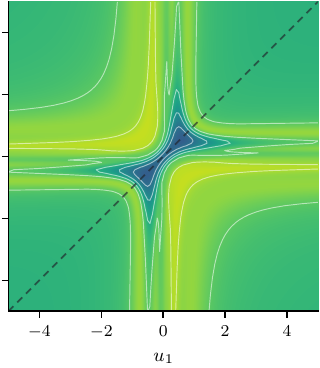}
\caption{\(\ell=2\)}
\end{subfigure}
\hspace{0.012\textwidth}
\begin{subfigure}[t]{0.265\textwidth}
\centering
\includegraphics[width=\linewidth]{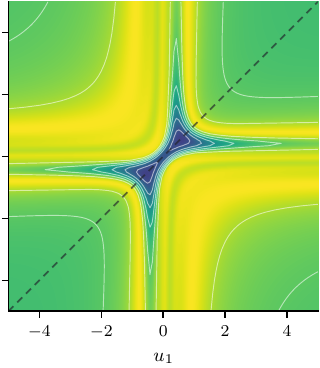}
\caption{\(\ell=3\)}
\end{subfigure}
\hspace{0.010\textwidth}
\begin{minipage}[t]{0.056\textwidth}
\centering
\raisebox{2.0em}[0pt][0pt]{\includegraphics[width=\linewidth]{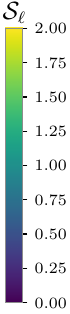}}
\end{minipage}
\caption{The real-rapidity slice shows how the $M=2$ off-shell entropy approaches its maximum along unbounded directions. The plots show $\mathcal{S}_\ell(u_1,u_2)$ for the XXX$_{1/2}$ chain at $L=6$, \(M=2\), and \(\ell=1,2,3\), with a common color scale; white curves are iso-entropy contours and the dashed line marks $u_1=u_2$.}
\label{fig:offshell_xxx_heatmap}
\end{figure}

\paragraph{Two magnons.}
For $M=2$, Figure~\ref{fig:offshell_xxx_heatmap} shows the entropy function $\mathcal{S}_\ell(u_1,u_2)$ on the \((u_1,u_2)\) plane for $L=6$ at $\ell=1,2,3$, with a common color scale. The middle-cut panel $\ell=3$ has the widest range: lower level sets are closed curves surrounding local minima, while higher level sets become unbounded and extend to $|u_j|\to\infty$. The maximum value $\mathcal{S}_{\max}^{\rm off}(6,2;3)=1.99$ is approached along asymptotic rays where at least one rapidity diverges, consistent with the fact that the global maximum from Table~\ref{tab:offshell_xxx} is not in general attained at finite real rapidities.

\paragraph{Three magnons.}
For $M=3$, the real-slice iso-entropy level sets are two-dimensional surfaces in $\mathbb{R}^3$. Near the off-shell maximum $\mathcal{S}_{\max}^{\rm off}(6,3;3)=2.87$, the real level set separates into six visible connected components associated with the $S_3$ permutations of the optimizing rapidities. As the target entropy decreases, these connected components deform and merge within the real slice; the observed connected-component count reaches a maximum near $\mathcal{S}\approx 2.0$ and then collapses to a single visible surface around $\mathcal{S}\approx 1.5$. These real-slice components provide concrete component pairs for the complex-connectivity test below, but their disconnection in \(\mathbb{R}^3\) is not by itself a statement about the full complex level set.

\begin{figure}[h!]
\centering
\begin{subfigure}[t]{0.31\textwidth}
\centering
\includegraphics[width=\textwidth]{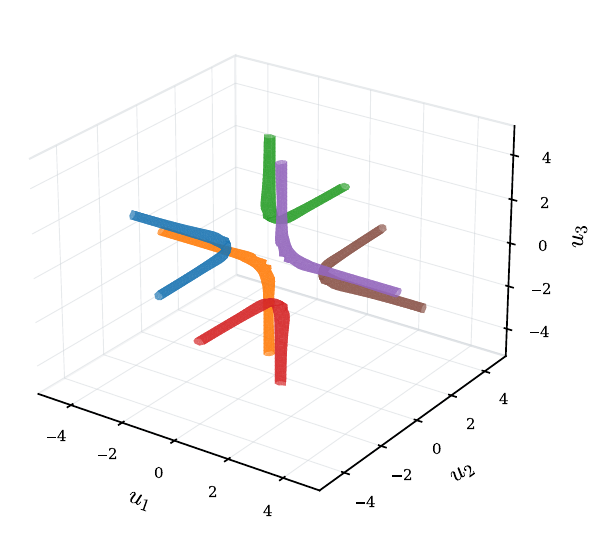}
\caption{$\mathcal{S}=2.70$}
\label{fig:offshell_xxx_iso_high}
\end{subfigure}
\hfill
\begin{subfigure}[t]{0.31\textwidth}
\centering
\includegraphics[width=\textwidth]{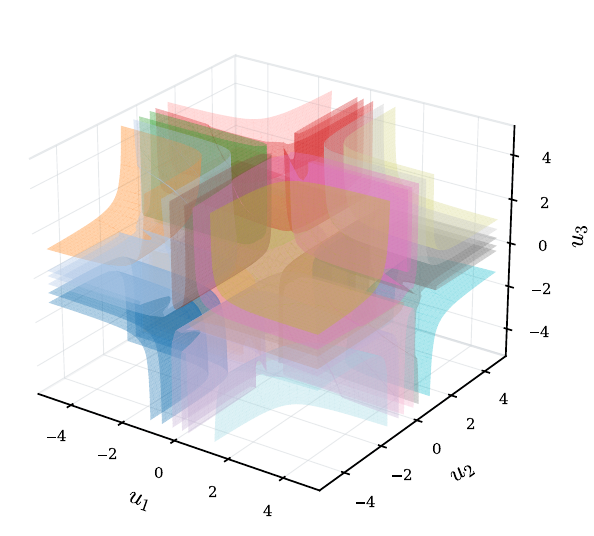}
\caption{$\mathcal{S}=2.00$}
\label{fig:offshell_xxx_iso_peak}
\end{subfigure}
\hfill
\begin{subfigure}[t]{0.31\textwidth}
\centering
\includegraphics[width=\textwidth]{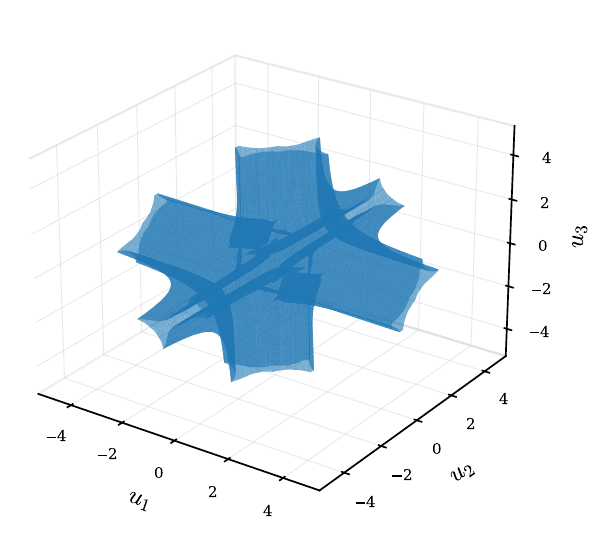}
\caption{$\mathcal{S}=1.50$}
\label{fig:offshell_xxx_iso_low}
\end{subfigure}
\caption{Real-slice iso-entropy surfaces provide the component pairs used in the complex-connectivity test. The panels show $\mathcal{S}_{\ell=3}(u_1,u_2,u_3)=\text{const}$ for $L=6$ and \(M=3\), restricted to $(u_1,u_2,u_3)\in\mathbb{R}^3$. At \(\mathcal{S}=2.70\), the six visible components are the \(S_3\) permutation orbits of the optimum; near \(\mathcal{S}=2.00\), the component count is close to its observed peak; by \(\mathcal{S}=1.50\), the visible real-slice surface has merged into one component. These are real-slice visualization features, not a topological claim about the full complex rapidity level set.}
\label{fig:offshell_xxx_iso}
\end{figure}

To test whether these real-slice components are separated only by the visualization slice, we searched for a complex path \(u_j(t)\in\mathbb{C}\), \(t\in[0,1]\), connecting points on different visible real components while keeping the entropy fixed. The path was represented by \(31\) sample points; a candidate was accepted only if the node errors were below \(10^{-3}\), interpolation between adjacent nodes stayed within \(5\times10^{-3}\) of the target entropy, and the rapidities remained in the bounded region \(\max_j|u_j|\le8\). The search used several initialization families, but no accepted bridge was found for the tested component pairs at \(L=6\), \(M=3\), and \(\ell=3\). Thus the real-slice disconnection in Figure~\ref{fig:offshell_xxx_iso} is not merely removed by complex directions in this bounded numerical test, although this remains numerical evidence rather than a proof of the full complex topology.

\subsection{Summary and on-shell/off-shell comparison}

The XXX$_{\frac{1}{2}}$ off-shell results have two main extremal mechanisms. For the maximum, the partition expansion in \eqref{eq:cutBS} gives the fixed-\(M\) upper bound \(M\), and the optimizer approaches this bound at fixed \(M\); when \(M\) scales with \(L\), the maximum instead shows volume-law growth. For the minimum, the mechanism is different: as rapidities approach the singular values \(u_j=\pm\ri/2\), zeros in the partition coefficients can leave a single product component after normalization, so \(\mathcal{S}_{\min}^{\rm off}\to0\). The real-slice plots and bounded complex-path search address a separate issue, namely the geometry of equal-entropy rapidity sets.

This structure also clarifies the comparison with the on-shell spectrum. At $L=8$, the on-shell and off-shell maxima are already close: the gap is at most about \(0.1\) across \(M=1,\ldots,4\). The main difference is at the minimum, where compact on-shell Bethe roots remain entangled but off-shell rapidities can reach the product-state locus at \(u_j\to\pm\ri/2\). Thus, for XXX$_{\frac{1}{2}}$, the Bethe equations mostly exclude the low-entanglement product region, while the off-shell maximum supplies the simpler organizing structure: fixed \(M\) partition upper-bound saturation and half-filled volume-law growth.

\section{The XXX$_s$ spin chain}
\label{sec:highSpin}

The XXX$_s$ chain tests which parts of the XXX$_{\frac{1}{2}}$ picture survive when each site carries a larger compact representation.

\subsection{Model and Bethe states}
\paragraph{Hilbert space and Hamiltonian.} We study a periodic spin chain of length $L$, where the local Hilbert space at site $n$ is $\mathbb{C}^{2s+1}$, carrying a spin-$s$ representation of $\mathfrak{su}(2)$. The integrable higher-spin XXX Hamiltonian can be written as a polynomial in the two-site spin coupling, following the standard ABA construction \cite{Sklyanin:1979pfu,Takhtajan:1979iv,Faddeev:1996iy}:
\begin{align}
\label{eq:HXXXs}
H_{\text{XXX}_s}=\sum_{j=1}^L h_{j,j+1},\qquad h_{j,j+1}=Q_{2s}(\vec{S}_j\cdot\vec{S}_{j+1})
\end{align}
where
\begin{align}
Q_{2s}(x)=\sum_{j=1}^{2s}h(j)\prod_{l=0\atop l\ne j}^{2s}\frac{x-x_l}{x_j-x_l},\qquad x_l=\frac{1}{2}l(l+1)-s(s+1)
\end{align}
and $h(j)$ is the $j$-th harmonic number,
\begin{align}
h(j)=\sum_{k=1}^j\frac{1}{k}\,.
\end{align}
The operator $\vec{S}_j\cdot\vec{S}_{j+1}$ can be written as
\begin{align}
\vec{S}_j\cdot\vec{S}_{j+1}=\frac{1}{2}(S_j^+S_{j+1}^- +S_j^- S_{j+1}^+)+S_j^zS_{j+1}^z\,.
\end{align}
The spin operators $S_j^{\alpha}$ are $(2s+1)\times (2s+1)$ matrices. In the standard representation theoretic notation, states of the spin-$s$ representation are labeled by $|s,m\rangle$ with $m=-s,-s+1,\ldots,s$, and satisfy
\begin{equation}
\begin{aligned}
S^+|s,m\rangle=&\sqrt{s(s+1)-m(m+1)}|s,m+1\rangle\,,\\
S^-|s,m\rangle=&\sqrt{s(s+1)-m(m-1)}|s,m-1\rangle\,,\\
S^z|s,m\rangle=& m|s,m\rangle\,.
\end{aligned}
\end{equation}
In the spin chain language, it is more convenient to define the local vacuum at site $j$ as the highest-weight state $|0\rangle_j \equiv |s,s\rangle_j$, satisfying $S^+_j|0\rangle_j = 0$. The remaining basis states $|n\rangle_j$ (with $n=0,1,\dots,2s$) are generated by acting with $S^-_j$:
\begin{equation}
\begin{aligned}
S^-_j|n\rangle_j=&\sqrt{(2s-n)(n+1)}|n+1\rangle_j\,,\\
S^+_j|n\rangle_j=&\sqrt{(2s-n+1)n}|n-1\rangle_j\,,\\
S^z_j|n\rangle_j=&(s-n)|n\rangle_j\,.
\end{aligned}
\end{equation}
The Hilbert space of the spin chain is then spanned by the orthonormal basis
\begin{align}
|n_1,n_2,\ldots,n_L\rangle,\qquad n_j=0,1,\ldots,2s,
\label{eq:xxxs_occupation_basis}
\end{align}
with $\langle m_1,\ldots,m_L|n_1,\ldots,n_L\rangle = \prod_{j=1}^L \delta_{m_j,n_j}$.

\paragraph{Bethe states.} We use ABA to construct the eigenstates of the XXX$_s$ spin chain. The Lax operator takes the same form as \eqref{eq:defLax}, with the spin operators specialized to the spin-$s$ representation\footnote{Here we always take the auxiliary space to be $\mathbb{C}_a^2$. The dimension of the quantum space and the auxiliary space are different for $s\ge 1$. Note that we cannot derive the Hamiltonian of the XXX$_s$ spin chain with this choice of auxiliary space; nevertheless, we can still use the corresponding monodromy matrix to construct the eigenstates.}. We denote the Lax operator by $\mathbf{L}_{an}^{(s)}(u)$, which acts on a given state $|m\rangle_n$ as
\begin{align}
\mathbf{L}_{an}^{(s)}(u)|m\rangle_n=\left(\begin{array}{cc} \left(u+\ri(s-m)\right)|m\rangle_n & \ri\sqrt{(2s-m)(m+1)}|m+1\rangle_n\\
\ri\sqrt{(2s-m+1)m}|m-1\rangle_n & \left(u-\ri(s-m)\right)|m\rangle_n  \end{array}\right)_a
\end{align}
 We denote the corresponding $B$-operator by $\mathbf{B}^{(s)}(u)$ and the Bethe state is constructed by
 \begin{align}
 \label{eq:BStatess}
 |\{u_j\}\rangle^{(s)}=\mathbf{B}^{(s)}(u_1)\ldots \mathbf{B}^{(s)}(u_M)|0\rangle\,,
 \end{align}
 where the pseudovacuum is $|0\rangle=|0\rangle_1\otimes|0\rangle_2\otimes\cdots\otimes|0\rangle_L$. The Bethe states \eqref{eq:BStatess} are eigenstates of the Hamiltonian if the corresponding rapidities $\{u_j\}$ satisfy the following BAE
\begin{align}
\label{eq:BAEspin-s}
\left(\frac{u_j+\ri s}{u_j-\ri s}\right)^L=\prod_{k=1\atop k\ne j}^M\frac{u_j-u_k+\ri}{u_j-u_k-\ri},\qquad j=1,\ldots,M\,.
\end{align}
The overall normalization of the Hamiltonian is not needed in the entropy analysis below: all on-shell data use the Bethe vectors determined by \eqref{eq:BAEspin-s}, and the off-shell analysis uses the same ABA vectors with unconstrained rapidities.

\paragraph{Special solutions and completeness.} Finding all physical solutions of the BAE \eqref{eq:BAEspin-s} is even more subtle than in the XXX$_{\frac{1}{2}}$ case. Two types of special solutions require particular attention. The first kind of solution is the \emph{singular solution}, which is similar to the XXX$_{\frac{1}{2}}$ case and takes the following form
\begin{align}
\label{eq:singularSpin-s}
\{-\ri s,-\ri(s-1),\ldots,\ri s,u_{2s+2},\ldots,u_{M}\}
\end{align}
namely the Bethe roots contain an exact $(2s+1)$-string. In addition to singular solutions, there are also \emph{strange solutions}, namely physical solutions with repeated roots, such as $\{u,u,u_3,\ldots,u_M\}$. If the repeated roots appear within the string $\{-\ri s,\ldots,\ri s\}$, the solution is called a \emph{strange singular solution}; otherwise, it is a \emph{strange regular solution}. Characterizing which singular and strange solutions are physical is subtle; we use the higher-spin completeness and physicality criteria of \cite{Hao:2013rza,Hou:2023ndn}.\par

Constructing the eigenstates corresponding to the singular solutions, as in the XXX$_{\frac{1}{2}}$ case \cite{Nepomechie:2013mua,Nepomechie:2014hma}, also requires careful regularization. We first consider the singular solutions without repeated roots; the strange singular solutions are more subtle and will be considered separately. We regularize the singular solution \eqref{eq:singularSpin-s} by
\begin{align}
u_j^{\epsilon}=\ri (s+1-j)+\epsilon+c_j\,\epsilon^L,\qquad j=1,\ldots,2s+1\,,
\end{align}
where $\epsilon$ is a small parameter, and $c_1,\ldots,c_{2s+1}$ are constants independent of $\epsilon$. They are not completely fixed, but satisfy the following conditions, which can be derived by requiring the unwanted terms in the ABA process to vanish at order $\epsilon^L$.
\begin{align}
\label{eq:recurseC}
c_2-c_1=&\,-\frac{(2s+1)}{(2\ri s)^{L-1}}\prod_{j=2s+2}^M\frac{u_j-\ri(s+1)}{u_j-\ri(s-1)}\,,\\\nonumber
c_{k+1}-c_{k}=&\,-(c_{k}-c_{k-1})\left(\frac{1-k}{2s+1-k}\right)^{L-1}\left(\frac{2s+2-k}{k}\right)\prod_{j=2s+2}^M\frac{u_j-\ri(s+2-k)}{u_j-\ri(s-k)}\,,\\\nonumber
c_{2s+1}-c_{2s}=&\,\frac{(2s+1)}{(-2\ri s)^{L-1}}\prod_{j=2s+2}^M\frac{u_j+\ri(s+1)}{u_j+\ri(s-1)}\,.
\end{align}
For the solution to be physical, the rest of the Bethe roots must satisfy
\begin{align}
\left[(-1)^{2s}\prod_{k=2s+2}^M\left(\frac{u_k+\ri s}{u_k-\ri s}\right)\right]^L=1\,.
\end{align}
As a convention, we can choose $c_1=0$ and fix the rest of $c_j$ using the recursion relations \eqref{eq:recurseC}. Using this regularization, the eigenstate is given by
\begin{align}
\lim_{\epsilon\to 0}\frac{1}{\epsilon^L}\mathbf{B}^{(s)}(u_1^{\epsilon})\ldots \mathbf{B}^{(s)}(u_{2s+1}^{\epsilon})
\mathbf{B}^{(s)}(u_{2s+2})\ldots \mathbf{B}^{(s)}(u_M)|0\rangle\,.
\end{align}

Furthermore, taking the $\mathrm{XXX}_1$ spin chain as an example, there exists another type of singular configuration characterized by repeated roots within a singular string:
\begin{equation}
    \label{eq:1_2_1}
    \left\{\ri, 0, 0, -\ri, \dots\right\}.
\end{equation}
The regularization scheme for this specific arrangement of Bethe roots can be formulated as follows:
\begin{align}
    \label{eq:regularization_121}
    &u^{\epsilon}_1 = \ri + \epsilon, \\
    &u^{\epsilon}_{(2,1)} = 0 + \epsilon + c_{2,1}\epsilon^{L/2}, \\
    &u^{\epsilon}_{(2,2)} = 0 + \epsilon + c_{2,2}\epsilon^{L/2}, \\
    &u^{\epsilon}_3 = -\ri + \epsilon + c_{3,1}\epsilon^{L}.
\end{align}
The action of the transfer matrix $\mathrm{T}(v)$ on the Bethe state is given by
\begin{align}
    \label{eq:ABA}
    &\mathrm{T}(v)\mathbf{B}^{(s)}(u_1)\dots \mathbf{B}^{(s)}(u_M)|0\rangle \nonumber \\ 
    &= t(v)\mathbf{B}^{(s)}(u_1)\dots \mathbf{B}^{(s)}(u_M)|0\rangle \nonumber \\
    &\quad + \sum_k \mathrm{F}^{(s)}_{k}(v,\bar{u})\mathbf{B}^{(s)}(v)\mathbf{B}^{(s)}(u_1)\dots \widehat{\mathbf{B}}^{(s)}(u_k)\dots\mathbf{B}^{(s)}(u_M)|0\rangle,
\end{align}
where $\widehat{\mathbf{B}}^{(s)}(u_k)$ means the absence of the corresponding operator and the coefficient $\mathrm{F}^{(s)}_k(v,\bar{u})$ is given by
\begin{align}
    \mathrm{F}^{(s)}_k(v,\bar{u}) = \frac{\ri}{v - u_k} \left[ \left({u_k + \ri s} \right)^L\prod_{j \neq k}^{M} \frac{u_k - u_j - \ri}{u_k - u_j} - \left({u_k - \ri s} \right)^L \prod_{j \neq k}^{M} \frac{u_k - u_j + \ri}{u_k - u_j} \right].
\end{align}
The regulator coefficients for a general spin-$s$ chain and arbitrary multiplicity can be determined by imposing the condition:
\begin{equation}
    \label{eq:higherMultipicity}
    \sum_{j=1}^{r_k} \mathrm{F}^{(s)}_{(k,j)}(v,\bar{u}) \prod_{\substack{i=1 \\ i \neq j}}^{r_k} c_{k,i} \sim \epsilon^{\frac{L}{r_k} + 1},
\end{equation}
where $k$ identifies the specific repeated root of multiplicity $r_k$, and $j$ enumerates the distinct individual roots branching from it.
Further details are provided in Appendix~\ref{app:GReg}. For the specific configuration presented above, the constraint associated with the repeated roots becomes:
\begin{equation}
    c_{2,1}\mathrm{F}^{(1)}_{(2,2)}(v,\bar{u}) + c_{2,2}\mathrm{F}^{(1)}_{(2,1)}(v,\bar{u}) \sim \epsilon^{L/2+1}.
\end{equation}
This condition yields an algebraic equation involving the coefficients $\{c_{2,1}c_{2,2}, c_{3,1}\}$. As a concrete illustration, for a 4-magnon strange singular solution in an $\mathrm{XXX}_1$ spin chain of length $L=6$, this reduces to the system:
\begin{align}
    &7 c_{2,1} c_{2,2} + \ri c_{3,1} = 0, \\
    &16 c_{2,1} c_{2,2} - 3 = 0.
\end{align}

\subsection{Results for on-shell Bethe states}
\label{subsec:onshell_xxxs}
We now present the on-shell primary Bethe state data for the higher-spin XXX$_s$ chain, focusing on \(s=1\) and \(s=\frac{3}{2}\). The full scan contains several chain lengths; we keep one representative entropy plot and one representative extrema table for each spin in the main text: Figure~\ref{fig:S1L8} and Table~\ref{tab:S1L8} for the XXX$_1$ chain, and Figure~\ref{fig:S32L6} and Table~\ref{tab:S32L6} for the XXX$_{3/2}$ chain. The remaining summary tables are collected in Appendix~\ref{app:extraEE}.

In these tables, $\mathcal{N}_R$ and $\mathcal{N}_S$ count regular and singular states, while $\mathcal{S}_{\text{min}/\text{max}}$ and $\{I_j\}_{\text{min}/\text{max}}$ give the extremal entropies and the corresponding Bethe quantum numbers. When an extremum comes from a singular solution, the rapidities are listed directly.

\subsubsection{Spin-1 chain}
\begin{figure}[h!]
\centering
\includegraphics[width=0.9\textwidth]{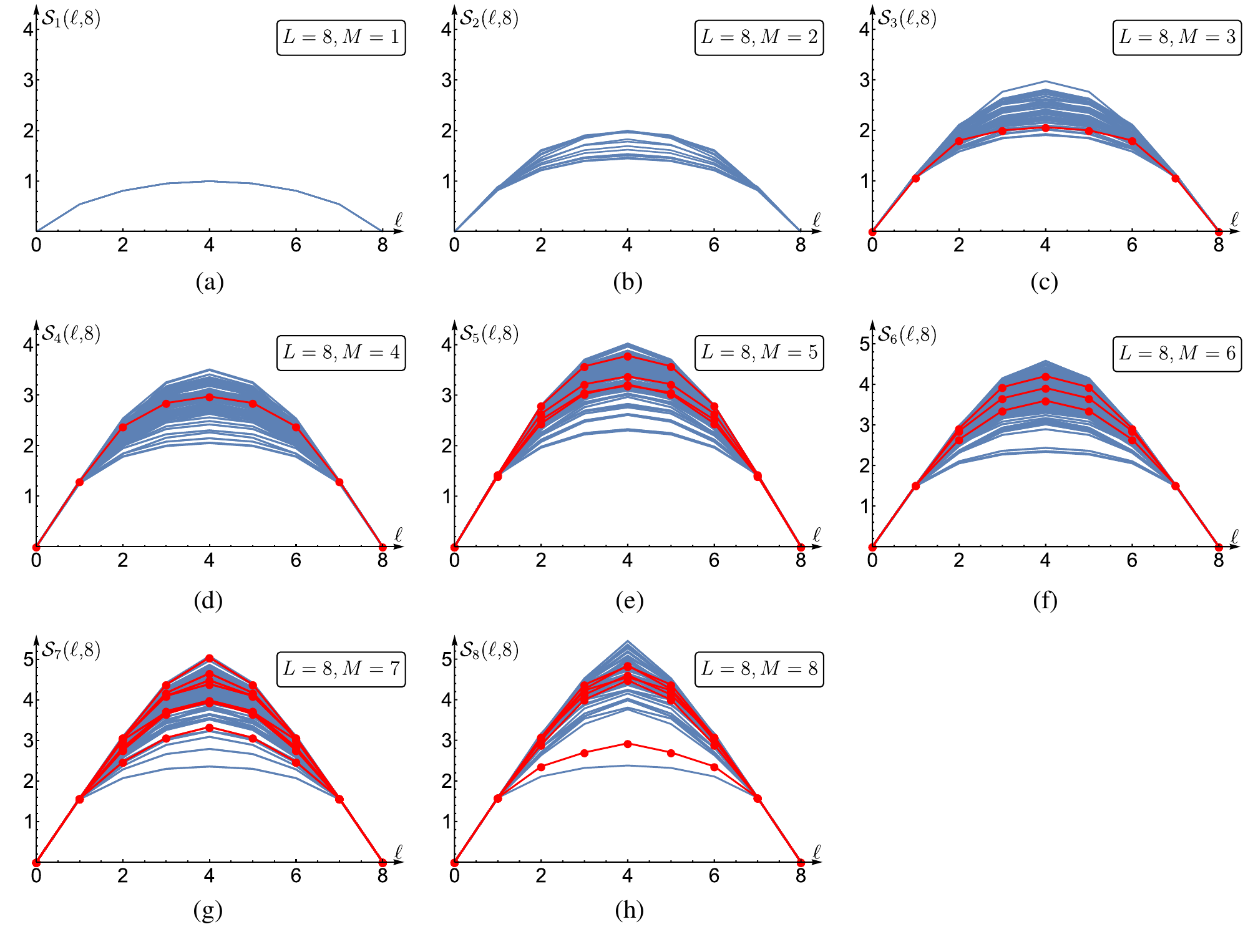}
\caption{The entanglement entropy of primary states of the XXX$_1$ chain at $L=8$. The blue lines correspond to regular solutions, while red lines correspond to singular solutions. The result shows a wider entropy range than XXX$_{\frac{1}{2}}$ because the local Hilbert space is larger. }
\label{fig:S1L8}
\end{figure}

The representative XXX$_1$ data at $L=8$ are shown in Figure~\ref{fig:S1L8} and summarized in Table~\ref{tab:S1L8}. The lower-length summary tables, $L=4,\ldots,7$, can be found in Appendix~\ref{app:extraEE}.

\begin{table}[h!]
\centering
\small
\begin{adjustbox}{max width=\textwidth}
\begin{tabular}{|c|c|c|c|c|c|c|}
  \hline
  $M$ & $\mathcal{N}_R$ & $\mathcal{N}_S$ & $\mathcal{S}_{\text{min}}$ & $\{I_j\}_{\text{min}}$ & $\mathcal{S}_{\text{max}}$ & $\{I_j\}_{\text{max}}$\\
  \hline
  1 & 7 & 0 & 1  & / & 1 & / \\
  \hline
 2 & 28 & 0 & 1.44746 & $\{-\frac{1}{2}, \frac{1}{2}\}$ & 1.99937 & $\{-\frac{5}{2}, \frac{3}{2}\}_{p}$\\
 \hline
 3 & 75 & 1 & 1.90387  & $\{-1, 0, 1\}$ & 2.97346 & $\{-2, 0, 2\}$ \\
 \hline
 4 & 153 & 1 & 2.04935 & $\{-\frac{1}{2}, -\frac{1}{2}, \frac{1}{2}, \frac{1}{2}\}$ & 3.51027 & $\{-\frac{3}{2}, -\frac{3}{2}, \frac{3}{2}, \frac{3}{2}\}_{p}$ \\
   \hline
5 & 234 & 4 & 2.30034 & $\{0, 0, 0, 0,0\}$ & 4.0209 & $\{-2, -1, 0, 1, 2\}$ \\
   \hline
6 & 277 & 3 & 2.33389 & $\{-\frac{1}{2}, -\frac{1}{2}, -\frac{1}{2}, \frac{1}{2}, \frac{1}{2}, \frac{1}{2}\}$ & 4.57923 & $\{-\frac{7}{2}, -\frac{5}{2}, -\frac{3}{2}, -\frac{1}{2}, -\frac{1}{2}, \frac{1}{2}\}_{p}$ \\
   \hline
7 & 230 & 18 & 2.35415 & $\{0, 0, 0, 0, 0, 0, 0\}$ & 5.07287 & $\left\{-3, -3, -2, -2,-2, 0, 1\right\}_{p}$ \\
   \hline
8 & 85 & 6 & 2.38015 & $\{-\frac{1}{2}, -\frac{1}{2}, -\frac{1}{2}, -\frac{1}{2}, \frac{1}{2}, \frac{1}{2}, \frac{1}{2}, \frac{1}{2}\}$ & 5.44666 & $\{-\frac{7}{2}, -\frac{5}{2}, -\frac{5}{2}, -\frac{3}{2}, -\frac{1}{2}, -\frac{1}{2}, \frac{1}{2}, \frac{1}{2}\}_{p}$ \\
   \hline
\end{tabular}
\end{adjustbox}
\caption{Data for on-shell Bethe states of XXX$_1$ chain at $L=8$. The minima remain tied to compact or repeated central mode numbers, while maxima can involve broader and singular configurations.}
\label{tab:S1L8}
\end{table}

The XXX$_1$ data show that increasing the local representation enlarges the accessible entanglement relative to the XXX$_{\frac{1}{2}}$ chain. This is expected because each site can carry three local states rather than two. The extrema, however, remain strongly organized by similar patterns. The minimum is usually obtained from compact Bethe quantum number configurations near the center, such as $\{-\frac12,\frac12\}$, $\{-1,0,1\}$, or repeated central patterns allowed by the spin-$1$ representation. By contrast, the maximum tends to involve broader or more irregular Bethe quantum number distributions, and at larger $M$ it can be influenced by singular and strange solutions.

\subsubsection{\texorpdfstring{Spin-\(\frac{3}{2}\) chain}{Spin-3/2 chain}}
For XXX$_{\frac{3}{2}}$ we keep the $L=6$ scan in the main text, shown in Figure~\ref{fig:S32L6} and Table~\ref{tab:S32L6}. The corresponding $L=4$ and $L=5$ summary tables are given in Appendix~\ref{app:extraEE}.

\begin{figure}[h!]
\centering
\includegraphics[width=0.9\textwidth]{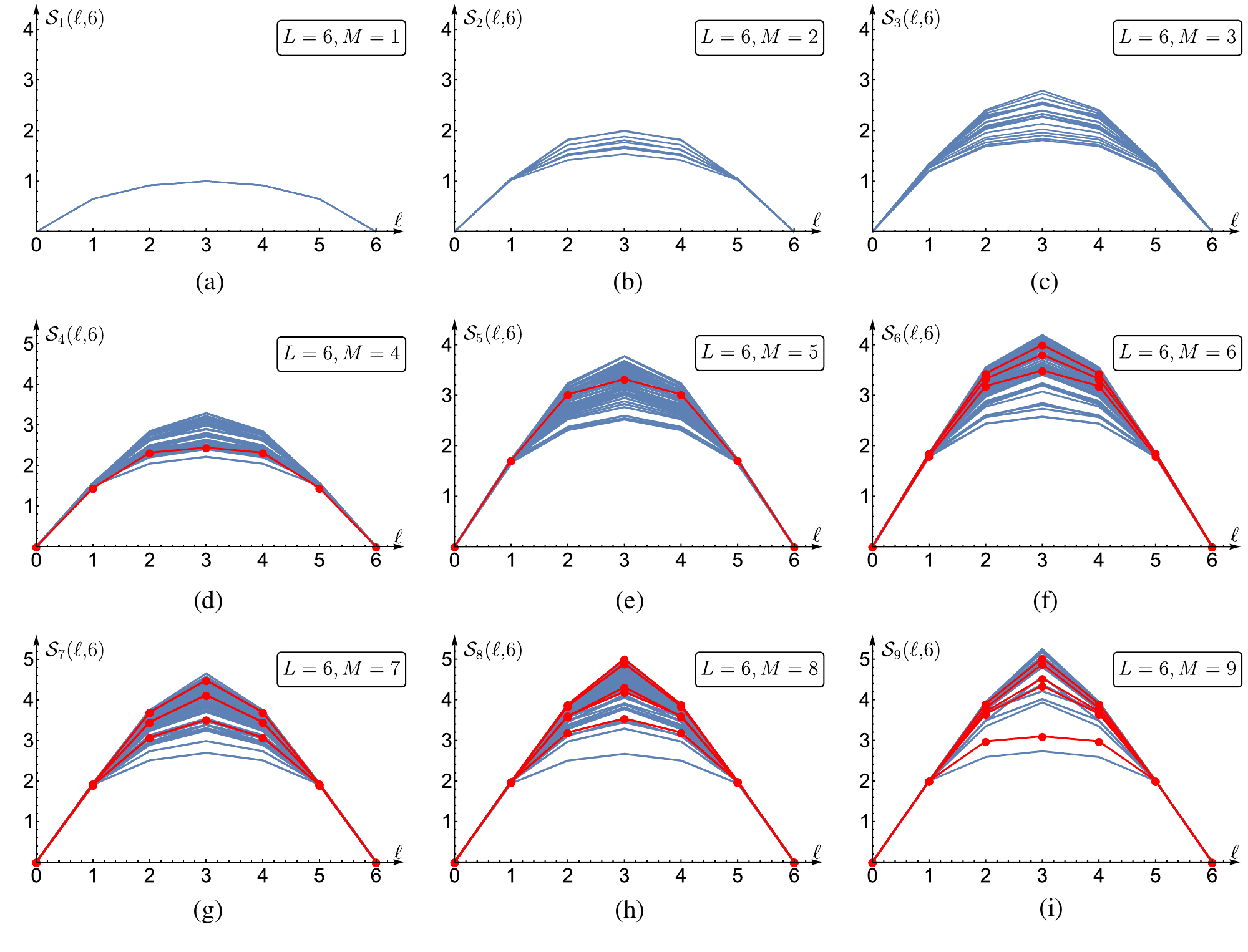}
\caption{The XXX$_{3/2}$ on-shell scan at $L=6$ shows representation-specific extrema, including shifted compact low-entropy blocks. Each curve is one Bethe state.}
\label{fig:S32L6}
\end{figure}

\begin{table}[h!]
\centering
\small
\begin{adjustbox}{max width=\textwidth}
\begin{tabular}{|c|c|c|c|c|c|c|}
  \hline
  $M$ & $\mathcal{N}_R$ & $\mathcal{N}_S$ & $\mathcal{S}_{\text{min}}$ & $\{I_j\}_{\text{min}}$ & $\mathcal{S}_{\text{max}}$ & $\{I_j\}_{\text{max}}$\\
  \hline
  1 & 5 & 0 & 1  & / & 1 & / \\
  \hline
 2 & 15 & 0 & 1.53365 & $\{-\frac{5}{2}, -\frac{3}{2}\}_{p}$ & 1.99992 & $\{-\frac{3}{2}, \frac{1}{2}\}_{p}$\\
 \hline
 3 & 35 & 0 & 1.80517  & $\{-1, 0, 1\}$ & 2.78996 & $\{-1, -1, 1\}_{p}$ \\
 \hline
 4 & 64 & 0 & 2.21682 & $\{-\frac{7}{2}, -\frac{5}{2}, -\frac{3}{2}, -\frac{1}{2}\}_{p}$ & 3.28914 & $\{-\frac{3}{2}, -\frac{1}{2}, \frac{1}{2}, \frac{1}{2}\}_{p}$ \\
   \hline
5 & 100 & 1 & 2.51667 & $\{-4, -3, -2, -1, 0\}_{p}$ & 3.7701 & $\{-1, -1, 0, 0, 1\}_{p}$ \\
   \hline
6 & 117 & 3 & 2.57222 & $\{-\frac{3}{2}, -\frac{1}{2}, -\frac{1}{2}, \frac{1}{2}, \frac{1}{2}, \frac{3}{2}\}$ & 4.19356 & $\{-\frac{1}{2}, -\frac{1}{2}, -\frac{1}{2}, \frac{1}{2}, \frac{1}{2}, \frac{1}{2}\}$ \\
   \hline
7 & 124 & 6 & 2.69058 & $\{-1, -1, 0, 0, 0, 1, 1\}$ & 4.64598 & $\{-1, -1, -1, 0, 1, 1, 1\}$ \\
   \hline
8 & 92 & 10 & 2.66643 & $\{-\frac{3}{2}, -\frac{1}{2}, -\frac{1}{2}, -\frac{1}{2}, \frac{1}{2}, \frac{1}{2},\frac{1}{2},\frac{3}{2}\}$ &5.00669  & $\begin{aligned}\{&-2.535\ri, -1.5\ri, -0.898, -0.5\ri,\\ &0.5\ri, 0.898, 1.5\ri, 2.535\ri\}\end{aligned}$ \\
   \hline
9 & 29 & 5 & 2.73044 & $\{-2, -1, -1, 0, 0, 0, 1, 1, 2\}$ &  5.2462& $\{-2, -2, -1, -1, -1, 0, 0, 0,1\}$ \\
   \hline
\end{tabular}
\end{adjustbox}
\caption{At $L=6$, XXX$_{3/2}$ extrema include shifted compact minima and high-entropy repeated-root or string-like configurations.}
\label{tab:S32L6}
\end{table}

The XXX$_{\frac{3}{2}}$ chain further confirms the role of the local representation. Since each site can accommodate up to three lowering excitations, the fixed-$M$ sectors contain more ways of distributing magnons across the cut than in the XXX$_{\frac{1}{2}}$ chain. This increases the absolute scale of the entropy and makes repeated quantum numbers, singular strings, and complex-root configurations more visible in the extrema. A new feature relative to XXX$_{\frac{1}{2}}$ and XXX$_1$ is that the minimum entropy state need not be a Fermi sea pattern centered on the origin: at low $M$, Tables~\ref{tab:S32L6}, \ref{tab:S32L4}, and \ref{tab:S32L5} display compact Bethe quantum number blocks (often two-strings tagged ``$p$'') that are shifted along the mode-number axis, including $\{-\frac{5}{2},-\frac{3}{2}\}_p$ at $(L,M)=(6,2)$ and $\{-4,-3,-2,-1,0\}_p$ at $(L,M)=(6,5)$. The block remains compact and string-organized, but the ``centered'' part of the XXX$_{\frac{1}{2}}$/XXX$_1$ pattern is no longer universal. The maximum can therefore be produced not only by widely separated real Bethe quantum numbers, but also by configurations whose rapidities form non-trivial strings or contain repeated roots.

\subsubsection{On-shell summary}
\begin{figure}[h!]
\centering
\begin{subfigure}[t]{0.32\textwidth}
\centering
\includegraphics[width=\linewidth]{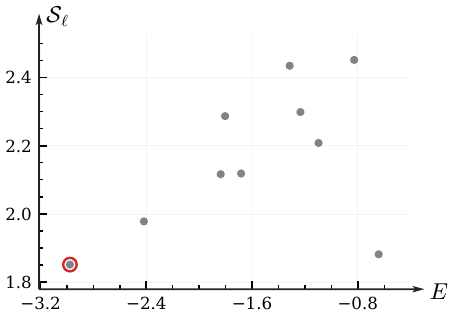}
\caption{\(M=3\)}
\end{subfigure}
\hfill
\begin{subfigure}[t]{0.32\textwidth}
\centering
\includegraphics[width=\linewidth]{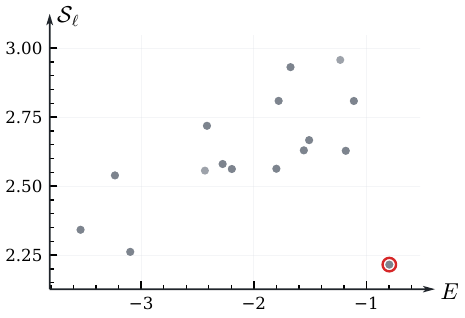}
\caption{\(M=4\)}
\end{subfigure}
\hfill
\begin{subfigure}[t]{0.32\textwidth}
\centering
\includegraphics[width=\linewidth]{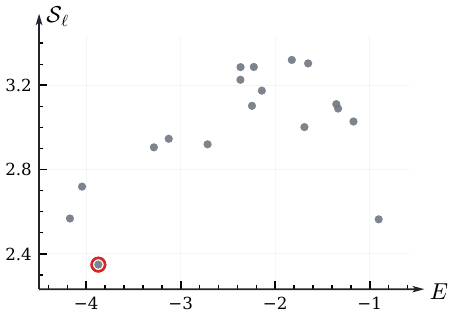}
\caption{\(M=5\)}
\end{subfigure}
\caption{{Energy--entropy scatter plots for finite on-shell XXX$_{\frac{3}{2}}$ Bethe states at fixed \(L=5\) and \(\ell=2\). Each point is one finite-energy state in the indicated \(M\)-sector. In (a), the minimum-entropy state is also the energy minimum; in (b), it is also the energy maximum; in (c), it has intermediate energy.}}
\label{fig:s32_l5_energy_entropy_triplet}
\end{figure}

The higher-spin on-shell scans preserve the compact chain pattern but add representation-specific structure. Many low-entanglement states are still compact and coherent. 
However, for the on-shell compact chains, the state with minimal half-cut entropy need not coincide with a fixed energy extremum across magnon sectors. Figure~\ref{fig:s32_l5_energy_entropy_triplet} shows this explicitly for the XXX$_{\frac{3}{2}}$ chain at fixed \(L=5\) and \(\ell=2\). Keeping the spin and length fixed while varying only \(M\), the minimum-entropy state can coincide with the energy minimum, coincide with the energy maximum, or lie at an intermediate energy.
High entanglement states use the larger local Hilbert space through broader mode-number support, repeated roots, and singular or string-like configurations. Increasing $s$ therefore changes both the entropy scale and the types of Bethe-root structures that can realize extrema.

\FloatBarrier

\subsection{Results for off-shell Bethe states}
\label{subsec:offshell_xxxs}

We now apply the optimizer of Section~\ref{sec:offshell} to the higher-spin compact chains, focusing on XXX$_1$ and XXX$_{\frac{3}{2}}$. The higher-spin data preserve the two-regime maximum structure seen for XXX$_{\frac{1}{2}}$: at fixed \(M\), the off-shell maxima approach the partition upper bound set by the rapidity partitions in \eqref{eq:detailedSum}, while for \(M\sim L\) the maxima grow linearly with \(L\) for large $L$. Higher spin changes the entropy scale and the product-state rapidity locus. The on/off-shell maximum gaps remain modest in the sampled sectors, whereas the off-shell minima again reach product states through compact boundary strings.

There is one common small-\(M\) fact used below. At \(M=1\), Eq.~\eqref{eq:detailedSum} has at most two partitions, so the reduced density matrix has at most two nonzero eigenvalues for every local spin \(s\) and every \(L\ge2\). Hence \(\mathcal{S}_{\max}^{\rm off}(L,M=1;\ell)\le1\). This upper bound is attainable by choosing the rapidity so that the two Schmidt weights are equal.

For later comparisons we also use the compact chain upper bound
\begin{align}
\mathcal{S}_\ell\le \log\min(2^M,d^\ell,d^{L-\ell}),\qquad d=2s+1 .
\label{eq:xxxs_compact_upper_bound}
\end{align}
The factor \(2^M\) follows from \eqref{eq:detailedSum}: each term is labelled by a subset of the \(M\) rapidities assigned to subchain \(A\), and the compact-chain restriction on the allowed values of \(|\alpha|\) only reduces this number. The factors \(d^\ell\) and \(d^{L-\ell}\) are the Hilbert-space dimensions on the two sides of the cut.

\subsubsection{XXX$_1$ off-shell results}

\begin{figure}[h!]
\centering
\includegraphics[width=0.82\textwidth]{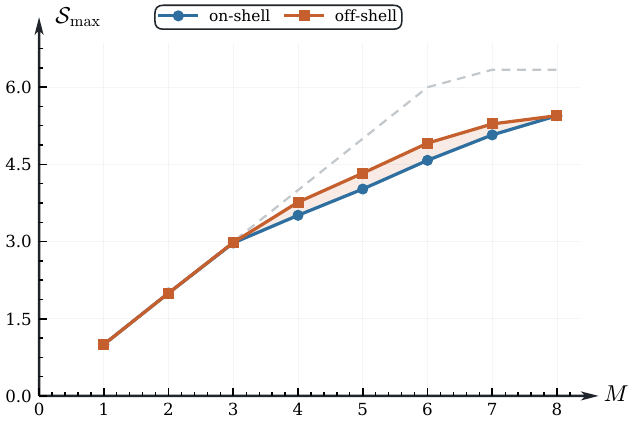}
\caption{XXX$_1$ off-shell maxima modestly exceed the on-shell maxima at \(L=8\). The two coincide at \(M=1\) and at \(M=8\), where the on-shell singular-pair maximum is itself in the off-shell parameter manifold. Red squares are off-shell maxima, blue circles are on-shell maxima, and the dashed line is the upper bound \eqref{eq:xxxs_compact_upper_bound} with \(d=3\) and \(\ell=4\).}
\label{fig:offshell_xxxs_onoff}
\end{figure}

\paragraph{Maximum.}
Table~\ref{tab:offshell_xxxs} collects the representative \(L=8\), \(\ell=4\) on/off-shell comparison for XXX$_1$, used in Figure~\ref{fig:offshell_xxxs_onoff}.

\begin{table}[h!]
\centering
\begin{tabular}{|c|c|c|c|c|c|c|c|c|}
  \hline
  $M$ & 1 & 2 & 3 & 4 & 5 & 6 & 7 & 8 \\
  \hline
  $\mathcal{S}_{\max}^{\rm off}$ & 1.00 & 2.00 & 2.98 & 3.77 & 4.33 & 4.91 & 5.29 & 5.45 \\
  $\mathcal{S}_{\max}^{\rm on}$  & 1.00 & 2.00 & 2.97 & 3.51 & 4.02 & 4.58 & 5.07 & 5.45 \\
  gap & 0.00 & 0.00 & 0.01 & 0.26 & 0.31 & 0.33 & 0.21 & 0.00 \\
  \hline
\end{tabular}
\caption{XXX$_1$ on/off-shell comparison at \(L=8\), \(\ell=4\). The off-shell row reports the optimized off-shell values; \(M=1\) is fixed by the analytic upper bound \(\mathcal{S}\le1\), and \(M=8\) is witnessed by the on-shell singular-pair rapidities themselves.}
\label{tab:offshell_xxxs}
\end{table}

For the representative length \(L=8\), Figure~\ref{fig:offshell_xxxs_onoff} compares the on-shell maxima with the off-shell search over \(M=1,\ldots,8\). The off-shell maximum meets or exceeds the on-shell maximum in every sector. The two coincide at \(M=1\) and at \(M=8\), where the on-shell singular-pair rapidities themselves give the off-shell value \(\mathcal{S}=5.4467\). The largest relative enhancement occurs at intermediate \(M=4,\ldots,6\), where the off-shell maximum is about \(7\%\)--\(8\%\) above the on-shell maximum. Every off-shell value stays below the upper bound \eqref{eq:xxxs_compact_upper_bound} with \(d=3\).

At the matched \(L=8\), \(M=4\), \(\ell=4\) sector, XXX$_1$ reaches \(\mathcal{S}_{\max}^{\rm off}=3.77\), compared with \(3.47\) for XXX$_{1/2}\). Both values should be compared with the same partition upper bound \(M=4\). Thus XXX$_1$ reaches about \(94\%\) of this upper bound, while XXX$_{\frac{1}{2}}\) reaches about \(87\%\).\footnote{For four spin-\(1\) sites, a completely generic state could have entropy as large as \(4\log 3\). This bound is larger than the partition upper bound \(M=4\) in the \(L=8\), \(M=4\), \(\ell=4\) comparison, so it is not the active upper bound for the off-shell Bethe vector discussed here.}

The qualitative scaling evidence comes from representative filling \(M=L\), where the available XXX$_1$ values grow from \(\mathcal{S}_{\max}^{\rm off}=1.58\) at \(L=2\) to about \(5.8\) by \(L=10\). These lengths show growth rather than a fixed-\(M\)-type constant limit, but they are too short to determine a stable volume-law coefficient.

\paragraph{Minimum.}
The XXX$_1$ off-shell minimum reaches \(\mathcal{S}_{\min}^{\rm off}\to0\) at compact boundary-string rapidities. The simplest example is \(\{\ri,0\}\), which gives a state proportional to \(|0,\ldots,0,2\rangle\) in the occupation basis \eqref{eq:xxxs_occupation_basis}, with the last site lowered twice. This is a product vector and therefore has Schmidt rank one across every cut. For larger \(M\), the same construction fills boundary sites by repeating the local compact string. The general compact boundary-string construction is given in Appendix~\ref{subsec:compact_spin_product_strings}.

\subsubsection{XXX$_{\frac{3}{2}}$ off-shell results}

\begin{figure}[h!]
\centering
\includegraphics[width=0.28\textwidth]{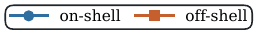}\\[0.15em]
\begin{subfigure}[t]{0.31\textwidth}
\centering
\includegraphics[width=\linewidth]{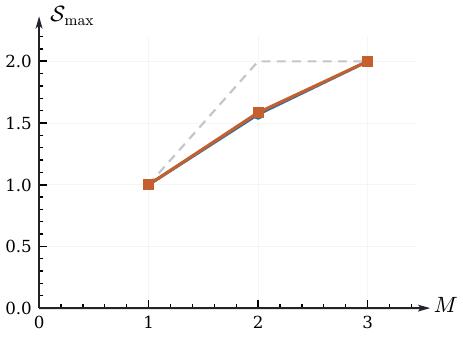}
\caption{\(L=2,\ \ell=1\)}
\end{subfigure}
\hfill
\begin{subfigure}[t]{0.31\textwidth}
\centering
\includegraphics[width=\linewidth]{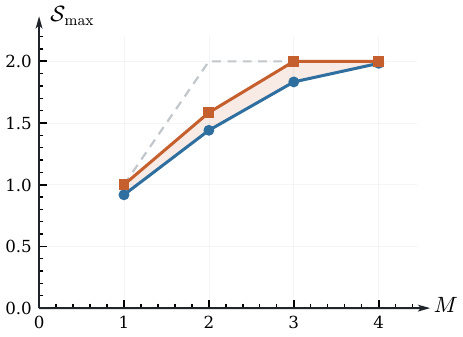}
\caption{\(L=3,\ \ell=1\)}
\end{subfigure}
\hfill
\begin{subfigure}[t]{0.31\textwidth}
\centering
\includegraphics[width=\linewidth]{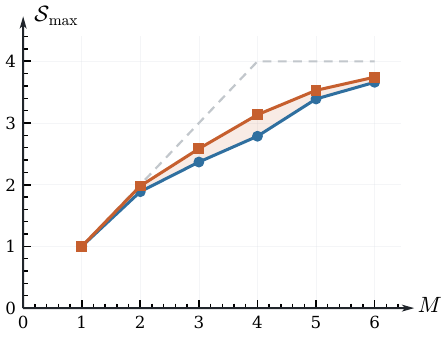}
\caption{\(L=4,\ \ell=2\)}
\end{subfigure}

\vspace{0.25em}
\begin{subfigure}[t]{0.31\textwidth}
\centering
\includegraphics[width=\linewidth]{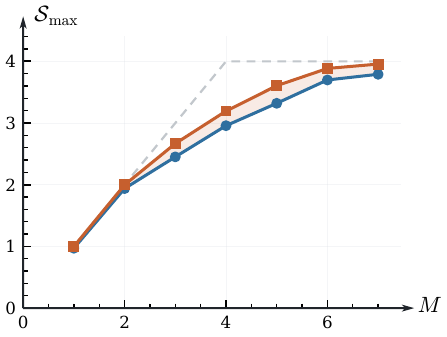}
\caption{\(L=5,\ \ell=2\)}
\end{subfigure}
\hspace{0.08\textwidth}
\begin{subfigure}[t]{0.31\textwidth}
\centering
\includegraphics[width=\linewidth]{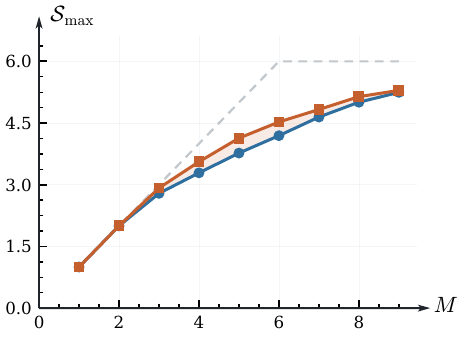}
\caption{\(L=6,\ \ell=3\)}
\end{subfigure}
\caption{XXX$_{3/2}$ on-shell maxima lie close to the sampled off-shell maxima. The five panels cover \(L=2,3,4,5,6\) at \(\ell=\lfloor L/2\rfloor\) and \(M=1,\ldots,\lfloor 3L/2\rfloor\). Blue points are on-shell maxima, red points are off-shell optima, and the dashed line is the upper bound \eqref{eq:xxxs_compact_upper_bound} with \(d=4\).}
\label{fig:offshell_xxx32}
\end{figure}

\paragraph{Maximum.}
The XXX$_{\frac{3}{2}}$ chain gives the same qualitative comparison across several lengths. Table~\ref{tab:offshell_xxx32} reports \(\mathcal{S}_{\max}^{\rm off}\) for \(L=2,\ldots,6\) at \(\ell=\lfloor L/2\rfloor\), over the magnon range \(M=1,\ldots,\lfloor 3L/2\rfloor\) covered by the on-shell scan of Section~\ref{subsec:onshell_xxxs}. The representative-filling entries \(M=\lfloor3L/2\rfloor\) grow from \(2.00\) at \(L=2,3\) to \(5.29\) at \(L=6\), again supporting volume-law growth without fixing a reliable coefficient. The off-shell maximum meets or exceeds the on-shell maximum in every sampled sector, as shown in Figure~\ref{fig:offshell_xxx32}. The relative enhancement remains modest: the largest sampled enhancement is about \(13\%\), and the visible intermediate-sector enhancements for \(L=5,6\) are about \(8\%\)--\(10\%\). The enhancement narrows again near the largest sampled \(M\). Every off-shell value stays below the upper bound \eqref{eq:xxxs_compact_upper_bound} with \(d=4\).

\begin{table}[h!]
\centering
\begin{tabular}{|c|c|c|c|c|c|c|c|c|c|}
  \hline
  $L \backslash M$ & 1 & 2 & 3 & 4 & 5 & 6 & 7 & 8 & 9 \\
  \hline
  2 & 1.00 & 1.58 & 2.00 &      &      &      &      &      &      \\
  3 & 1.00 & 1.58 & 2.00 & 2.00 &      &      &      &      &      \\
  4 & 1.00 & 1.98 & 2.58 & 3.13 & 3.53 & 3.74 &      &      &      \\
  5 & 1.00 & 2.00 & 2.67 & 3.19 & 3.61 & 3.89 & 3.96 &      &      \\
  6 & 1.00 & 2.00 & 2.92 & 3.56 & 4.13 & 4.53 & 4.83 & 5.14 & 5.29 \\
  \hline
\end{tabular}
\caption{Sampled XXX$_{3/2}$ off-shell maxima modestly exceed the corresponding on-shell values while staying below the upper bound \eqref{eq:xxxs_compact_upper_bound} with \(d=4\). Entries use \(\ell=\lfloor L/2\rfloor\) and \(M=1,\ldots,\lfloor 3L/2\rfloor\).}
\label{tab:offshell_xxx32}
\end{table}

\paragraph{Minimum.}
The XXX$_{\frac{3}{2}}$ minimum is the spin-\(\frac{3}{2}\) boundary-string analogue of the XXX$_1$ construction. In the same occupation basis \eqref{eq:xxxs_occupation_basis}, the rapidities \(\{3\ri/2,\ri/2\}\) give \(|0,\ldots,0,2\rangle\), again producing a product vector with zero entanglement across every cut. More generally, a spin-\(s\) boundary site can absorb \(2s\) lowerings through the local string \(\{\ri s,\ri(s-1),\ldots,\ri(1-s)\}\), with the sign-reversed string localizing at the opposite boundary. Appendix~\ref{subsec:compact_spin_product_strings} gives the general construction and the explicit XXX$_{3/2}$ examples.

\subsection{Summary and on-shell/off-shell comparison}

The higher-spin compact chains preserve the two extremal mechanisms found for XXX$_{\frac{1}{2}}$, with spin-dependent changes in the rapidity realization. For the maximum, the fixed-\(M\) organizing principle is still the partition upper bound \(M\) coming from \eqref{eq:detailedSum}. The XXX$_1$ data at \(L=8\) and the sampled XXX$_{\frac{3}{2}}$ data remain close to this upper-bound structure, while the representative-filling data grow with \(L\), as in the XXX$_{\frac{1}{2}}$ volume-law regime. The available higher-spin lengths are not sufficient to extract a reliable slope. The main difference from XXX$_{\frac{1}{2}}$ is the entropy scale: a higher-spin site can carry more lowering excitations, so the same fixed-\(M\) mechanism can produce larger absolute entropies.

For the minimum, the common feature is \(\mathcal{S}_{\min}^{\rm off}\to0\): in all compact chains, off-shell rapidities can produce product vectors. The difference is the rapidity locus. In XXX$_{\frac{1}{2}}$ this occurs at repeated singular rapidities \(u=\pm\ri/2\), whereas for XXX$_1$ and XXX$_{\frac{3}{2}}$ it occurs through compact boundary strings, as described in Appendix~\ref{subsec:compact_spin_product_strings}. The on/off-shell comparison follows the same pattern as in XXX$_{\frac{1}{2}}$: the maxima in Table~\ref{tab:offshell_xxxs}, Figure~\ref{fig:offshell_xxxs_onoff}, Table~\ref{tab:offshell_xxx32}, and Figure~\ref{fig:offshell_xxx32} are only modestly above the on-shell maxima, while the sharper distinction is at the minimum, where the Bethe equations exclude the off-shell product-state loci.

\section{The non-compact $SL(2,\mathbb{R})$ spin chain}
\label{sec:nonComp}
In this section, we turn to the investigation of entanglement entropy of the non-compact $SL(2,\mathbb{R})$ chain. To our knowledge, the entanglement entropy of the non-compact $SL(2,\mathbb{R})$ chain has not been studied before. Our analysis below reveals many new features specific to the non-compact chains.

\subsection{Model and Bethe states}

\paragraph{Hilbert space and Hamiltonian.} We again consider a length-$L$ spin chain with periodic boundary conditions. The spin chain is non-compact because the local Hilbert space at each site is infinite dimensional. Physically, this means each site may accommodate an unlimited number of excitations. The $SL(2,\mathbb{R})$ generators can be expressed in terms of bosonic oscillators as
\begin{align}
\label{eq:SpinOsc}
\mathbf{S}^-=a,\qquad \mathbf{S}^+=a^{\dagger}(a^{\dagger}a+1),\qquad \mathbf{S}^z=-(a^{\dagger}a+\tfrac12)
\end{align}
with the usual bosonic commutation relation $[a,a^{\dagger}]=1$. The local basis states are
\begin{align}
\label{eq:sl2_local_basis}
|n\rangle\equiv\frac{(\mathbf{S}^+)^n}{n!}|0\rangle,\qquad n=0,1,2,\ldots\,.
\end{align}
The state $|0\rangle$ is the Fock vacuum of the oscillator and satisfies $\mathbf{S}^-|0\rangle=0$. The generators act on these states as
\begin{align}
\mathbf{S}^+|m\rangle=(m+1)|m+1\rangle,\quad \mathbf{S}^-|m\rangle=m|m-1\rangle,\quad \mathbf{S}^z|m\rangle=-(m+\tfrac{1}{2})|m\rangle\,.
\end{align}
Dual states are defined by
\begin{align}
\langle n|=\langle 0|\frac{(\mathbf{S}^-)^n}{n!},\qquad \langle 0|\mathbf{S}^+=0\,.
\end{align}
The Hamiltonian is given by
\begin{align}
H=\sum_{j=1}^L h_{j,j+1}
\end{align}
where the Hamiltonian density $h_{j,j+1}$ acts on two nearest neighboring sites as
\begin{equation}
\begin{aligned}
h_{j,j+1}|m\rangle_j\otimes|n\rangle_{j+1}=&\left(h(m)+h(n)\right)|m\rangle_j\otimes|n\rangle_{j+1}\\
&-\sum_{k=1}^m\frac{1}{k}|m-k\rangle_j\otimes|n+k\rangle_{j+1}\\
&-\sum_{k=1}^n\frac{1}{k}|m+k\rangle_j\otimes |n-k\rangle_{j+1}\,.
\end{aligned}
\end{equation}
with $h(m)$ being the $m$-th harmonic number. Although less common in condensed matter physics, this Hamiltonian plays a significant role in high-energy theory, describing certain classes of operators in $\mathcal{N}=4$ Super-Yang-Mills theory \cite{Beisert:2004ry} and QCD \cite{Braun:2003rp}. In this context, magnons correspond to covariant derivatives, allowing for an unlimited number of excitations. More recently, the $SL(2,\mathbb{R})$ spin chain has also appeared in the study of stochastic particle processes \cite{Frassek:2019vjt,Frassek:2019isa}.\par

Owing to the non-compactness of the spin chain, certain configurations that are atypical for compact models become accessible. For example, one may consider a two-site chain with a large number of magnons $M\gg1$. This is known as the large spin limit. Here the ``spin'' refers to the Lorentz spin, which is directly related to the number of magnons (or covariant derivatives). In this regime, the entanglement entropy displays universal features absent in compact spin chains.\par

\paragraph{Bethe ansatz.} Eigenstates of the non-compact chain can be constructed via ABA. The vacuum $|0\rangle$ now corresponds to the lowest-weight state, annihilated by the $B$-operator. The Lax operator retains the standard form, with spin operators given in \eqref{eq:SpinOsc}. Its action on a local state reads
\begin{align}
\mathbf{L}_{an}(u)|m\rangle_n=\left(\begin{array}{cc}(u-\ri(m+\tfrac{1}{2}))|m\rangle_n & \ri m|m-1\rangle_n\\ \ri(m+1)|m+1\rangle_n & (u+\ri(m+\tfrac{1}{2}))|m\rangle_n  \end{array}\right)\,.
\end{align}
The Bethe states are constructed by acting with $C$-operators
\begin{align}
|\{u_j\}\rangle^{(-\frac{1}{2})}=\mathbf{C}^{(-\frac{1}{2})}(u_1)\ldots\mathbf{C}^{(-\frac{1}{2})}(u_M)|0\rangle
\end{align}
where the superscript ``$-\tfrac{1}{2}$'' indicates that the non-compact chain can be interpreted as the spin $s=-\tfrac{1}{2}$ chain. For such a state to be an eigenstate of the Hamiltonian, the rapidities must satisfy
\begin{align}
\label{eq:BAESL2}
\left(\frac{u_j-\tfrac{\ri}{2}}{u_j+\tfrac{\ri}{2}}\right)^L=\prod_{k=1\atop k\ne j}^M\frac{u_j-u_k+\ri}{u_j-u_k-\ri}\,.
\end{align}
Although these equations resemble those of compact models, their solutions exhibit qualitatively different properties. Specifically, all Bethe roots are real and uniquely correspond to a set of Bethe quantum numbers. The absence of singular solutions simplifies the construction of Bethe states, requiring no additional regularization.

\subsection{Results for on-shell Bethe states}
Before presenting results for different chain lengths and magnon numbers, we provide some general remarks concerning features unique to the non-compact chain. First, we note that although the Hilbert space dimension at each site is infinite, for a fixed magnon number $M$ and subsystem size $\ell$, the dimension of the Hilbert space of the subsystem $A$ is given by 
\begin{align}
\dim\mathcal{H}_A={M+\ell\choose \ell}\,.
\end{align}
For example, with $\ell=2$ and $M=2$, the subsystem $A$ can be spanned by the basis states 
\begin{align}
\{|0,0\rangle_A,|0,1\rangle_A,|1,0\rangle_A,|2,0\rangle_A,|1,1\rangle_A,|0,2\rangle_A\}\,.
\end{align}
For fixed $\ell$ and large $M$, the entanglement entropy is bounded by $\log(\dim\mathcal{H}_A)\sim \ell\log M$.\par 

The basis states can be further classified by their total magnon number. Using this structure, we may also define symmetry-resolved entanglement entropy.

\subsubsection{\texorpdfstring{\(\bm{L=2}\)}{L=2}}
\begin{figure}[h!]
\centering
\includegraphics[scale=0.7]{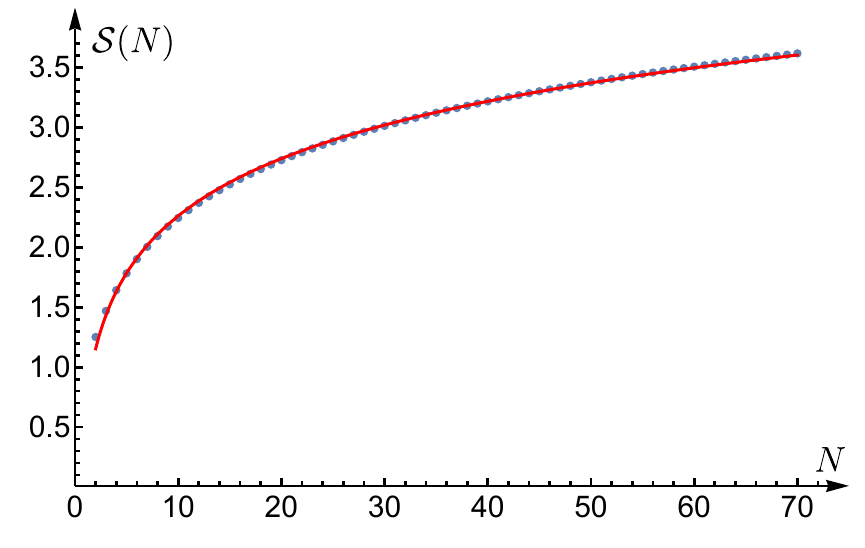}
\caption{The unique $L=2$ on-shell $SL(2,\mathbb{R})$ state has logarithmic single-site entropy. Blue points are numerical data for \(\mathcal{S}^{\rm on}(2,M;1)\), and the red curve is \(a\log M+b\).}
\label{fig:twist2}
\end{figure}

For the minimal chain length $L=2$, there exists a unique solution for each magnon number $M$. The only non-trivial subsystem size is $\ell_A=1$. We denote by $\mathcal{S}^{\rm on}(2,M;1)$ the bipartite entropy of this unique on-shell state. For short spin chains, direct computation is more efficient because $\dim\mathcal{H}_A=M+1$ grows only linearly with $M$. We calculate the entanglement entropy up to $M=70$, and the results reveal a clear asymptotic behavior. From the data, we identify the following logarithmic dependence on $M$
\begin{align}
\mathcal{S}^{\rm on}(2,M;1)=a\log M+b\,.
\end{align}
Numerical fitting yields the coefficients:
\begin{align}
a=0.477685,\qquad b=0.675394\,.
\end{align}
The fitted curve together with the original data is shown in Figure~\ref{fig:twist2}.
As noted earlier, for $\ell=1$, we have $\text{dim}\mathcal{H}_A=M+1$ and the reduced density matrix is automatically diagonal; the corresponding off-shell diagonal form is given in Section~\ref{subsec:offshell_sl2}.
\begin{figure}[h!]
\centering
\includegraphics[scale=0.4]{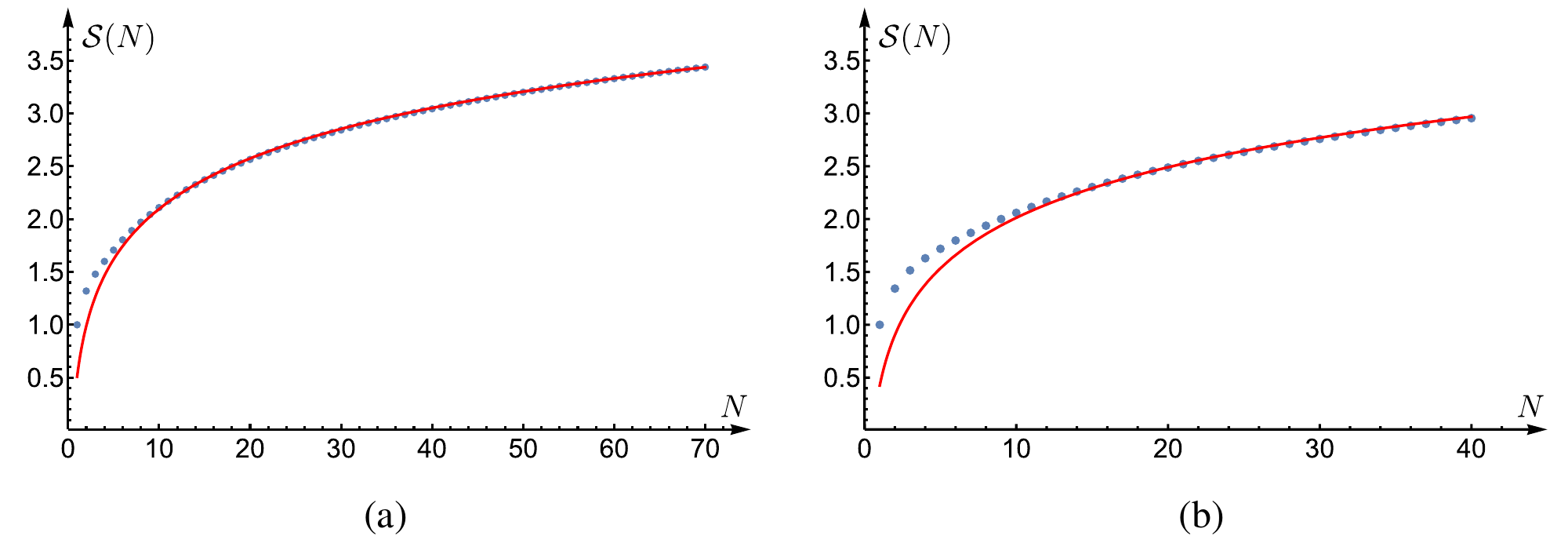}
\caption{The $L=4,6$ on-shell $SL(2,\mathbb{R})$ data are compatible with logarithmic fits over the sampled range. (a) \(\mathcal{S}^{\rm on}(4,M;2)\), fitted by \(a\log M+b\) up to \(M_{\max}=70\). (b) \(\mathcal{S}^{\rm on}(6,M;2)\), fitted by \(a\log M+b\) up to \(M_{\max}=40\). In both cases, $a\approx 0.48$ while the constant pieces are different.}
\label{fig:twist46}
\end{figure}

Thus the on-shell \(SL(2,\mathbb{R})\) data are well described by fits of the form
\begin{align}
\mathcal{S}^{\rm on}(L,M;\ell)\simeq a_{L,\ell}\log M+b_{L,\ell}
\end{align}
over the available range of \(M\). The \(L=2,\ell=1\) fit above is the most stable one, with \(a_{2,1}=0.477685\) and \(b_{2,1}=0.675394\). For the larger lengths shown in Figure~\ref{fig:twist46}, the same fitting form is useful, but the fitted slope \(a_{L,\ell}\) fluctuates with \(L\), with the cut, and with the fitting window. We therefore use the logarithmic form as an empirical description of the finite data, not as evidence for a universal on-shell coefficient.

\subsection{Results for off-shell Bethe states}
\label{subsec:offshell_sl2}

For the non-compact off-shell problem, the first issue is the Schmidt-rank upper bound at fixed \(M\). As in the compact chains, the partition expansion gives at most \(2^M\) rapidity assignments across the cut. In the \(SL(2,\mathbb{R})\) chain this is not the only relevant count, because a single site can carry any nonnegative number of excitations. The sharper comparison is the fixed charge Hilbert space rank obtained by resolving the cut into subsystem excitation number blocks. We denote this rank by \(R(L,\ell,M)\).

The fixed-charge upper bound follows directly from the block decomposition in \eqref{eq:detailedSum}. If the left subsystem contains \(m\) excitations, its \(SL(2,\mathbb{R})\) oscillator basis has
\[
\binom{m+\ell-1}{\ell-1}
\]
states. The right subsystem must then contain \(M-m\) excitations and has
\[
\binom{M-m+L-\ell-1}{L-\ell-1}
\]
states. The Schmidt rank in this block is bounded by the smaller of these two dimensions, and summing over the orthogonal excitation-number blocks gives
\begin{align}
\label{eq:sl2_rank_bound}
R(L,\ell,M)=\sum_{m=0}^{M}
\min\!\left[
\binom{m+\ell-1}{\ell-1},
\binom{M-m+L-\ell-1}{L-\ell-1}
\right].
\end{align}
In particular, \(R(L,1,M)=M+1\), so the single-site upper bound is \(\mathcal{S}\le\log(M+1)\).

With this upper bound in place, the finite-root off-shell data show three features: for \(\ell=1\), the maximum nearly saturates \(\log(M+1)\); for \(\ell=2,3\), the maximum grows with \(M\) but remains below \(\log R(L,\ell,M)\); and the minimum again reaches \(\mathcal{S}_{\min}^{\rm off}\to0\), now through half-integer imaginary towers.

\subsubsection{Maximum off-shell entropy}

The maximum off-shell entropy separates sharply by the size of the cut. For a single-site cut, \(\ell=1\), the finite-root optimum is almost the full fixed-charge upper bound \(\log(M+1)\). For the higher cuts sampled here, \(\ell=2,3\), the finite-root optimum still grows with \(M\), but it remains visibly below the fixed-charge rank bound \(\log R(L,\ell,M)\). Table~\ref{tab:offshell_sl2} displays this contrast at selected values of \(M\).\footnote{The entries use the finite-root convention of Appendix~\ref{subsec:offshell_chain_specific}. Candidates with rapidities larger than the cutoff \(|u_j|\le100\) are kept as diagnostics but are excluded from the reported maxima, because large-rapidity candidates can approach the fixed-charge upper bound without representing a finite-root optimum.}

\begin{table}[h!]
\centering
\begin{tabular}{|c|c|c|c|c|c|c|c|c|}
  \hline
  $L \backslash M$ & 1 & 2 & 3 & 5 & 10 & 20 & 30 & 42 \\
  \hline
  2 & 1.00 & 1.58 & 2.00 & 2.58 & 3.46 & 4.39 & 4.95 & 5.43 \\
  \hline
  3 & 1.00 & 1.58 & 2.00 & 2.58 & 3.46 & 4.39 & 4.95 & 5.42 \\
  \hline
  4 & 1.00 & 1.97 & 2.54 & 3.36 & 4.70 & 6.16 & 7.17 & 8.03 \\
  \hline
  6 & 1.00 & 1.99 & 2.93 & 4.18 & 5.79 & 7.71 & 8.94 & 10.07 \\
  \hline
\end{tabular}
\caption{\(SL(2,\mathbb{R})\) finite-root off-shell maxima nearly saturate the single-site upper bound \(\log(M+1)\) but remain below \(\log R(L,\ell,M)\) for \(\ell=2,3\). The \(L=6\) data extend to \(M=42\).}
\label{tab:offshell_sl2}
\end{table}

\begin{figure}[h!]
\centering
\begin{subfigure}[t]{0.49\textwidth}
\centering
\includegraphics[width=\textwidth]{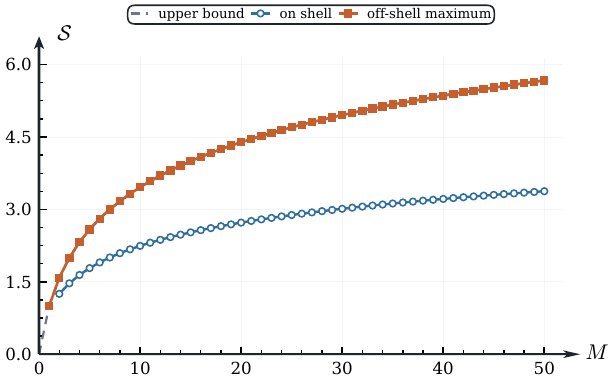}
\caption{\(L=2,\ell=1\).}
\end{subfigure}
\hfill
\begin{subfigure}[t]{0.49\textwidth}
\centering
\includegraphics[width=\textwidth]{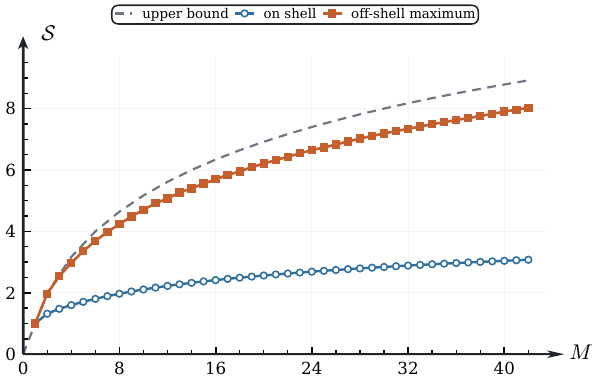}
\caption{\(L=4,\ell=2\).}
\end{subfigure}
\caption{\(SL(2,\mathbb{R})\) off-shell maxima sit well above the corresponding on-shell values. In (a), the off-shell data reach the upper bound \(\log(M+1)\), so the dashed bound is hidden by the off-shell curve. In (b), the off-shell maximum remains below the fixed-charge upper bound \(\log R(4,2,M)\), but still lies far above the on-shell curve.}
\label{fig:offshell_sl2_l2_onoff}
\end{figure}

\paragraph{Single-site cuts \(\ell=1\).}
For a single-site cut, \eqref{eq:sl2_rank_bound} gives \(R(L,1,M)=M+1\), hence \(\mathcal{S}\le\log(M+1)\). The finite-root off-shell maxima nearly saturate this upper bound for both \(L=2\) and \(L=3\). The on-shell \(L=2\) states behave differently: the fit in Figure~\ref{fig:twist2} grows as \(\mathcal{S}^{\rm on}(2,M;1)\approx0.477685\,\log M+0.675394\), while the off-shell maximum follows \(\log(M+1)\). At \(M=50\), for example, the on-shell value is \(3.38\), whereas the off-shell value is \(5.67\). Figure~\ref{fig:offshell_sl2_l2_onoff}(a) shows this gap directly.

The \(L=2\) and \(L=3\) off-shell rows in Table~\ref{tab:offshell_sl2} agree to the displayed precision. We treat this as an observed \(\ell=1\) feature of the sampled data, not as a proof that the two optimization problems are identical.

\paragraph{Higher cuts \(\ell=2,3\).}
For \(L=4\), \(\ell=2\), the fixed-charge rank bound \eqref{eq:sl2_rank_bound} becomes
\begin{align}
\label{eq:sl2_l4_rank_bound}
R(4,2,M)
=\sum_{m=0}^{M}\min(m+1,M-m+1)
=(\lfloor M/2\rfloor+1)(\lceil M/2\rceil+1).
\end{align}
This is the relevant rank bound for an \(M\)-magnon global state. Figure~\ref{fig:offshell_sl2_l2_onoff}(b) shows the main result: the finite-root off-shell curve stays below \(\log R(4,2,M)\), but remains far above the corresponding on-shell curve. In the sampled large-\(M\) range, the off-shell curve is about \(90\%\) of this fixed-charge upper bound.

For \(L=6\), \(\ell=3\), the same non-saturation is stronger: for large \(M\) in the sampled range, \(\mathcal{S}_{\max}^{\rm off}\) reaches about \(84.3\%\text{--}84.7\%\) of \(\log R(6,3,M)\). Thus the single-site cut is close to optimal in the fixed-charge space, while the \(\ell=2,3\) data behave more like the compact chains: the off-shell maximum grows, but it does not reach the full fixed-charge upper bound.

\begin{figure}[h!]
\centering
\includegraphics[width=\textwidth]{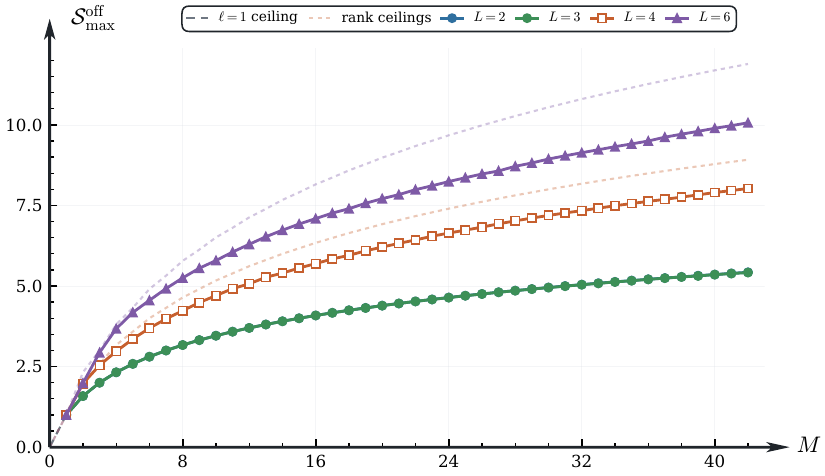}
\caption{\(SL(2,\mathbb{R})\) finite-root maxima nearly saturate the single-site upper bound but stay below the higher-cut fixed-charge upper bounds. Curves show \(\mathcal{S}_{\max}^{\rm off}\) versus \(M\) for \(L=2,3,4,6\), with \(\ell=\lfloor L/2\rfloor\), over the common range \(M=1,\dots,42\). The \(L=2\) curve coincides with the \(L=3\) curve to plotting resolution and is hidden beneath it.}
\label{fig:offshell_sl2_smax}
\end{figure}

Figure~\ref{fig:offshell_sl2_smax} summarizes the same split: the \(\ell=1\) curve tracks its upper bound closely, while the \(\ell=2,3\) curves stay below the corresponding fixed-charge upper bounds.

\subsubsection{Minimum off-shell entropy}

The non-compact chain also has off-shell product states, so \(\mathcal{S}_{\min}^{\rm off}\to0\). Here the product vectors are produced by half-integer imaginary towers. For example, the balanced \(L=2\), \(M=10\) tower
\[
\left\{-\frac{\ri}{2},-\frac{3\ri}{2},\ldots,-\frac{9\ri}{2}\right\}
\cup
\left\{\frac{\ri}{2},\frac{3\ri}{2},\ldots,\frac{9\ri}{2}\right\}
\]
gives a state proportional to \(|5,5\rangle=|5\rangle_1\otimes|5\rangle_2\) in the local basis \eqref{eq:sl2_local_basis}, hence has \(\mathcal{S}_{\ell=1}=0\). Appendix~\ref{subsec:sl2_product_towers} gives the general tower construction and shows that it produces product vectors across every contiguous bipartition.

The optimizer minima are consistent with this exact construction. Small nonzero numerical minima come from the finite numerical search and validation procedure discussed in Appendix~\ref{subsec:offshell_chain_specific}. Thus the off-shell minimum is zero, and the rapidity pattern realizing it is the half-integer tower described in Appendix~\ref{subsec:sl2_product_towers}.

\subsection{Summary and on-shell/off-shell comparison}

The \(SL(2,\mathbb{R})\) chain preserves the same maximum/minimum organization as the compact chains, but the upper bound controlling the maximum changes. In compact chains, the fixed-\(M\) maximum is organized by the partition upper bound \(M\). In the non-compact chain, a site can carry any number of excitations, so the sharper comparison is the fixed-charge rank bound \(\log R(L,\ell,M)\). This distinction is already visible in the off-shell data: for \(\ell=1\), \(R=M+1\) and the finite-root maximum nearly reaches \(\log(M+1)\); for the higher cuts \(\ell=2,3\), the maximum grows with \(M\) but remains below \(\log R(L,\ell,M)\), as shown by the \(L=4\) and \(L=6\) data.

The minimum is similar to the compact chains in outcome but different in rapidity realization. In all three families, off-shell rapidities can produce product vectors and hence \(\mathcal{S}_{\min}^{\rm off}\to0\). For XXX$_{\frac{1}{2}}$ the product locus is built from singular rapidities \(u=\pm\ri/2\); for XXX$_{s\ge1}$ it is built from compact boundary strings; for \(SL(2,\mathbb{R})\) it is built from half-integer imaginary towers. The on/off-shell comparison also has a stronger maximum-side effect than in the compact examples: at \(L=2,\ell=1\), the on-shell entropy grows as \(\mathcal{S}^{\rm on}(2,M;1)\approx0.48\log M+0.68\), whereas the off-shell maximum nearly reaches \(\log(M+1)\). For higher cuts, the finite-root off-shell family is already below the fixed-charge upper bound before the Bethe equations are imposed.

\section{Conclusions and Discussion}
\label{sec:conclude}

In this work, we have taken the first steps toward a systematic study of the quantum entanglement structures of Bethe states. Our investigation has yielded several new insights into the entanglement properties of Bethe states, particularly for off-shell Bethe states and those arising in non-compact spin chains. Additionally, this study opens up several avenues for further investigation.

Through our numerical analyses, we have made a number of noteworthy observations that call for analytical proofs and deeper physical understanding. The first concerns maximal entropy states. Unlike minimal entropy states, which exhibit a clear pattern in terms of Bethe quantum numbers, the Bethe roots corresponding to maximal entropy states do not appear to follow a similarly universal pattern. Whether this reflects a general property or merely our inability to discern such a pattern warrants further investigation. It would be desirable to gain a clearer understanding of which root distributions give rise to large quantum entanglement, both for on-shell and off-shell Bethe states.

For non-compact spin chains, numerical evidence suggests that for fixed $L$ in the large-$M$ limit, the entanglement entropy exhibits a logarithmic scaling $\mathcal{S}(L,M;\ell)\sim a\log M+b$ with a universal coefficient $a$. Additional numerical data would be helpful to consolidate this finding. If confirmed, a natural next step is to uncover the origin of this behavior and derive the universal constant analytically. In this context, the $SL(2,\mathbb{R})$ spin chain corresponds to the XXX$_s$ chain with $s=-\frac{1}{2}$. It would be interesting to see whether similar logarithmic behavior persists for other non-compact chains with negative spin, and how it depends on the spin value. Analytically continuing $s$ to negative real values, as a special limit, connects to continuum models such as the Lieb--Liniger model. Our current approach for computing entanglement entropy should be generalizable to the analysis of eigenstates of the Lieb--Liniger model.

A natural extension of the present work is to consider $q$-deformed XXZ-type spin chains. For such models, one can examine both generic $q$ and $q$ at roots of unity, where the rational $Q$-system has been established \cite{Bajnok:2019zub,Hou:2023jkr}. It would be interesting to see how $q$-deformation modifies quantum entanglement. At the same time, for $q$ at a root of unity, intriguing new features emerge. There exist solutions involving so-called Fabricius–McCoy strings \cite{Fabricius:2000yx,Fabricius:2000em}, which are ``transparent'' to other Bethe roots and scatter trivially. As a result, adding such a string does not alter the energy or higher conserved charges. Would this affect the entanglement entropy? Based on the analysis presented here, the answer is most likely yes, but precisely how it influences quantum entanglement remains an open question.

Beyond periodic boundary conditions, it is natural to ask how different boundary conditions affect quantum entanglement. Starting with simple integrability-preserving cases, one can study the effect of diagonal twists and open boundaries on entanglement. A key question is whether localized zero modes influence quantum entanglement. For special choices of boundary parameters, phenomena absent in the periodic case, such as spectrum-preserving deformations \cite{Jiang:2024xgx}, can occur. Understanding how boundary magnetic fields affect quantum entanglement and whether new phenomena arise would be a natural goal.

A more ambitious direction is the study of models with higher-rank symmetry, such as $SU(N)$- and $SU(M|N)$-invariant spin chains. The wave functions of such models can be constructed using the nested Bethe ansatz, involving multiple sets of rapidities — both momentum-carrying and auxiliary ones. The wave function is considerably more complex, and it would be very interesting to understand whether and how auxiliary Bethe roots affect quantum entanglement.

In the present work, we have analyzed only bipartite entanglement entropy. To gain a more complete understanding of quantum entanglement structures, other quantum information quantities, such as mutual information and symmetry-resolved entanglement entropy, should be studied. Indeed, we have obtained the full spectrum of the reduced density matrix for all primary on-shell Bethe states. Leveraging this, we can compute a broader range of entanglement-related quantities beyond the von Neumann and R\'enyi entropies.

\acknowledgments

Y.J. would like to thank Xiwen Guan for proposing a `homework' at the workshop ``Challenges in Integrability'', which led to the current work. The work of Y.J. is supported by the National Natural Science Foundation of China through Grant No.12575073. This work is also supported by the NSFC Grant No. 12247103. DlZ is supported in part by the UK Engineering and Physical Sciences Research Council grant number EP/Z000106/1, and the Royal Society under the grant URF\textbackslash R1\textbackslash 221310. We acknowledge the use of the Imperial College Research Computing Service (DOI: 10.14469/hpc/2232).

\appendix

\section{Izergin--Korepin formula}
\label{app:IZformula}
In this appendix, we summarize useful formulas for the scalar products of Bethe states. The scalar product is defined as
\begin{align}
S_M(\{v_j\},\{u_k\})\equiv\langle 0|\prod_{j=1}^M\mathbf{C}(v_j)\prod_{k=1}^M\mathbf{B}(u_k)|0\rangle
\end{align} 
This quantity can be computed using the Yangian algebra, \emph{i.e.}, the algebra of the $A$, $B$, $C$ and $D$ operators in the monodromy matrix. The result can be expressed explicitly, although it becomes rather involved when both sets $\{v_j\}$ and $\{u_k\}$ are off-shell. We use the shorthand notation of \eqref{eq:shorthandNotation}. In this appendix, single-argument functions act on rapidity differences, \(F(u,v)=F(u-v)\), so \(f(u,v)=f(u-v)\) is the same function as in the main text. The additional functions needed below are
\begin{align}
f(u)=\frac{u+\ri}{u},\qquad
g(u)=\frac{\ri}{u},\qquad
h(u)=\frac{f(u)}{g(u)}=\frac{u+\ri}{\ri},\qquad
t(u)=\frac{g^2(u)}{f(u)}=-\frac{1}{u(u+\ri)}\,.
\label{eq:IK_basic_functions}
\end{align}
For the ordered products in the prefactor,
\begin{align}
g_<^{\{u\}\{u\}}=\prod_{1\le i<j\le M}g(u_i-u_j),\qquad
g_>^{\{v\}\{v\}}=\prod_{1\le j<i\le M}g(v_i-v_j)\,.
\label{eq:IK_ordered_g}
\end{align}
The notation \(t^{\alpha\gamma}\) in a determinant is different from the product shorthand: it denotes the matrix \([t(u_i-v_j)]_{u_i\in\alpha,\,v_j\in\gamma}\). Thus \(\det t^{\alpha\gamma}\) is defined only when \(|\alpha|=|\gamma|\).
\begin{align}
\label{eq:IKsum}
S_M(\{v_j\},\{u_k\})=g_<^{\{u\}\{u\}}g_>^{\{v\}\{v\}}
\sum_{\substack{\alpha\cup\bar{\alpha}=\{u\}\\ \gamma\cup\bar{\gamma}=\{v\}\\ |\alpha|=|\gamma|}}
(-1)^{P_{\alpha}+P_{\gamma}}d^{\alpha}a^{\bar{\alpha}}a^{\gamma}d^{\bar{\gamma}}
h^{\alpha\gamma}h^{\bar{\gamma}\bar{\alpha}}h^{\alpha\bar{\alpha}}h^{\bar{\gamma}\gamma}\det t^{\alpha\gamma}\det t^{\bar{\gamma}\bar{\alpha}}
\end{align}
where \(a(u)\) and \(d(u)\) are the pseudovacuum eigenvalues of \(\mathbf{A}(u)\) and \(\mathbf{D}(u)\), specialized to the spin chain under consideration; for XXX$_s$ they are \(a_s(u)\) and \(d_s(u)\) in \eqref{eq:defad}. In \eqref{eq:IKsum}, the factor $(-1)^{P_{\alpha}}$ is the sign of the permutation of the ordered set $\{u\}$ which gives $\alpha\cup\bar{\alpha}$, and $(-1)^{P_{\gamma}}$ is defined analogously from the ordered set $\{v\}$ giving $\gamma\cup\bar{\gamma}$.
Using the relation $\mathbf{C}(u^*)=-[\mathbf{B}(u)]^{\dagger}$ where $u^*$ is the complex conjugate of $u$, we can write the inner product of two Bethe states as
\begin{align}
\langle\{v\}|\{u\}\rangle=(-1)^M S_M\left(\{v^*\},\{u\}\right)\,.
\end{align}

\paragraph{Recursion relations} The scalar product $S_M(\{v\},\{u\})$ satisfies recursion relations that can also be used to compute it numerically in a straightforward manner. There are two kinds of such relations. The first one has been known for some years \cite{Korepin:1982gg}.
We apply a different recursion relation, which was derived in \cite{Escobedo:2010xs} and given by
\begin{align}
S_M(\{v_1,\ldots,v_M\},\{u_1,\ldots,u_M\})=\sum_n b_n S_{M-1}(\{v_1,\ldots,\hat{v}_n,\ldots,v_M\},\{\hat{u}_1,u_2,\ldots,u_M\})\\\nonumber
-\sum_{n<m}c_{n,m}\,S_{M-1}(\{u_1,v_1,\ldots,\hat{v}_n,\ldots,\hat{v}_m,\ldots,v_M\},\{\hat{u}_1,u_2,\ldots,u_M\})\,.
\end{align}
 where $\hat{u}_j$ and $\hat{v}_k$ mean that the corresponding rapidity is missing. Here $b_n$ and $c_{n,m}$ are given by
 \begin{align}
 b_n=g(u_1-v_n)a(v_n)d(u_1)\prod_{j\ne n} f(u_1-v_j)f(v_j-v_n)+(u_1\leftrightarrow v_n)
 \end{align}
 and
 \begin{align}
 c_{n,m}=g(u_1-v_n)g(u_1-v_m)a(v_m)d(v_n)f(v_n-v_m)\prod_{j\ne n,m}^M f(v_n-v_j)f(v_j-v_m)+(n\leftrightarrow m)\,.
 \end{align}

\section{The Gram--Schmidt process}
\label{app:GSprocess}
In this appendix, we review the Gram--Schmidt orthogonalization procedure and provide an explicit construction for the matrices $\mathbf{L}^{(n)}$ and $\mathbf{R}^{(n)}$.
\paragraph{Gram--Schmidt process} We begin with a set of linearly independent vectors $\{|a_i\rangle\}_{i=1}^M$ that are generally not orthonormal. From these we construct an orthonormal set $\{|e_i\rangle\}_{i=1}^M$ as follows:
\begin{itemize}
\item Define the first orthonormal vector
\begin{align}
|e_1\rangle=\frac{|a_1\rangle}{\sqrt{\langle a_1|a_1\rangle}}
\end{align}
\item For the second vector, define
\[
|e'_2\rangle = |a_2\rangle - |e_1\rangle\langle e_1|a_2\rangle
             = |a_2\rangle - \frac{\langle a_1|a_2\rangle}{\langle a_1|a_1\rangle}|a_1\rangle.
\]
This satisfies \(\langle e_1|e'_2\rangle = 0\). After normalization,
\[
|e_2\rangle = \frac{|e'_2\rangle}{\sqrt{\langle e'_2|e'_2\rangle}}.
\]

\item For the third vector,
\[
|e'_3\rangle = |a_3\rangle - |e_1\rangle\langle e_1|a_3\rangle - |e_2\rangle\langle e_2|a_3\rangle,
\qquad 
|e_3\rangle = \frac{|e'_3\rangle}{\sqrt{\langle e'_3|e'_3\rangle}}.
\]

\item Iterating the process, for \( m \ge 1 \) define
\[
|e'_{m+1}\rangle = |a_{m+1}\rangle - \sum_{i=1}^{m} |e_i\rangle\langle e_i|a_{m+1}\rangle,
\qquad
|e_{m+1}\rangle = \frac{|e'_{m+1}\rangle}{\sqrt{\langle e'_{m+1}|e'_{m+1}\rangle}},
\]
until \( |e_M\rangle \) is obtained.
\end{itemize}
We also require the expansion of the original states $|a_i\rangle$ in the new orthonormal basis $\{|e_i\rangle\}$. For convenience, define
\begin{align}
\alpha_{ij}\equiv\langle a_i|a_j\rangle\,,
\end{align}
so that $\alpha_{ji}=\alpha^*_{ij}$ and the diagonal entries $\alpha_{ii}$ are real. To find $\langle a_j|e_n\rangle$ in terms of $\alpha_{ij}$, note first that
\begin{align}
\langle a_j|e_n\rangle=\frac{\langle a_j|e'_n\rangle}{\sqrt{\langle e'_n|e'_n\rangle}}\,.
\end{align}
Using the recurrence
\begin{align}
|e'_{m+1}\rangle=|a_{m+1}\rangle-\sum_{k=1}^m\frac{\langle e'_k|a_{m+1}\rangle}{\langle e'_k|e'_k\rangle}|e'_k\rangle\,,
\end{align}
we obtain
\begin{align}
\label{eq:overlapRec}
\langle a_j|e'_{m+1}\rangle=\alpha_{j,m+1}-\sum_{k=1}^m\frac{\langle a_j|e'_k\rangle\langle e'_k|a_{m+1}\rangle}{\langle e'_k|e'_k\rangle}\,.
\end{align}
Introduce the shorthand notation
\begin{align}
\beta_{jn}=\langle a_j|e'_n\rangle,\qquad \beta^*_{jn}=\langle e'_n|a_j\rangle,\qquad \gamma_n=\langle e'_n|e'_n\rangle
\end{align}
Then
\begin{align}
|e'_{m+1}\rangle=|a_{m+1}\rangle-\sum_{k=1}^m\frac{\beta^*_{m+1,k}}{\gamma_k}|e'_k\rangle
\end{align}
and
\begin{align}
\beta_{j,m+1}=\alpha_{j,m+1}-\sum_{k=1}^m\frac{\beta^*_{m+1,k}}{\gamma_k}\beta_{j,k}
\end{align}
The norm $\gamma_{m+1}$ is given by
\begin{align}
\gamma_{m+1}=&\,\left(\langle a_{m+1}|-\sum_{j=1}^m\frac{\beta_{m+1,j}}{\gamma_j}\langle e'_j|\right)
\left(|a_{m+1}\rangle-\sum_{k=1}^m\frac{\beta^*_{m+1,k}}{\gamma_k}|e'_k\rangle\right)\\\nonumber
=&\,\alpha_{m+1,m+1}-\sum_{j=1}^m\frac{|\beta_{m+1,j}|^2}{\gamma_j}
\end{align}
Finally,
\begin{align}
\langle a_j|e_n\rangle=\frac{\beta_{jn}}{\sqrt{\gamma_n}}
\end{align}
Closed-form expressions for $\beta_{jn}$ and $\gamma_n$ in terms of $\alpha_{ij}$ are increasingly lengthy as $n$ grows. Nevertheless, given numerical $\alpha_{ij}$, the recurrence relations allow efficient computation of $\beta_{jn}$ and $\gamma_n$. In practice, the main computational cost lies in constructing the overlaps $\alpha_{ij}$ themselves.
\paragraph{First few calculations} Explicit calculations for the initial steps are:
\begin{itemize}
\item $m=0$
\begin{align}
\beta_{j,1}=\alpha_{j,1},\qquad \beta_{j,1}^*=\alpha_{1,j},\qquad \gamma_1=\alpha_{1,1}
\end{align}
\item For \(m=1\):
\begin{align*}
\beta_{j,2} &= \alpha_{j,2} - \frac{\beta_{2,1}^*}{\gamma_1} \beta_{j,1},\\
\beta_{j,2}^* &= \alpha_{2,j} - \frac{\beta_{2,1}}{\gamma_1} \beta_{j,1}^*,\\
\gamma_2 &= \alpha_{2,2} - \frac{|\beta_{2,1}|^2}{\gamma_1}.
\end{align*}
\end{itemize}
The iteration can be continued straightforwardly: at each step $m$, we compute $\beta_{j,m}$, $\beta^*_{j,m}$ and $\gamma_m$, then proceed to $\beta_{j,m+1}$, $\beta^*_{j,m+1}$ and $\gamma_{m+1}$ until the desired order $M$ is reached. The original states expand in the orthonormal basis as
\begin{align}
|a_j\rangle=\sum_{k=1}^M\frac{\beta^*_{j,k}}{\sqrt{\gamma_k}}|e_k\rangle
\end{align}
The main-text convention in \eqref{eq:HLRdecomp} is
\(|a_j^{(n)}\rangle\!\rangle_A=\sum_k L_{jk}^{(n)}|e_k^{(n)}\rangle\!\rangle_A\):
the row index \(j\) labels the original non-orthonormal subchain state, and the column index \(k\) labels the orthonormal Gram--Schmidt vector. Comparing with the expansion above gives
\begin{align}
\label{eq:Lmatrix_entries}
L_{jk}^{(n)}=\frac{\beta^*_{j,k}}{\sqrt{\gamma_k}}\,,
\end{align}
with no additional transpose in the definition of \(\mathbf{L}^{(n)}\). Here $\beta_{j,k}$ and $\gamma_k$ are obtained from the recurrences above with the overlaps $\alpha_{ij}$ taken to be those of the $A$-subchain Bethe states, $\alpha_{ij}={}_A\langle\!\langle a_i^{(n)}|a_j^{(n)}\rangle\!\rangle_A$. The matrix $\mathbf{R}^{(n)}$ is given by the same row/column convention and the same formula~\eqref{eq:Lmatrix_entries}, with $\alpha_{ij}$ replaced by the corresponding $B$-subchain overlaps ${}_B\langle\!\langle\bar{a}_i^{(n)}|\bar{a}_j^{(n)}\rangle\!\rangle_B$.

\section{Invariance of Entanglement Entropy}
\label{app:EEinvariance}

\subsection{Global spin flip}

The global spin-flip operator for the XXX$_{\frac{1}{2}}$ spin chain is a tensor product of single-site unitaries,
$U_{\text{flip}}=\bigotimes_{n=1}^L \sigma_n^x$.
For any bipartition of the chain into subsystems $(A,B)$, it factorizes as
\beq
U_{\text{flip}} = U_A \otimes U_B, \qquad
U_A=\bigotimes_{n\in A}\sigma_n^x,\ \ U_B=\bigotimes_{n\in B}\sigma_n^x .
\eeq
If the full state is $\rho$, then under the global spin flip it transforms as
\beq
\rho \ \mapsto\ \rho'=(U_A\otimes U_B)\,\rho\, (U_A^\dagger\otimes U_B^\dagger).
\eeq
Taking the partial trace over $B$ gives
\beq
\rho'_A=\mathrm{Tr}_B(\rho')
= U_A\,\mathrm{Tr}_B(\rho)\,U_A^\dagger
=U_A\rho_A U_A^\dagger,
\eeq
where we used the cyclicity of the trace under unitary conjugation on subsystem $B$. Therefore $\rho_A$ and $\rho^{'}_A$ are unitarily similar and have the same spectrum, so any entanglement measure that depends only on the eigenvalues of $\rho_A$, in particular the von Neumann entropy, remains unchanged:
\(\mathcal{S}(\rho_A)=\mathcal{S}(\rho^{'}_A)\).

\subsection{Global $\mathbf{\Pi}$ or $\mathbf{\Pi} \mathcal{T}$}

For the Lax operator, define the conjugation matrix
\begin{equation}
\mathcal{C} = \ri \sigma_y
=\begin{pmatrix}
0 & 1 \\
-1 & 0
\end{pmatrix},
\qquad \mathcal{C}^2=-\mathbb{I}.
\end{equation}
A direct computation gives
\begin{equation}
\mathcal{C}\, \mathbf{L}_{an}(u)^{t_a}\, \mathcal{C}^{-1}
= -\mathbf{L}_{an}(-u),
\end{equation}
where $t_a$ denotes transposition in the auxiliary space.

We now introduce the reflection operator $\mathbf{\Pi}$, which reverses the ordering of sites along the chain. Its action on the monodromy matrix is
\beq
\mathbf{\Pi}\, \mathbf{M}_a(u)\, \mathbf{\Pi}^{-1} = \mathbf{L}_{aL}(u) \cdots \mathbf{L}_{a1}(u)\, .
\eeq
Using the definition of the monodromy matrix and applying the identity above site by site, we obtain
\begin{equation}
\mathbf{\Pi}\, \mathbf{M}_a(u)\, \mathbf{\Pi}^{-1}
= (-1)^L \mathcal{C}\,\mathbf{M}_a(-u)^{t_a}\,\mathcal{C}^{-1}
= (-1)^L
\begin{pmatrix}
\mathbf{D}(-u) & -\mathbf{B}(-u)\\
-\mathbf{C}(-u) & \mathbf{A}(-u)
\end{pmatrix}_a .
\end{equation}
In particular,
\begin{equation}
\mathbf{\Pi}\mathbf{B}(u)\mathbf{\Pi}^{-1}=(-1)^{L+1}\mathbf{B}(-u).
\end{equation}

The pseudovacuum state is invariant under $\mathbf{\Pi}$. Therefore, the off-shell Bethe state  
\beq
|\{u_j\}\rangle =  \mathbf{B}(u_1) \cdots  \mathbf{B}(u_M) |0\rangle\, ,
\eeq
transforms as
\beq
\mathbf{\Pi} |\{u_j\}\rangle = (-1)^{M(L+1)} |\{-u_j\}\rangle\, .
\eeq

\paragraph{Complex conjugation of the Lax operator.}
Since $S^z, S^\pm$ have real matrix elements in the computational basis,
taking the complex conjugate of the Lax operator gives
\begin{equation}
\mathbf{L}_{an}(u)^* 
= \begin{pmatrix} u^* - \ri S_n^z & -\ri S_n^- \\ -\ri S_n^+ & u^* + \ri S_n^z \end{pmatrix}
= -\mathbf{L}_{an}(-u^*)\,.
\end{equation}
Applying this identity site by site to the monodromy matrix gives
\begin{equation}
\mathbf{M}_a(u)^* 
= \prod_{n=1}^L \mathbf{L}_{an}(u)^* 
= \prod_{n=1}^L \bigl[-\mathbf{L}_{an}(-u^*)\bigr] 
= (-1)^L\, \mathbf{M}_a(-u^*)\,.
\end{equation}
Extracting the $(1,2)$ auxiliary-space element:
\begin{equation}
\mathbf{B}(u)^* = (-1)^L\, \mathbf{B}(-u^*)\,.
\end{equation}

\paragraph{Action of $\mathbf{\Pi}\mathcal{T}$ on off-shell Bethe states.}
Since the pseudovacuum is real, $|0\rangle^* = |0\rangle$, applying the
conjugation identity $M$ times gives
\begin{equation}
\overline{|\{u_j\}\rangle} 
= \mathbf{B}(u_1)^* \cdots \mathbf{B}(u_M)^*\,|0\rangle
= (-1)^{ML}\,|\{-u_j^*\}\rangle\, .
\end{equation}
Applying $\mathbf{\Pi}$ and using the parity result:
\begin{equation}
\mathbf{\Pi}\,\overline{|\{u_j\}\rangle}
= (-1)^{ML}\,\mathbf{\Pi}\,|\{-u_j^*\}\rangle
= (-1)^{ML}\cdot(-1)^{M(L+1)}\,|\{u_j^*\}\rangle
= (-1)^{M}\,|\{u_j^*\}\rangle\,.
\end{equation}

\paragraph{Entanglement entropy symmetry.}

Write the Bethe state in the Hilbert space,
\beq
|\{u_j\}\rangle = \sum_{s_1,\ldots,s_L} c_{s_1\cdots s_L}\,|s_1\cdots s_L\rangle\, .
\eeq
where $s_j = 0,1$. For the bipartition of the first $\ell$ sites from the remaining $L-\ell$ sites,
define the Schmidt matrix
\begin{equation}
\mathsf{M}^{(\ell)}_{I,J} = c_{IJ}\,,
\qquad I\in\{0,1\}^{\otimes \ell},\quad J\in\{0,1\}^{\otimes(L-\ell)},
\end{equation}
and for the complementary bipartition
\begin{equation}
\mathsf{N}^{(L-\ell)}_{I',J'} = c_{I'J'}\,,
\qquad I'\in\{0,1\}^{\otimes(L-\ell)},\quad J'\in\{0,1\}^{\otimes \ell}.
\end{equation}
We prove $\mathcal{S}_{\ell}=\mathcal{S}_{L-\ell}$ under either of two conditions on the rapidities.

\paragraph{Case 1: $\{-u_k\} = \{u_k\}$.}

The state is a $\mathbf{\Pi}$-eigenstate, $\mathbf{\Pi}|\Psi\rangle = (-1)^{M(L+1)}|\Psi\rangle$,
which in components reads
\begin{equation}
c_{s_L\cdots s_1} = (-1)^{M(L+1)}\,c_{s_1\cdots s_L}\,.
\end{equation}
Setting $s_1\cdots s_\ell = J'^R$ and $s_{\ell+1}\cdots s_L = I'^R$
(where $X^R$ denotes the bit-reversed index) in the relation above:
\begin{equation}
\mathsf{N}^{(L-\ell)}_{I'J'} = c_{I'J'} 
= (-1)^{M(L+1)}\,c_{J'^R I'^R}
= (-1)^{M(L+1)}\,\mathsf{M}^{(\ell)}_{J'^R,\,I'^R}\,.
\end{equation}
In matrix form, with $P_\ell$ and $P_{L-\ell}$ the bit-reversal permutation matrices
(which are unitary and real):
\begin{equation}
\mathsf{N}^{(L-\ell)} = (-1)^{M(L+1)}\,P_{L-\ell}\,\bigl(\mathsf{M}^{(\ell)}\bigr)^T P_\ell\,.
\end{equation}
Since permutation matrices are unitary and transposition preserves singular values,
\begin{equation}
\mathrm{SVD}\bigl(\mathsf{N}^{(L-\ell)}\bigr) = \mathrm{SVD}\bigl(\mathsf{M}^{(\ell)}\bigr)
\implies \mathcal{S}_{\ell}=\mathcal{S}_{L-\ell}\,. 
\end{equation}

\paragraph{Case 2: $\{u_k^*\} = \{u_k\}$.}

The state satisfies $\mathbf{\Pi}\,\overline{|\Psi\rangle} = (-1)^M|\Psi\rangle$,
which in components reads
\begin{equation}
\overline{c_{s_L\cdots s_1}} = (-1)^M\,c_{s_1\cdots s_L}\,.
\end{equation}
The same index relabeling as in Case 1 gives
\begin{equation}
\mathsf{N}^{(L-\ell)}_{I'J'} = c_{I'J'}
= (-1)^M\,\overline{c_{J'^R I'^R}}
= (-1)^M\,\overline{\mathsf{M}^{(\ell)}_{J'^R,\,I'^R}}\,,
\end{equation}
hence in matrix form:
\begin{equation}
\mathsf{N}^{(L-\ell)} = (-1)^M\,P_{L-\ell}\,\overline{\bigl(\mathsf{M}^{(\ell)}\bigr)^T}\,P_\ell\,.
\end{equation}
The overall phase $(-1)^M$ does not affect singular values;
complex conjugation and transposition each preserve singular values;
$P_{L-\ell}$ and $P_\ell$ are unitary.  Therefore
\begin{equation}
\mathrm{SVD}\bigl(\mathsf{N}^{(L-\ell)}\bigr) = \mathrm{SVD}\bigl(\mathsf{M}^{(\ell)}\bigr)
\implies \mathcal{S}_{\ell}=\mathcal{S}_{L-\ell}\,.
\end{equation}

\paragraph{On-shell Bethe states.}
We show that every non-degenerate on-shell Bethe state satisfies
the condition of Case~2 and hence has \(\mathcal{S}_{\ell}=\mathcal{S}_{L-\ell}\).

The Bethe equations are
\begin{equation}
\left(\frac{u_k+\ri/2}{u_k-\ri/2}\right)^L
= \prod_{j\neq k}\frac{u_k-u_j+\ri}{u_k-u_j-\ri}\,,
\qquad k=1,\ldots,M\,.
\end{equation}
Taking the complex conjugate of both sides and inverting:
\begin{equation}
\left(\frac{u_k^*+\ri/2}{u_k^*-\ri/2}\right)^L
= \prod_{j\neq k}\frac{u_k^*-u_j^*+\ri}{u_k^*-u_j^*-\ri}\,.
\end{equation}
This is exactly the Bethe equations for the set $\{u_k^*\}$,
so $\{u_k^*\}$ is also a valid solution whenever $\{u_k\}$ is.
This is precisely the hypothesis of Case~2, and the conclusion
\(\mathcal{S}_{\ell}=\mathcal{S}_{L-\ell}\) follows immediately.

\section{Off-shell Bethe-state construction and optimizer implementation} \label{sec:TechDetailOffShell}

\subsection{Bethe state construction via Lax propagation}
\label{subsec:bethe_state_construction}

\paragraph{Brute force viewpoint and why it is impractical.}
A conceptually straightforward implementation would be to represent the operator $\mathbf{B}(u)$ as an explicit linear operator on a $d$-dimensional state space,
\begin{equation}
\mathbf{B}(u)\ \in\ \mathrm{End}(\mathcal{V})\simeq \mathbb{C}^{d\times d},
\end{equation}
and then construct the off-shell Bethe vector by repeated matrix multiplications
\begin{equation}
|\{u_j\}\rangle = \mathbf{B}(u_1)\cdots \mathbf{B}(u_M)\,|0\rangle.
\end{equation}
However, storing even a single dense matrix of size $d\times d$ is impractical for the system sizes of interest: memory scales as $\mathcal{O}(d^2)$ and a single multiplication costs $\mathcal{O}(d^2)$ operations. Here $d$ may be either $2^L$ (the full Hilbert space dimension) or $\binom{L}{M}$ (the dimension of the magnetization subspace, see Section~\ref{subsec:magnon_subspace}), both of which grow rapidly with~$L$.

\paragraph{Key idea: propagate the product.}
Recall that $\mathbf{B}(u)$ is the $(1,2)$ entry of the monodromy matrix $\mathbf{M}_a(u)$, which is an ordered product of local Lax operators:
\begin{equation}
\mathbf{M}_a(u)=\mathbf{L}_{a1}(u)\,\mathbf{L}_{a2}(u)\cdots\mathbf{L}_{aL}(u).
\end{equation}
Therefore $\mathbf{B}(u)\,|\psi\rangle$ can be obtained by \emph{propagating the action of this product} on a state $|\psi\rangle$, without ever constructing the full operator $\mathbf{B}(u)$.

\paragraph{Step 1: introduce a running partial product acting on $|\psi\rangle$.}
Define the partial monodromy
\begin{equation}
\mathbf{M}_a^{(n)}(u)=\mathbf{L}_{a1}(u)\cdots\mathbf{L}_{an}(u),\qquad n=0,1,\ldots,L,
\end{equation}
with $\mathbf{M}_a^{(0)}(u)=\mathbf{1}_a$. When $\mathbf{M}_a^{(n)}(u)$ acts on $|\psi\rangle$, it produces an auxiliary-space doublet whose components live in the physical space. We package the result as a $2\times 2$ matrix in auxiliary indices whose entries are \emph{state vectors}:
\begin{equation}
\mathbf{M}_a^{(n)}(u)\,|\psi\rangle \;\equiv\;
\begin{pmatrix}
V_{11}^{(n)} & V_{12}^{(n)}\\
V_{21}^{(n)} & V_{22}^{(n)}
\end{pmatrix},
\qquad
V_{\alpha\beta}^{(n)}\in\mathcal{V}.
\end{equation}
At $n=0$ this is simply
\begin{equation}
V_{11}^{(0)}=|\psi\rangle,\qquad V_{22}^{(0)}=|\psi\rangle,\qquad V_{12}^{(0)}=0,\qquad V_{21}^{(0)}=0.
\end{equation}

\paragraph{Step 2: update rule from multiplying by one more Lax operator.}
Because
\begin{equation}
\mathbf{M}_a^{(n)}(u)=\mathbf{M}_a^{(n-1)}(u)\,\mathbf{L}_{an}(u),
\end{equation}
the vectors $V_{\alpha\beta}^{(n)}$ follow from right-multiplying the $V$-matrix by $\mathbf{L}_{an}(u)$ in auxiliary space. Substituting the Lax matrix entries from~\eqref{eq:defLax}, each of which acts only on site~$n$, we obtain the recursion
\begin{equation}
\begin{pmatrix}
V_{11}^{(n)} & V_{12}^{(n)}\\
V_{21}^{(n)} & V_{22}^{(n)}
\end{pmatrix}
=
\begin{pmatrix}
V_{11}^{(n-1)} & V_{12}^{(n-1)}\\
V_{21}^{(n-1)} & V_{22}^{(n-1)}
\end{pmatrix}
\begin{pmatrix}
u+\ri\,\mathbf{S}_n^z & \ri\,\mathbf{S}_n^- \\
\ri\,\mathbf{S}_n^+ & u-\ri\,\mathbf{S}_n^z
\end{pmatrix}.
\label{eq:lax_update_matrix}
\end{equation}
Since the $V$-matrix entries act on sites $1,\ldots,n{-}1$ while the Lax entries act on site~$n$, these operators commute and the matrix product is well-defined.
Explicitly, each new entry is a linear combination of the previous ones with one-site operators:
\begin{align}
V_{11}^{(n)} &= (u+\ri\,\mathbf{S}_n^z)\,V_{11}^{(n-1)}+\ri\,\mathbf{S}_n^+\,V_{12}^{(n-1)},\label{eq:lax_comp_V11}\\
V_{12}^{(n)} &= \ri\,\mathbf{S}_n^-\,V_{11}^{(n-1)}+(u-\ri\,\mathbf{S}_n^z)\,V_{12}^{(n-1)},\label{eq:lax_comp_V12}\\
V_{21}^{(n)} &= (u+\ri\,\mathbf{S}_n^z)\,V_{21}^{(n-1)}+\ri\,\mathbf{S}_n^+\,V_{22}^{(n-1)},\\
V_{22}^{(n)} &= \ri\,\mathbf{S}_n^-\,V_{21}^{(n-1)}+(u-\ri\,\mathbf{S}_n^z)\,V_{22}^{(n-1)}.
\end{align}
The algorithm starts from $(V_{\alpha\beta}^{(0)})$ and applies the above update for $n=1,2,\ldots,L$. The only operations involved are applications of \emph{single-site} operators to $d$-dimensional state vectors.

\paragraph{Step 3: read off the $(1,2)$ entry after the full product.}
After all $L$ local factors have been incorporated, $\mathbf{M}_a^{(L)}(u)=\mathbf{M}_a(u)$. Since $\mathbf{B}(u)$ is by definition the $(1,2)$ entry of $\mathbf{M}_a(u)$, its action on $|\psi\rangle$ is
\begin{equation}
\mathbf{B}(u)\,|\psi\rangle \;=\; V_{12}^{(L)}.
\end{equation}
This achieves the desired result with cost $\mathcal{O}(L \times d)$ per evaluation of $\mathbf{B}(u)\,|\psi\rangle$, rather than the $\mathcal{O}(d^2)$ cost of forming $\mathbf{B}(u)$ as a dense matrix. Constructing an $M$-magnon state then costs $\mathcal{O}(M\times L\times d)$, which is optimal up to constant factors when the full state vector must be stored explicitly.

The recursion \eqref{eq:lax_update_matrix} is valid in any invariant subspace of the spin chain; the choice of working space (the full Hilbert space with $d=2^L$, or the $M$-magnon sector with $d=\binom{L}{M}$) is discussed next.

\subsection{Restriction to the magnetization subspace}
\label{subsec:magnon_subspace}

In this and the following subsection we present the XXX$_{\frac{1}{2}}$ construction explicitly because the index combinatorics are simplest there. The higher-spin XXX$_s$ and non-compact $SL(2,\mathbb{R})$ adaptations require only the substitutions of the magnon-sector and Schmidt-block dimensions stated in the paragraph at the end of this subsection. The Lax recursion of Section~\ref{subsec:bethe_state_construction}, the per-column streaming, the adaptive block-diagonal SVD of Section~\ref{subsec:entropy_svd}, and the autograd machinery of Section~\ref{subsec:complex_autodiff} all carry over verbatim once those dimensions are substituted.

\paragraph{U(1) symmetry and subspace restriction.}
The operator $\mathbf{B}(u)$ maps the $M$-magnon sector $\mathcal{H}_{M}$ to the $(M+1)$-magnon sector $\mathcal{H}_{M+1}$. Starting from the pseudovacuum $|0\rangle\in\mathcal{H}_0$ and applying $\mathbf{B}(u)$ a total of $M$ times, the resulting Bethe state resides in $\mathcal{H}_{M}$. The dimension of $\mathcal{H}_{M}$ is
\begin{equation}
\dim\mathcal{H}_{M}=\binom{L}{M},
\end{equation}
which is much smaller than $2^L$ near half filling.

Working in the $M$-magnon subspace rather than the full Hilbert space is the largest single improvement: memory and computational cost both decrease proportionally.

\paragraph{Basis enumeration.}
The $M$-magnon basis states are in bijection with the $M$-element subsets of $\{1,\ldots,L\}$: each subset specifies which sites host a down-spin. These $\binom{L}{M}$ states are enumerated combinatorially and indexed in a fixed canonical order. No filtering of the full $2^L$-dimensional space is necessary.

\paragraph{Sparse operator application in the magnon basis.}
The Lax recursion \eqref{eq:lax_update_matrix} requires the action of three one-site operators $\mathbf{S}_n^z$, $\mathbf{S}_n^-$, and $\mathbf{S}_n^+$ on states in $\mathcal{H}_{M}$. Within the magnon basis, each of these operators is sparse:
\begin{itemize}
\item $\mathbf{S}_n^z$ is \emph{diagonal}: its eigenvalue on a basis state depends only on whether site $n$ carries an up- or down-spin. No change of magnon number occurs.
\item $\mathbf{S}_n^-$ maps a basis state in $\mathcal{H}_{M}$ to $\mathcal{H}_{M+1}$ by flipping spin-up $\to$ spin-down at site~$n$, provided site~$n$ is currently up; otherwise the result is zero. Each basis state maps to at most one target state.
\item $\mathbf{S}_n^+$ similarly maps $\mathcal{H}_{M}\to\mathcal{H}_{M-1}$ by flipping spin-down $\to$ spin-up at site~$n$, with the same sparsity.
\end{itemize}
These maps are implemented via precomputed index tables: for each site~$n$ and magnon number~$M$, a table records the target basis index (in $\mathcal{H}_{M\pm 1}$) and a Boolean flag indicating whether the action is nonzero. Applying $\mathbf{S}_n^\pm$ to a state vector then only requires adding each nonzero contribution to its target basis component, with no matrix--vector multiplication required.

\paragraph{Per-column streaming.}
The index tables for all $L$ sites can be large (scaling as $\binom{L}{m}\times L$), and for the largest system sizes studied ($L\ge 26$), storing all tables simultaneously on the GPU exceeds available memory. We therefore transfer only the index column for the current site~$n$ to the GPU, use it in the Lax update, and release it before processing site~$n{+}1$. This site-by-site transfer, which we call \emph{per-column streaming}, reduces the persistent GPU memory for index tables from $\mathcal{O}(\binom{L}{m}\times L)$ to $\mathcal{O}(\binom{L}{m})$.

\paragraph{Complexity.}
The cost of a single application $\mathbf{B}(u)\,|\psi\rangle$ in the $m$-magnon subspace is $\mathcal{O}(L\times\binom{L}{m})$. Constructing the full $M$-magnon Bethe state requires $M$ such applications (with $m=0,1,\ldots,M{-}1$), giving a total cost of $\mathcal{O}(M\times L\times\binom{L}{M})$, where $\binom{L}{M}$ dominates the sum.

\paragraph{Generalization to higher-spin and non-compact chains.}
The construction above is stated for XXX$_{\frac{1}{2}}$, where the $M$-magnon sector has dimension $\binom{L}{M}$. The same Lax-propagation recursion and per-site streaming apply to the other models treated in this paper, with only the subspace dimension replaced. For the XXX$_s$ chain of Section~\ref{sec:highSpin}, each site admits up to $2s$ magnons and the $M$-magnon sector has dimension
\begin{align}
\label{eq:dim_xxxs_M}
\dim\mathcal{H}_M^{(s)} = \bigl|\{(n_1,\ldots,n_L)\in\mathbb{Z}_{\ge 0}^{L}:\,n_j\le 2s,\;\textstyle\sum_j n_j=M\}\bigr|\,,
\end{align}
which reduces to $\binom{L}{M}$ at $s=\frac{1}{2}$ and to the number of weak compositions of $M$ with at most $L$ parts each bounded by $2s$ in general. For the non-compact $SL(2,\mathbb{R})$ chain of Section~\ref{sec:nonComp}, each site hosts an unbounded number of oscillator quanta. The exact fixed-\(M\) sector dimension is the untruncated weak-composition count
\begin{align}
\label{eq:dim_sl2_M}
\dim\mathcal{H}_M^{\text{SL(2)}} = \binom{M+L-1}{L-1}\,,
\end{align}
which is a polynomial of degree $L-1$ in $M$. The local oscillator cutoff \(M_{\mathrm{trunc}}=M+L\) used in the implementation is a cutoff for intermediate or full-space oscillator representations; it does not change \eqref{eq:dim_sl2_M}, because in an \(M\)-magnon state no site can carry more than \(M\) quanta. The block-diagonal SVD of Section~\ref{subsec:entropy_svd} adapts directly: the block labeled by $m_A$ magnons in $A$ has dimensions $\binom{\ell_A}{m_A}\binom{L-\ell_A}{M-m_A}$ for XXX$_{\frac{1}{2}}$, the analogous bounded-composition counts for XXX$_s$, and $\binom{m_A+\ell_A-1}{\ell_A-1}\binom{M-m_A+\ell_B-1}{\ell_B-1}$ in the non-compact case. The remaining machinery is unchanged up to the substitution of $\binom{L}{M}$ by the model-appropriate dimension. This includes per-column streaming of spin-flip index tables, the adaptive SVD strategy, and gradient checkpointing.

\subsection{Entanglement entropy via block-diagonal SVD}
\label{subsec:entropy_svd}

\paragraph{Reduced density matrix and Schmidt decomposition.}
Fix a bipartition $A\cup B$ with $|A|=\ell_A$ and $|B|=\ell_B=L-\ell_A$ (in our numerics we mostly take $\ell_A=L/2$).
Given a normalized pure state
\begin{equation}
|\psi\rangle=\sum_{\sigma_A,\sigma_B}\psi_{\sigma_A\sigma_B}\,|\sigma_A\rangle\otimes|\sigma_B\rangle,
\qquad
\sum_{\sigma_A,\sigma_B}|\psi_{\sigma_A\sigma_B}|^2=1,
\end{equation}
the reduced density matrix on $A$ is
\begin{equation}
\hat{\rho}_A=\mathrm{Tr}_B\!\left(|\psi\rangle\langle\psi|\right),
\qquad
(\hat{\rho}_A)_{\sigma_A,\tau_A}=\sum_{\sigma_B}\psi_{\sigma_A\sigma_B}\,\psi^{*}_{\tau_A\sigma_B}.
\label{eq:rhoA_def_components}
\end{equation}
Regarding the amplitudes as a rectangular matrix $\Psi_{\sigma_A,\sigma_B}\equiv \psi_{\sigma_A\sigma_B}$, the partial trace becomes
\begin{equation}
\hat{\rho}_A=\Psi\,\Psi^\dagger.
\label{eq:rhoA_psipsidag}
\end{equation}
The eigenvalues of $\hat{\rho}_A$ are the squared singular values of $\Psi$: performing an SVD $\Psi = U\,\mathrm{diag}(s_k)\,V^\dagger$ yields $\lambda_k = s_k^2$, and the von Neumann entropy is
\begin{equation}
\mathcal{S}_\ell=-\sum_{k}\lambda_k\log\lambda_k
=-\sum_{k} s_k^2\log(s_k^2).
\label{eq:entropy_from_svd}
\end{equation}

\paragraph{Block-diagonal decomposition by magnon number.}
As discussed in Section~\ref{sec:EEint} (cf.\ \eqref{eq:detailedSum}--\eqref{eq:decomposeState}), the U(1) magnetization symmetry implies that states with different magnon numbers in subsystem~$A$ are orthogonal. Consequently, the wavefunction matrix $\Psi$ decomposes into blocks labeled by $m_A$, the number of magnons in subsystem~$A$:
\begin{equation}
\Psi = \bigoplus_{m_A=m_A^{\min}}^{m_A^{\max}} \Psi^{(m_A)},\qquad
\Psi^{(m_A)} \in \mathbb{C}^{\binom{\ell_A}{m_A}\times \binom{\ell_B}{M-m_A}},
\label{eq:block_diagonal_psi}
\end{equation}
where $m_A$ ranges over $\max(0,M{-}\ell_B)\le m_A \le \min(M,\ell_A)$. The singular values of $\Psi$ are the union of the singular values of the individual blocks $\Psi^{(m_A)}$, so the SVD can be performed \emph{independently} for each block, which is cheaper than a single SVD of the full matrix.

\paragraph{Adaptive SVD strategy.}
The block dimensions $\binom{\ell_A}{m_A}\times \binom{\ell_B}{M{-}m_A}$ vary widely as $m_A$ ranges over its allowed values, so we choose the method separately for each block. For blocks whose smaller dimension is $\le 500$ we compute the singular values with a dense SVD. For $500<\min(\dim)\le 2000$ we instead diagonalize the smaller of the two Gram matrices $\Psi^{(m_A)}(\Psi^{(m_A)})^\dagger$ and $(\Psi^{(m_A)})^\dagger\Psi^{(m_A)}$, whose eigenvalues are the squared singular values; this is still exact. Only for $\min(\dim)>2000$ do we fall back to an approximate low-rank SVD, projecting the block onto a randomly sampled $k=500$-dimensional subspace and applying two power iterations to improve the dominant singular directions. This last branch is not exercised by any result in the present work: across all models, magnon sectors, and cuts entering the tabulated data, the largest Schmidt block has minimum dimension $462$ (the XXX$_{\frac{1}{2}}$ off-shell sector $L=22$, $M=10$ at the central cut), so every displayed entropy is computed by the exact dense-SVD branch. The randomized branch is an infrastructure provision for $L\gtrsim 28$ runs that lie beyond the data presented here.

\paragraph{Numerical stabilizers.}
In exact arithmetic, some $\lambda_k$ can vanish; then $\lambda_k\log\lambda_k$ is zero by continuity, but floating-point evaluation of $\log(0)$ is problematic. We therefore introduce a small floor:
\begin{equation}
\lambda_k \leftarrow \max(\lambda_k,\varepsilon),
\label{eq:lambda_clamp}
\end{equation}
and renormalize so that $\sum_k\lambda_k=1$ after clamping. These regularizations do not affect the entropy at the level of displayed digits, but they keep gradients finite during optimization.

\subsection{Gradient-based optimization and automatic differentiation}
\label{subsec:complex_autodiff}

Section~\ref{sec:offshell} gives the conceptual optimization strategy: Adam, automatic differentiation, random restarts, and state renormalization. The XXX$_{\frac{1}{2}}$ implementation used the following default run-level settings: \(R=30\) random restarts, matching Section~\ref{subsec:algorithm_summary_physics}, step sizes \(\eta_{\min}=0.03\) and \(\eta_{\max}=0.01\), initial rapidity sets sampled with \(|u_j|\sim0.8\), rejection of runs that return undefined numerical values \texttt{NaN}, and identification of permutation- or complex-conjugation-related duplicates. This subsection describes some details for the implementation including: complex-parameter handling via Wirtinger calculus, gradient checkpointing for memory-bounded autograd at large \(L\), and \(L\)-dependent maximum step counts.

\paragraph{Complex parameters and Wirtinger calculus.}
The rapidities $u_j\in\mathbb{C}$ are stored directly as complex-valued variational parameters. Since the objective $\mathcal{S}_\ell$ is real-valued, the gradient with respect to complex parameters is well-defined via Wirtinger calculus. The automatic differentiation framework handles the bookkeeping internally and produces the correct complex gradient, which corresponds to steepest descent on $\mathbb{R}^{2M}$.

\paragraph{Gradient checkpointing.}
A naive implementation of automatic differentiation stores all intermediate state vectors from the forward pass (all $V_{\alpha\beta}^{(n)}$ at each site $n$ and for each application of $\mathbf{B}(u_m)$). For the system sizes of interest, this accumulated graph can exceed GPU memory. We apply \emph{gradient checkpointing} to avoid this problem: intermediate states are discarded during the forward pass and recomputed on the fly during the backward pass. This roughly doubles the computation time but greatly reduces peak memory, so that the full Bethe-state construction can be differentiated.

\paragraph{Numerical stabilizers.}
The state-renormalization safeguard is given in Section~\ref{sec:offshell} (equation~\eqref{eq:renorm}). The entropy evaluation additionally uses the clamped eigenvalues~\eqref{eq:lambda_clamp} to avoid $\log(0)$ singularities in the gradient.

\paragraph{Optimization settings: $L$-dependent adjustments.}
The only \(L\)-dependent adjustment concerns the maximum number of Adam update steps. For \(L\le20\), a fixed limit of \(100\) Adam steps is sufficient because every restart that does not produce \texttt{NaN} converges within that window. For \(21\le L\le23\), the limit is \(300\) steps. For \(L=24\), the same \(300\)-step budget is used, with gradient checkpointing enabled. For \(L\ge25\), the run terminates when \(|\Delta\mathcal{S}_\ell|/|\mathcal{S}_\ell|<10^{-5}\) for \(100\) consecutive steps, with an upper limit of \(5000\) steps, and checkpointing is also enabled. Checkpointing is a memory-control choice independent of the step-budget rule.

The XXX$_{s\geq 1}$ and $SL(2,\mathbb{R})$ chains require additional per-restart refinements beyond the recipe above; these will be discussed separately in Appendix~\ref{subsec:offshell_chain_specific}.

\subsection{Permutation symmetry and result collection}
\label{subsec:dedup_symmetry}

For the XXX$_s$ chains the operators $\mathbf{B}(u_j)$ commute, hence the off-shell Bethe vector is invariant under permutations of the rapidities. Consequently, the entropy $\mathcal{S}_\ell(\{u_j\})$ has the same symmetry: distinct rapidity orderings arising from different random initializations can represent the same physical state. In a gradient-based search with many restarts, it is therefore common to converge repeatedly to the same extremum with the $u_j$ appearing in different orders.

To identify genuinely distinct extrema, one can put each converged configuration in a standard form by sorting the rapidities lexicographically and rounding to a tolerance to absorb small numerical differences. For the system sizes studied in this work, comparing the converged entropy values directly provides a simpler and equally effective criterion for distinguishing solutions; the sorting-based standard form is used as a secondary check when needed.

\subsection{XXX$_{\frac{1}{2}}$ off-shell diagnostics}
\label{app:offshell_xxx_diagnostics}

This subsection collects two diagnostics for the XXX$_{\frac{1}{2}}$ off-shell optimization that support the presentation choices in Section~\ref{subsec:offshell_xxx} but are not central results.

\begin{figure}[h!]
\centering
\includegraphics[width=0.6\textwidth]{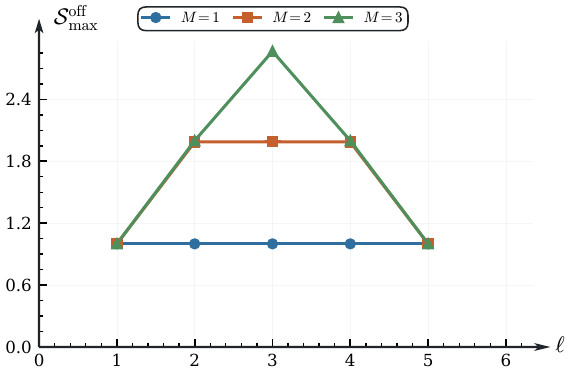}
\caption{The XXX$_{\frac{1}{2}}$ off-shell maximum is largest near the centre cut. The plot shows \(\mathcal{S}_{\max}^{\rm off}(L{=}6,M;\ell)\) versus subsystem size \(\ell\) for \(M=1,2,3\).}
\label{fig:offshell_xxx_ell_arc_L6}
\end{figure}

\paragraph{$\ell$-dependence at fixed $(L,M)$.}
Before restricting to the symmetric bipartition \(\ell=\lfloor L/2\rfloor\), one can scan the subsystem size \(\ell=1,\ldots,L-1\) at fixed \(L\) and \(M\). Figure~\ref{fig:offshell_xxx_ell_arc_L6} shows the resulting arc for \(L=6\) and \(M=1,2,3\). The maximum is symmetric under \(\ell\leftrightarrow L-\ell\), peaks at the centre cut, and drops to the edge-cut upper bound \(\mathcal{S}\le 1\) at \(\ell=1\) and \(\ell=L-1\). The optimizing rapidity set is not generally independent of \(\ell\): evaluating the centre-cut optimum on the edge cuts can lose up to \(\approx0.5\) relative to re-optimizing at that cut.

\begin{figure}[h!]
\centering
\includegraphics[width=0.7\textwidth]{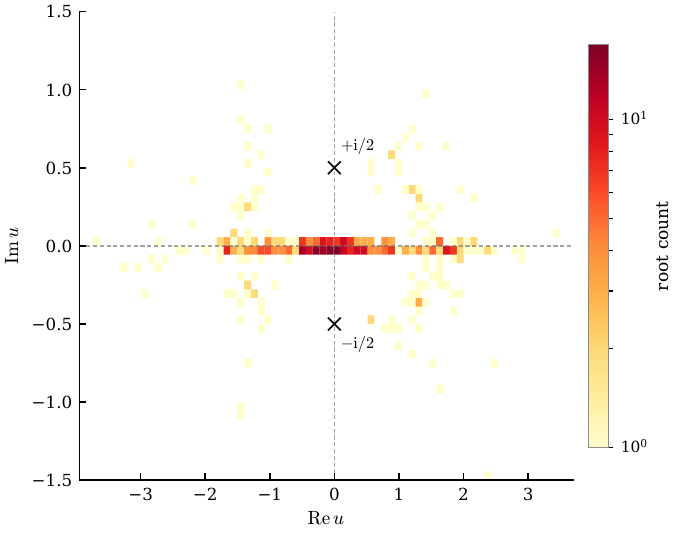}
\caption{Maximizing XXX$_{\frac{1}{2}}$ off-shell rapidities avoid the product-state singular points. The heatmap aggregates the rank-1 optimizing rapidities for \(\mathcal{S}_{\max}^{\rm off}\) across the sectors surveyed in Table~\ref{tab:offshell_xxx}, shown in the central window \(-1.5\le\mathrm{Im}\,u\le1.5\); black crosses mark \(u=\pm\mathrm{i}/2\).}
\label{fig:offshell_xxx_roots_smax}
\end{figure}

\paragraph{Distribution of maximizing rapidities.}
The rapidities that realize \(\mathcal{S}_{\max}^{\rm off}\) are also useful as an optimizer diagnostic. Figure~\ref{fig:offshell_xxx_roots_smax} aggregates the root clouds across the XXX$_{\frac{1}{2}}$ sectors in Table~\ref{tab:offshell_xxx}. The plot shows that the maximizing roots are not driven to the product-state singular locus \(u=\pm\mathrm{i}/2\); this supports the separation between the maximum mechanism and the minimum mechanism in Section~\ref{subsec:offshell_xxx}. Because the figure aggregates many \((L,M)\) sectors, it should not be read as a universal root-density law.

\subsection{Computational resources and scaling}
\label{subsec:cpu_gpu_batching}

\paragraph{CPU versus GPU.}
For small system sizes ($L\lesssim 12$), CPU execution is competitive because the state vectors fit in processor cache and the overhead of GPU data transfer is not justified. For $L>12$, the rapidly growing subspace dimension $\binom{L}{M}$ makes GPU acceleration necessary: both the Lax propagation and block-diagonal SVD reduce to vectorized operations on large arrays, which map well onto GPU hardware.

The implementation is written in \texttt{PyTorch} and supports both CPU and NVIDIA GPU (\texttt{CUDA}) backends. Small system sizes can be evaluated on CPU, while the larger runs use CUDA-enabled GPUs.

\paragraph{Memory scaling.}
The dominant memory cost is the state vector in the $M$-magnon subspace: $\binom{L}{M}$ complex numbers stored in single precision (8 bytes each). During the Lax propagation, four auxiliary vectors ($V_{11}, V_{12}, V_{21}, V_{22}$) of the same dimension are maintained. The table below gives representative values:
\begin{center}
\begin{tabular}{ccrrl}
\hline
$L$ & $M$ & $\binom{L}{M}$ & State vector & Estimated peak GPU \\
\hline
20 & 10 & 184\,756 & 1.4\,MB & $<1$\,GB \\
24 & 12 & 2\,704\,156 & 21\,MB & 2--3\,GB \\
26 & 13 & 10\,400\,600 & 79\,MB & 6--11\,GB \\
28 & 14 & 40\,116\,600 & 306\,MB & 10--15\,GB \\
\hline
\end{tabular}
\end{center}

\paragraph{Memory reduction techniques.}
Two techniques, described in Sections~\ref{subsec:magnon_subspace} and~\ref{subsec:complex_autodiff} respectively, are needed to reach the largest system sizes. Per-column streaming reduces the persistent GPU memory for the spin-flip index tables from $\mathcal{O}(\binom{L}{M}\times L)$ to $\mathcal{O}(\binom{L}{M})$ by transferring only one site's data at a time. Gradient checkpointing trades computation for memory by recomputing intermediate states during the backward pass instead of storing them. Together, these two techniques reduce peak GPU memory by a factor of three to five compared to a straightforward implementation, which brings $L=28$ within reach of a single GPU with 40--80\,GB of memory.

\subsection{\texorpdfstring{Chain-specific optimizer refinements for XXX$_{s\geq 1}$ and \(SL(2,\mathbb{R})\)}{Chain-specific optimizer refinements for XXXs>=1 and SL(2,R)}}
\label{subsec:offshell_chain_specific}

The optimizer of the previous subsections is sufficient for the XXX$_{\frac{1}{2}}$ chain of Section~\ref{subsec:offshell_xxx}. That optimizer uses Adam on a complex-valued rapidity vector with state renormalization, random restarts, and an $L$-dependent maximum number of update steps. For the XXX$_{s\geq 1}$ chains of Section~\ref{subsec:offshell_xxxs} and the non-compact $SL(2,\mathbb{R})$ chain of Section~\ref{subsec:offshell_sl2}, this procedure is not sufficient, and the additional refinements are described below.

\subsubsection{\texorpdfstring{Why the XXX$_{\frac{1}{2}}$ recipe is insufficient}{Why the XXX1/2 recipe is insufficient}}
The rapidity-space entropy function depends qualitatively on the local Hilbert space dimension \(d=2s+1\). For XXX$_{\frac{1}{2}}$ (\(d=2\)), the fixed \(100\)-step Adam limit used for \(L\le20\) in Appendix~\ref{subsec:complex_autodiff} is enough for every non-\texttt{NaN} restart to converge within the window: at each sampled \((L,M)\), the search is dominated by one local extremum and independent random restarts differ only at the numerical-precision level. For XXX$_1$ (\(d=3\)), the same single-stage procedure is no longer reliable. At \(L=4,M=2\), five random seeds give entropies \(\{1.862,\,1.991,\,1.992,\,1.990,\,1.992\}\), a worst-versus-best gap of \(0.13\). At \(L=6,M=3\), the corresponding values \(\{2.858,\,2.860,\,2.882,\,2.870,\,2.829\}\) still show a non-trivial \(0.05\) spread.

The non-compact \(SL(2,\mathbb{R})\) chain has a different numerical issue. The finite-root search is anisotropic in the real and imaginary rapidity directions; if the same step size is used in all directions, the imaginary components can move too quickly before the real part has settled. The \(SL(2,\mathbb{R})\) search therefore first optimizes mainly along the real directions and then gradually allows the imaginary directions to move, while the XXX$_1$ search does not use this real-axis prior because the reported \(L=8\) XXX$_1$ optima require genuinely complex rapidities in several sectors.

\subsubsection{\texorpdfstring{XXX$_1$ procedure}{XXX1 procedure}}
The XXX$_1$ entries of Table~\ref{tab:offshell_xxxs} use the \(L=8\) off-shell search in the spin-$1$ magnon sector. The spin-$1$ claims in the main text are restricted to this \(L=8\) comparison; a multi-\(L\) spin-$1$ scaling law is not reported.

For \(M=2,\ldots,6\), the search used warm-started random restarts followed by double-precision entropy evaluation of the best trial rapidity set. The \(M=7\) sector stabilized after \(85\) accepted restarts. The \(M=8\) entry is included by taking the on-shell singular-pair maximizing rapidities themselves as an off-shell witness, and the \(M=1\) entry is the analytic upper-bound value \(\mathcal{S}=1\).

\subsubsection{\texorpdfstring{XXX$_{\frac{3}{2}}$ procedure}{XXX3/2 procedure}}
The XXX$_{\frac{3}{2}}$ entries of Table~\ref{tab:offshell_xxx32} are computed directly in the full \(d^L=4^L\) Hilbert space, without magnetization-subspace projection. The entanglement entropy is evaluated from the Schmidt-block Gram matrices, with a fallback dense evaluation if the primary path fails. Each restart uses single-stage Adam with an adaptive entropy-change stopping rule after a short warm-up, followed by a short local refinement on a real/imaginary split parameterization.

Two numerical criteria reject artifacts that would otherwise survive the fallback evaluation. First, any entropy above \(\log\min(2^M,d^\ell,d^{L-\ell})\) is discarded up to the displayed numerical tolerance. Second, a run is marked as \texttt{NaN} when the Bethe state's Frobenius norm drops below \(10^{-6}\). The search starts with \(30\) restarts and adds groups of \(20\) restarts until the cumulative best changes by less than \(10^{-3}\) between groups, with at most three added groups. The \(L=6,M=8\) sector was later reconverged with a \(90\)-restart override after the initial run reached this group limit. With this override, all \(29\) comparison sectors entering Table~\ref{tab:offshell_xxx32} are stable.

\subsubsection{\texorpdfstring{\(SL(2,\mathbb{R})\) search, reprocessing, and ordering stability}{SL(2,R) search, reprocessing, and ordering stability}}
The \(SL(2,\mathbb{R})\) rapidity search is a heuristic numerical search whose trial rapidity sets are accepted only after independent re-evaluation. Each restart begins from two equivalent constructions of the Bethe vector, obtained by applying the \(C\)-operators in the two site orderings, and keeps the better initial state. The search then first stabilizes the dominant real rapidity profile, gradually releases the imaginary directions, and finally applies a short local refinement. The purpose of this staged procedure is not to define a different variational problem, but to avoid trajectories in which the imaginary rapidity components move too aggressively before the real profile has settled.

The per-step entropy inside the search is computed from a Gram matrix with a small diagonal term added, \(G\to G+\epsilon\,\mathbf{1}\) with \(\epsilon=10^{-6}\). This stabilizes the gradient near degenerate Schmidt eigenvalues but is unsafe as a final entropy measurement. At large rapidity magnitude, the single-precision \(SL(2,\mathbb{R})\) state degrades into numerical noise; the shifted Gram matrix can then acquire a near-uniform spectrum, and the reported entropy spuriously snaps to \(\log R(L,\ell,M)\). Every trial rapidity set is therefore re-evaluated independently: the Bethe state is rebuilt in double precision with scale-stable normalization, and the block-diagonal Schmidt entropy is computed with no Gram shift. A trial rapidity set enters the reported finite-root data only if this re-evaluation has unit reduced-density-matrix trace, no construction overflow, no spurious negative eigenvalue, and \(|u_j|\le100\). Large-rapidity trial points that appear to saturate the rank bound are retained only as diagnostics of how the variational family behaves as $|u_j|\to\infty$, not as physical optima.

For \(L\le4\), the entropy-evaluation artifact is confined to \(|u|\gtrsim10^4\), far outside the finite-root region, so a single-precision search combined with double-precision measurement is sufficient. For \(L\ge5\), the Lax factors are degree-\(L\) polynomials in \(u\), and single-precision cancellation becomes relevant already at \(|u|\sim10^2\)--\(10^3\), inside the legitimate optimization band. The \(L=6\) run therefore uses complex double precision throughout state construction, entropy evaluation, gradient computation, and measurement. At \(L=6\), dense monodromy construction at \(M=1,2,3,8\) agrees with the final entropy evaluation to \(\sim10^{-13}\).

The \(L=6\) finite-root data are converged across \(M=1,\ldots,42\). The entropy is monotone in \(M\); the forward warm-start chain, an independent backward chain, and four targeted additional searches agree within the run-to-run variation of the optimizer; a \(60\)-restart cold-start ensemble at \(M=10,12,14,16\) never exceeds the accepted chain by more than \(5\times10^{-3}\); and the maximizing rapidities satisfy \(|u_j|\lesssim36\), well inside the cutoff. An independent on-shell comparison value obtained by solving the \(SL(2,\mathbb{R})\) Bethe equations in the symmetric ground-state sector lies \(3\)--\(7\) bits below the off-shell maxima at every studied \(M\), ruling out gross optimizer failure.

Three numerical safeguards are used in the final evaluation. First, if the Hermitian eigensolver fails to converge on a near-degenerate Schmidt-block Gram matrix, the evaluation retries after adding a small relative diagonal shift (\(\leq10^{-6}\,\mathrm{tr}/n\), entropy-neutral) and, if needed, uses SVD on the block instead. Second, the backward warm-start chain propagates higher-\(M\) optima downward and improves \(M\in[25,30]\) by \(0.16\)--\(0.26\) over the forward-only chain. Third, the sequential numerical product \(\mathbf{C}(u_1)\mathbf{C}(u_2)\cdots \mathbf{C}(u_M)|0\rangle\) is evaluated after sorting rapidities by ascending \(|u|\). Although the \(\mathbf{C}(u_j)\) commute exactly, finite-precision evaluations with different rapidity orderings can differ by up to \(\sim0.5\) at \(L=6,M=30\) and \(\sim0.4\) at \(L=4,M=40\). Sorting by ascending \(|u|\) removes this order dependence at the tested points. The reported values always use the standard ascending-\(|u|\) order.

\section{Product-state limits for off-shell minima}
\label{app:offshell_product_state_proofs}

This appendix proves the product-state limits used in the discussion of off-shell minima. In all three model families, the relevant rapidities force the Bethe vector onto a single product vector, and the Schmidt rank is therefore one.

\subsection{XXX$_{\frac{1}{2}}$: one-sided and mixed approaches to singular rapidities}
\label{subsec:xxx_half_product_approaches}

We start with the XXX$_{\frac{1}{2}}$ product-state limits discussed in Section~\ref{subsec:offshell_xxx}. The one-sided singular rapidity already proves the off-shell minimum, while the mixed singular-rapidity statement explains the additional near-zero minima found by the optimizer.

We use the Lax recursion of Appendix~\ref{subsec:bethe_state_construction}. In particular, the \((1,2)\) component obeys
\begin{align}
V_{12}^{(n)}(u)
=\ri\,\mathbf{S}_n^-\,V_{11}^{(n-1)}(u)
+(u-\ri\,\mathbf{S}_n^z)\,V_{12}^{(n-1)}(u),
\label{eq:xxx_half_recursion_input}
\end{align}
with \(V_{12}^{(0)}=0\) and \(V_{11}^{(0)}=|\psi\rangle\). For XXX$_{\frac{1}{2}}$, \(S^z|\!\uparrow\rangle=\frac12|\!\uparrow\rangle\), \(S^z|\!\downarrow\rangle=-\frac12|\!\downarrow\rangle\), and \(S^-|\!\downarrow\rangle=0\).

\paragraph{One-sided singular rapidity.}
Let
\begin{align}
|R_p\rangle&=\mathbf{S}_{L-p+1}^-\cdots \mathbf{S}_{L}^-|0\rangle,\qquad
|L_p\rangle=\mathbf{S}_{1}^-\cdots \mathbf{S}_{p}^-|0\rangle,
\end{align}
with \(|R_0\rangle=|L_0\rangle=|0\rangle\). Both states are the so-called domain wall states. We first show that, for \(0\le p<L\),
\begin{align}
\mathbf{B}(\ri/2)|R_p\rangle&=c^+_{L,p}\,|R_{p+1}\rangle,\qquad c^+_{L,p}\ne0,
\label{eq:xxx_half_right_step}\\
\mathbf{B}(-\ri/2)|L_p\rangle&=c^-_{L,p}\,|L_{p+1}\rangle,\qquad c^-_{L,p}\ne0.
\label{eq:xxx_half_left_step}
\end{align}
For the first identity, apply \eqref{eq:xxx_half_recursion_input} to \(|R_p\rangle\) at \(u=\ri/2\). A term in which \(\mathbf{S}_n^-\) lowers a site \(n<L-p\) is later multiplied by \(u-\ri \mathbf{S}^z=0\) when the recursion passes an up spin to its right. A term in which \(\mathbf{S}_n^-\) acts inside the existing right boundary block vanishes because \(\mathbf{S}^-|\downarrow\rangle=0\). The only surviving term lowers site \(L-p\), and all subsequent factors \(u-\ri \mathbf{S}^z\) act on down spins and are nonzero. This proves \eqref{eq:xxx_half_right_step}. The second identity is the left-right mirror statement at \(u=-\ri/2\): the zero is now in the factor \(u+\ri \mathbf{S}^z\) entering the \(V_{11}\) propagation, so only the next available spin from the left survives.

Iterating \eqref{eq:xxx_half_right_step} and \eqref{eq:xxx_half_left_step} gives, for \(1\le M\le L\),
\begin{align}
\mathbf{B}(\ri/2)^M|0\rangle
&=c^+_{L,M}\,
\mathbf{S}_{L-M+1}^-\cdots \mathbf{S}_{L}^-|0\rangle,
\label{eq:xxx_half_one_sided_plus}\\
\mathbf{B}(-\ri/2)^M|0\rangle
&=c^-_{L,M}\,
\mathbf{S}_{1}^-\cdots \mathbf{S}_{M}^-|0\rangle,
\label{eq:xxx_half_one_sided_minus}
\end{align}
with nonzero constants \(c^\pm_{L,M}\). Both right-hand sides are product vectors, so their Schmidt rank is one across every bipartition and \(\mathcal{S}_\ell=0\).

\paragraph{Mixed singular rapidities.}
When roots approach both singular rapidities, the values \(u=\pm\ri/2\) are not by themselves enough to determine the normalized limiting state. We write the approaching rapidities as
\begin{align}
u_a^-=-\frac{\ri}{2}+\delta_a^-,
\qquad
u_b^+=\frac{\ri}{2}+\delta_b^+,
\qquad
\delta_a^-=\epsilon\,\xi_a^-,
\qquad
\delta_b^+=\epsilon\,\xi_b^+,
\qquad
\epsilon\to0.
\label{eq:xxx_half_singular_approach_definition}
\end{align}
Here \(\delta_a^\pm\) are the small displacements that are taken to zero. The constants \(\xi_a^\pm\) are held fixed and specify the relative direction of the approach to the singular rapidities.

The elementary mixed step is the two-root calculation. For \(p+q+2\le L\), define
\[
|\Omega_{p,q}\rangle
=\mathbf{S}_1^-\cdots\mathbf{S}_{p}^-
\mathbf{S}_{L-q+1}^-\cdots\mathbf{S}_{L}^-|0\rangle .
\]
A direct use of \eqref{eq:xxx_half_recursion_input} gives
\begin{align}
\mathbf{B}\!\left(-\frac{\ri}{2}+\epsilon\xi^-\right)
\mathbf{B}\!\left(\frac{\ri}{2}+\epsilon\xi^+\right)
|\Omega_{p,q}\rangle
=\epsilon\,c_{L,p,q}(\xi^- - \xi^+)\,
|\Omega_{p+1,q+1}\rangle+\mathcal{O}(\epsilon^2),
\label{eq:xxx_half_two_root_step}
\end{align}
with \(c_{L,p,q}\ne0\). The zeroth-order term vanishes because, after one root has localized on one boundary, the exact opposite singular rapidity must propagate through the occupied boundary block on the other side, where the relevant diagonal factor is zero. The first nonzero contribution is obtained by replacing exactly one such zero factor by its \(\epsilon\xi^\pm\) correction. The two possible replacements contribute with opposite signs, giving the factor \(\xi^- - \xi^+\); all remaining local factors are nonzero and are absorbed into \(c_{L,p,q}\).

Let \(k_-\) and \(k_+\) be the numbers of rapidities approaching \(-\ri/2\) and \(+\ri/2\), with \(k_-+k_+=M\le L\). If the two sets approach the singular rapidities in separated directions,
\begin{align}
\xi_a^-\ne \xi_b^+\qquad
\text{for all }a=1,\ldots,k_-\text{ and }b=1,\ldots,k_+,
\label{eq:xxx_half_separated_approach_condition}
\end{align}
then iterating the one-sided steps \eqref{eq:xxx_half_right_step}, \eqref{eq:xxx_half_left_step} and the mixed step \eqref{eq:xxx_half_two_root_step} gives the leading expansion, in Hilbert-space norm,
\begin{align}
\prod_{a=1}^{k_-}\mathbf{B}(u_a^-)
\prod_{b=1}^{k_+}\mathbf{B}(u_b^+)|0\rangle
=\epsilon^{k_-k_+}C_{L,k_-,k_+}(\xi)\,
|\Omega_{k_-,k_+}\rangle
+\mathcal{O}\!\left(\epsilon^{k_-k_++1}\right),
\label{eq:xxx_half_product_approach}
\end{align}
where
\begin{align}
|\Omega_{k_-,k_+}\rangle
=\mathbf{S}_1^-\cdots\mathbf{S}_{k_-}^-
\mathbf{S}_{L-k_++1}^-\cdots\mathbf{S}_{L}^-|0\rangle
\end{align}
and
\begin{align}
C_{L,k_-,k_+}(\xi)
=c_{L,k_-,k_+}
\prod_{a=1}^{k_-}\prod_{b=1}^{k_+}(\xi_a^--\xi_b^+),
\qquad c_{L,k_-,k_+}\ne0 .
\end{align}
Each opposite-sign pair contributes one factor from \eqref{eq:xxx_half_two_root_step}; roots of the same sign only extend the appropriate boundary block by \eqref{eq:xxx_half_right_step} or \eqref{eq:xxx_half_left_step}. Under \eqref{eq:xxx_half_separated_approach_condition}, \(C_{L,k_-,k_+}(\xi)\ne0\), and every component orthogonal to \( |\Omega_{k_-,k_+}\rangle\) is suppressed by at least one additional power of \(\epsilon\). Hence
\begin{align}
\lim_{\epsilon\to0}
\frac{
\prod_{a=1}^{k_-}\mathbf{B}(u_a^-)
\prod_{b=1}^{k_+}\mathbf{B}(u_b^+)|0\rangle}
{\left\|
\prod_{a=1}^{k_-}\mathbf{B}(u_a^-)
\prod_{b=1}^{k_+}\mathbf{B}(u_b^+)|0\rangle
\right\|}
=e^{\ri\phi}|\Omega_{k_-,k_+}\rangle
\end{align}
for an irrelevant phase \(e^{\ri\phi}\). The limiting state is therefore a product vector and has zero bipartite entropy.

\paragraph{Dependence on the approach.}
Equation~\eqref{eq:xxx_half_product_approach} is the result used in the main text: mixed singular configurations have zero entropy when the two sets of rapidities approach \(u=\pm\ri/2\) in separated directions satisfying \eqref{eq:xxx_half_separated_approach_condition}. The singular rapidities alone do not determine the normalized limit. If a mixed pair has \(\xi_a^-=\xi_b^+\), the leading product coefficient can vanish and subleading components can determine a non-product limiting state. Thus the optimizer-observed mixed minima should be interpreted as convergence to these separated product limits, not as a path-independent statement for every approach to \(u=\pm\ri/2\).

\subsection{XXX$_1$ and XXX$_{\frac{3}{2}}$: compact boundary strings}
\label{subsec:compact_spin_product_strings}

For the compact higher-spin chains it is useful to label a local basis vector by the number \(n=0,\ldots,2s\) of lowerings from the highest-weight state. Thus
\[
|n_1,\ldots,n_L\rangle_s
\]
denotes the basis vector whose \(j\)-th site has \(n_j\) lowerings. The one-site boundary string of length \(r\le 2s\) is
\begin{align}
\mathcal{V}^{(s)}_r(+)
=\{\ri s,\ri(s-1),\ldots,\ri(s-r+1)\},
\qquad
\mathcal{V}^{(s)}_r(-)=-\mathcal{V}^{(s)}_r(+).
\label{eq:compact_spin_boundary_string}
\end{align}
We set \(\mathcal{V}^{(s)}_0(\pm)=\varnothing\). For a general magnon number \(M=2s\,q+r\), \(0\le r<2s\), define the multisets
\begin{align}
\mathcal{U}^{(s)}_{M,+}
&=\underbrace{\mathcal{V}^{(s)}_{2s}(+)\cup\cdots\cup\mathcal{V}^{(s)}_{2s}(+)}_{q\ \mathrm{copies}}
\cup\mathcal{V}^{(s)}_r(+),\nonumber\\
\mathcal{U}^{(s)}_{M,-}
&=\underbrace{\mathcal{V}^{(s)}_{2s}(-)\cup\cdots\cup\mathcal{V}^{(s)}_{2s}(-)}_{q\ \mathrm{copies}}
\cup\mathcal{V}^{(s)}_r(-),
\label{eq:compact_spin_product_multisets}
\end{align}
where repeated rapidities are retained. Since the \(\mathbf{B}^{(s)}(u)\) commute, we may order the roots in the sequence displayed in \eqref{eq:compact_spin_boundary_string}. Define the right-boundary state
\[
|R_{q,r}\rangle_s
=|0,\ldots,0,r,\underbrace{2s,\ldots,2s}_{q\ {\rm sites}}\rangle_s,
\qquad 0\le r<2s,
\]
where the entry \(r\) is omitted when \(r=0\). Let \(|R_{q,2s}\rangle_s\equiv |R_{q+1,0}\rangle_s\). The elementary induction step is
\begin{align}
\mathbf{B}^{(s)}\!\left(\ri(s-r)\right)|R_{q,r}\rangle_s
=c^+_{q,r}\,|R_{q,r+1}\rangle_s,\qquad c^+_{q,r}\ne0,
\label{eq:compact_spin_right_step}
\end{align}
for \(0\le r<2s\), as long as the right boundary block has not filled the whole chain. To see this, use the same recursion as in \eqref{eq:xxx_half_recursion_input}, now with the spin-\(s\) local action \(\mathbf{S}^z|a\rangle=(s-a)|a\rangle\) and \(\mathbf{S}^-|a\rangle\propto |a+1\rangle\) for \(a<2s\). At \(u=\ri(s-r)\), the diagonal factor \(u-\ri \mathbf{S}^z\) vanishes on the boundary site with occupation \(r\). Therefore any term in which the lowering occurs to the left of that site is killed when the recursion crosses the boundary site. A lowering inside the already filled block vanishes because \(\mathbf{S}^-|2s\rangle=0\). The only surviving term lowers the boundary site from \(r\) to \(r+1\), giving \eqref{eq:compact_spin_right_step}. The left-boundary step follows from the sign-reversed zero in \(u+\ri \mathbf{S}^z\).

Iterating \eqref{eq:compact_spin_right_step} gives
\begin{align}
\prod_{u\in \mathcal{U}^{(s)}_{M,+}}\mathbf{B}^{(s)}(u)|0\rangle
\propto
|0,\ldots,0,r,\underbrace{2s,\ldots,2s}_{q\ {\rm sites}}\rangle_s ,
\label{eq:compact_spin_right_product}
\end{align}
where the entry \(r\) is also omitted when \(r=0\). The sign-reversed roots give
\begin{align}
\prod_{u\in \mathcal{U}^{(s)}_{M,-}}\mathbf{B}^{(s)}(u)|0\rangle
\propto
|\underbrace{2s,\ldots,2s}_{q\ {\rm sites}},r,0,\ldots,0\rangle_s .
\label{eq:compact_spin_left_product}
\end{align}
Both right-hand sides are product vectors and have zero entropy across every cut.

For example, the compact boundary-string construction gives, for XXX$_1$,
\begin{align*}
\mathbf{B}^{(1)}(\ri)\mathbf{B}^{(1)}(0)|0\rangle
&\propto |0,\ldots,0,2\rangle_1,\\
\mathbf{B}^{(1)}(\ri)\mathbf{B}^{(1)}(\ri)\mathbf{B}^{(1)}(0)|0\rangle
&\propto |0,\ldots,0,1,2\rangle_1,
\end{align*}
and, for XXX$_{\frac{3}{2}}$,
\begin{align*}
\mathbf{B}^{(\frac{3}{2})}(3\ri/2)\mathbf{B}^{(\frac{3}{2})}(\ri/2)|0\rangle
&\propto |0,\ldots,0,2\rangle_{\frac{3}{2}},\\
\mathbf{B}^{(\frac{3}{2})}(3\ri/2)\mathbf{B}^{(\frac{3}{2})}(\ri/2)
\mathbf{B}^{(\frac{3}{2})}(-\ri/2)|0\rangle
&\propto |0,\ldots,0,3\rangle_{\frac{3}{2}}.
\end{align*}
These examples are not additional assumptions; they are the first non-trivial cases of \eqref{eq:compact_spin_right_product}. They illustrate why the higher-spin product-state locus is a boundary string rather than repeated copies of a single singular rapidity.

\subsection{\texorpdfstring{\(SL(2,\mathbb{R})\): half-integer towers}{SL(2,R): half-integer towers}}
\label{subsec:sl2_product_towers}

The non-compact chain uses \(\mathbf{C}(u)\) rather than \(\mathbf{B}(u)\). Let \(M=p+q\), with \(p,q\ge0\), and define
\begin{align}
\label{eq:sl2_product_tower_roots}
\mathcal{U}_{p,q}
=\left\{-\frac{\ri}{2},-\frac{3\ri}{2},\ldots,-\frac{(2p-1)\ri}{2}\right\}
\cup
\left\{\frac{\ri}{2},\frac{3\ri}{2},\ldots,\frac{(2q-1)\ri}{2}\right\},
\end{align}
with either tower omitted if its length is zero. Then
\begin{align}
\prod_{u\in\mathcal{U}_{p,q}}\mathbf{C}(u)\,|0\rangle
\;\propto\;
|p,0,\ldots,0,q\rangle ,
\label{eq:sl2_product_tower_state}
\end{align}
where the basis is the local occupation basis \eqref{eq:sl2_local_basis}. Thus every contiguous bipartition has Schmidt rank one.

To prove this, write the local \(SL(2,\mathbb{R})\) Lax action as
\begin{align}
\mathbf{A}_n(u)|m\rangle&=(u-\ri(m+\tfrac12))|m\rangle,\nonumber\\
\mathbf{D}_n(u)|m\rangle&=(u+\ri(m+\tfrac12))|m\rangle,\nonumber\\
\mathbf{C}_n(u)|m\rangle&=\ri(m+1)|m+1\rangle .
\end{align}
At \(u=\ri(q+\tfrac12)\), the rightmost local factor has \(\mathbf{A}_L(u)|q\rangle=0\). The monodromy decomposition
\[
\mathbf{C}_{1\ldots L}(u)=\mathbf{C}_{1\ldots L-1}(u)\mathbf{A}_{L}(u)
+\mathbf{D}_{1\ldots L-1}(u)\mathbf{C}_{L}(u)
\]
therefore kills the first term on \(|p,0,\ldots,0,q\rangle\), while the second term raises only the last site. Thus
\begin{align}
\mathbf{C}\!\left(\ri(q+\tfrac12)\right)|p,0,\ldots,0,q\rangle
\propto |p,0,\ldots,0,q+1\rangle .
\label{eq:sl2_raise_right_boundary}
\end{align}
Similarly, at \(u=-\ri(p+\tfrac12)\), the first-site factor has \(\mathbf{D}_1(u)|p\rangle=0\), and the decomposition
\[
\mathbf{C}_{1\ldots L}(u)=\mathbf{C}_{1}(u)\mathbf{A}_{2\ldots L}(u)
+\mathbf{D}_{1}(u)\mathbf{C}_{2\ldots L}(u)
\]
gives
\begin{align}
\mathbf{C}\!\left(-\ri(p+\tfrac12)\right)|p,0,\ldots,0,q\rangle
\propto |p+1,0,\ldots,0,q\rangle .
\label{eq:sl2_raise_left_boundary}
\end{align}
Starting from the vacuum, repeated use of \eqref{eq:sl2_raise_left_boundary} and \eqref{eq:sl2_raise_right_boundary} proves \eqref{eq:sl2_product_tower_state}. Since the \(\mathbf{C}(u)\) commute, the ordering of the rapidities in \(\mathcal{U}_{p,q}\) is irrelevant.

For instance, the balanced \(L=2,M=10\) tower with \(p=q=5\) gives the single basis vector \(|5,5\rangle\). As a useful contrast, direct evaluation of the Lax recursion for the repeated set \(\{\ri/2,\ri/2,\ri/2\}\) at \(L=3,\ell=1\) gives a Bethe vector with Schmidt rank \(2\), not rank \(1\). This repeated set is not a tower of the form \eqref{eq:sl2_product_tower_roots}. The product-state minimum is therefore described by half-integer towers, not by repeated copies of a single rapidity.

\section{Extra Information for Entanglement Entropy}
\label{app:extraEE}
\begingroup
\setlength{\tabcolsep}{4pt}
\renewcommand{\arraystretch}{1.08}
\setlength{\floatsep}{8pt plus 2pt minus 2pt}
\setlength{\textfloatsep}{10pt plus 2pt minus 2pt}
\captionsetup[table]{font=small,skip=4pt}
\subsection[Spin 1/2]{Spin $\frac{1}{2}$}
\subsubsection*{Summary tables for the on-shell XXX$_{1/2}$ chain}
For completeness, we collect here the length-by-length summary tables from the on-shell spin-$\frac{1}{2}$ XXX scan that are not displayed in the main text. The representative $L=8$ table remains in Section~\ref{subsec:onshell_xxx}; the corresponding figures are omitted because Figure~\ref{fig:EEL8M3} already shows the qualitative behavior of the entropy curves.
Throughout these summary tables, $\mathcal{N}_R$ and $\mathcal{N}_S$ denote the numbers of regular and singular states, while $\mathcal{S}_{\text{min}/\text{max}}$ and $\{I_j\}_{\text{min}/\text{max}}$ denote the extremal entropies and the corresponding Bethe quantum numbers.

\begin{table}[!htbp]
\centering
\small
\begin{adjustbox}{max width=\textwidth}
\begin{tabular}{|c|c|c|c|c|c|c|}
  \hline
  $M$ & $\mathcal{N}_R$ & $\mathcal{N}_S$ & $\mathcal{S}_{\text{min}}$ & $\{I_j\}_{\text{min}}$ & $\mathcal{S}_{\text{max}}$ & $\{I_j\}_{\text{max}}$\\
  \hline
  1 & 5 & 0 & 1.0  & / & 1.0 & / \\
  \hline
 2 & 8 & 1 & 1.31293  & $\{-\tfrac{1}{2},\tfrac{1}{2}\}$ & 1.97987 & $\{-\frac{3}{2},\frac{1}{2}\}_p$ \\
 \hline
 3 & 4 & 1 & 1.38355  & $\{-1, 0, 1\}$ & 2.60796 & $\{-2, -1, 1\}_p$ \\
   \hline
\end{tabular}
\end{adjustbox}
\caption{Summary of the on-shell spin-$\frac{1}{2}$ XXX scan at $L=6$.}
\label{tab:L6}
\end{table}

\begin{table}[!htbp]
\centering
\small
\begin{adjustbox}{max width=\textwidth}
\begin{tabular}{|c|c|c|c|c|c|c|}
  \hline
  $M$ & $\mathcal{N}_R$ & $\mathcal{N}_S$ & $\mathcal{S}_{\text{min}}$ & $\{I_j\}_{\text{min}}$ & $\mathcal{S}_{\text{max}}$ & $\{I_j\}_{\text{max}}$\\
  \hline
  1 & 6 & 0 & 0.98522  & / & 0.98522 & / \\
  \hline
 2 & 14 & 0 & 1.31692  & $\{-1,0\}_p$ & 1.92417 & $\{-1,1\}$, $\{-2,2\}$\\
 \hline
 3 & 14 & 0 & 1.43921  & $\{-\frac{3}{2}, -\frac{1}{2}, \frac{1}{2}\}_p$ & 2.5809 & $\{-\frac{5}{2}, -\frac{3}{2}, \frac{3}{2}\}_p$ \\
   \hline
\end{tabular}
\end{adjustbox}
\caption{Summary of the on-shell spin-$\frac{1}{2}$ XXX scan at $L=7$.}
\label{tab:L7}
\end{table}

\begin{table}[!htbp]
\centering
\small
\begin{adjustbox}{max width=\textwidth}
\begin{tabular}{|c|c|c|c|c|c|c|}
  \hline
  $M$ & $\mathcal{N}_R$ & $\mathcal{N}_S$ & $\mathcal{S}_{\text{min}}$ & $\{I_j\}_{\text{min}}$ & $\mathcal{S}_{\text{max}}$ & $\{I_j\}_{\text{max}}$\\
  \hline
  1 & 8 & 0 & 0.99108  & / & 0.99108 & / \\
  \hline
 2 & 27 & 0 & 1.33843  & $\{-1,0\}_p$ & 1.96028 & $\{-2,3\}_p$\\
 \hline
 3 & 46 & 2 & 1.49869  & $\{-\frac{1}{2},\frac{1}{2},\frac{3}{2}\}_p$ & 2.86438 & $\{-\frac{3}{2},\frac{1}{2},\frac{5}{2}\}_p$ \\
 \hline
 4 & 42 & 0 & 1.56689  & $\{-1,0,1,2\}_p$ & 3.35103 & $\{-2, 0, 2, 3\}_p$ \\
   \hline
\end{tabular}
\end{adjustbox}
\caption{Summary of the on-shell spin-$\frac{1}{2}$ XXX scan at $L=9$.}
\label{tab:L9}
\end{table}

\begin{table}[!htbp]
\centering
\small
\begin{adjustbox}{max width=\textwidth}
\begin{tabular}{|c|c|c|c|c|c|c|}
  \hline
  $M$ & $\mathcal{N}_R$ & $\mathcal{N}_S$ & $\mathcal{S}_{\text{min}}$ & $\{I_j\}_{\text{min}}$ & $\mathcal{S}_{\text{max}}$ & $\{I_j\}_{\text{max}}$\\
  \hline
  1 & 9 & 0 & 1  & / & 1 & / \\
  \hline
 2 & 34 & 1 & 1.35113  & $\{-\frac{1}{2},\frac{1}{2}\}$ & 1.99584 & $\{-\frac{5}{2},\frac{7}{2}\}_p$\\
 \hline
 3 & 74 & 1 & 1.51966  & $\{-1, 0, 1\}$ & 2.96454 & $\{2,0,-2\}$ \\
 \hline
 4 & 86 & 4 & 1.60262 & $\{-\frac{3}{2}, -\frac{1}{2}, \frac{1}{2}, \frac{3}{2}\}$ & 3.53772 & $\{-\frac{7}{2}, -\frac{5}{2}, \frac{1}{2}, \frac{5}{2}\}_p$ \\
   \hline
5 & 38 & 4 & 1.62174 & $\{-2, -1, 0, 1, 2\}$ & 4.14249 & $\{-4, -2, 0, 2, 4\}$ \\
   \hline
\end{tabular}
\end{adjustbox}
\caption{Summary of the on-shell spin-$\frac{1}{2}$ XXX scan at $L=10$.}
\label{tab:L10}
\end{table}

\begin{table}[!htbp]
\centering
\small
\begin{adjustbox}{max width=\textwidth}
\begin{tabular}{|c|c|c|c|c|c|c|}
  \hline
  $M$ & $\mathcal{N}_R$ & $\mathcal{N}_S$ & $\mathcal{S}_{\text{min}}$ & $\{I_j\}_{\text{min}}$ & $\mathcal{S}_{\text{max}}$ & $\{I_j\}_{\text{max}}$\\
  \hline
  1 & 10 & 0 & 0.99403  & / & 0.99403 & / \\
  \hline
 2 & 44 & 0 & 1.34862 & $\{-1, 0\}_{p}$ & 1.97816 & $\{-4, 3\}_p$\\
 \hline
 3 & 110 & 0 & 1.52554  & $\{-\frac{3}{2}, -\frac{1}{2}, \frac{1}{2}\}_{p}$ & 2.91555 & $\{-\frac{7}{2},\frac{1}{2},\frac{5}{2}\}_{p}$ \\
 \hline
 4 & 165 & 0 & 1.62229 & $\{-2, -1, 0, 1\}_{p}$ & 3.79681 & $\{-3, -1, 1, 3\}_p$ \\
   \hline
5 & 132 & 0 & 1.66654 & $\{\frac{5}{2}, -\frac{3}{2}, -\frac{1}{2}, \frac{1}{2}, \frac{3}{2}\}_{p}$ & 4.01854 & $\{\frac{9}{2}, -\frac{7}{2}, -\frac{5}{2}, \frac{1}{2}, \frac{5}{2}\}_{p}$ \\
   \hline
\end{tabular}
\end{adjustbox}
\caption{Summary of the on-shell spin-$\frac{1}{2}$ XXX scan at $L=11$.}
\label{tab:L11}
\end{table}

\begin{table}[!htbp]
\centering
\small
\begin{adjustbox}{max width=\textwidth}
\begin{tabular}{|c|c|c|c|c|c|c|}
  \hline
  $M$ & $\mathcal{N}_R$ & $\mathcal{N}_S$ & $\mathcal{S}_{\text{min}}$ & $\{I_j\}_{\text{min}}$ & $\mathcal{S}_{\text{max}}$ & $\{I_j\}_{\text{max}}$\\
  \hline
  1 & 11 & 0 & 1  & / & 1 & / \\
  \hline
 2 & 53 & 1 & 1.35661& $\{-\frac{1}{2}, \frac{1}{2}\}$ & 1.99785 & $\{-\frac{9}{2}, \frac{7}{2}\}_p$\\
 \hline
 3 & 153 & 1 & 1.53764  & $\{-1, 0, 1\}$ & 2.97608 & $\{-2, 0, 2\}$ \\
 \hline
 4 & 270 & 5 & 1.642 & $\{-\frac{3}{2}, -\frac{1}{2}, \frac{1}{2}, \frac{3}{2}\}$ & 3.90822 & $\{-\frac{7}{2}, \frac{3}{2}, \frac{1}{2}, \frac{5}{2}\}_p$ \\
   \hline
5 & 292 & 5 & 1.69687 & $\{-2, -1, 0, 1, 2\}$ & 4.34992 & $\{-4, -3,-1,1,3\}_{p}$ \\
   \hline
6 & 122 & 10 & 1.70847 & $\{-\frac{5}{2}, -\frac{3}{2}, -\frac{1}{2}, \frac{1}{2}, \frac{3}{2}, \frac{5}{2}\}$ & 4.72472 & $\{-1.52, -0.95, -0.5\ri, 0.5\ri, 0.95, 1.52\}$ \\
   \hline
\end{tabular}
\end{adjustbox}
\caption{Summary of the on-shell spin-$\frac{1}{2}$ XXX scan at $L=12$; for $M=6$, the maximizing singular solution is shown by its rapidities.}
\label{tab:L12}
\end{table}

\begin{table}[!htbp]
\centering
\small
\begin{adjustbox}{max width=\textwidth}
\begin{tabular}{|c|c|c|c|c|c|c|}
  \hline
  $M$ & $\mathcal{N}_R$ & $\mathcal{N}_S$ & $\mathcal{S}_{\text{min}}$ & $\{I_j\}_{\text{min}}$ & $\mathcal{S}_{\text{max}}$ & $\{I_j\}_{\text{max}}$\\
  \hline
  1 & 12 & 0 & 0.99573  & / & 0.99573 & / \\
  \hline
 2 & 65 & 0 & 1.35426 & $\{-1,0\}_{p}$ & 1.98625 & $\{-5, 4\}_p$\\
 \hline
 3 & 208 & 0 & 1.53988  & $\{-\frac{3}{2}, -\frac{1}{2}, \frac{1}{2}\}_{p}$ & 2.94138 & $\{-\frac{7}{2}, \frac{1}{2}, \frac{5}{2}\}_{p}$ \\
 \hline
 4 & 429 & 0 & 1.65139 & $\{-2, -1, 0, 1\}_{p}$ & 3.87087 & $\{-3, -1, 1, 3\}$ \\
   \hline
5 & 572 & 0 & 1.71724 & $\{-\frac{5}{2}, -\frac{3}{2}, -\frac{1}{2}, \frac{1}{2}, \frac{3}{2}\}_{p}$ & 4.47927 & {$\{-\frac{7}{2}, -\frac{7}{2}, -\frac{1}{2}, \frac{3}{2}, \frac{7}{2}\}_{p}$} \\
   \hline
6 & 429 & 0 & 1.7485 & $\{-3, -2, -1, 0, 1, 2\}_{p}$ & 4.78575 & $\{-6, -5, -3, -2, 1, 3\}_{p}$ \\
   \hline
\end{tabular}
\end{adjustbox}
\caption{Summary of the on-shell spin-$\frac{1}{2}$ XXX scan at $L=13$.}
\label{tab:L13}
\end{table}

\FloatBarrier
\subsubsection*{Subsystem-size dependence for selected spin-$\frac{1}{2}$ sectors}
The following tables record the minimum and maximum entanglement entropy as the subsystem size is varied for selected sectors. The entries are unchanged from the scan data.
Here $\ell$ denotes the size of the left subsystem; the columns $\mathcal{S}_{\text{min}/\text{max}}$ and $\{I_j\}_{\text{min}/\text{max}}$ have the same meaning as above.

\begin{table}[!htbp]
\centering
\small
\begin{adjustbox}{max width=\textwidth}
\begin{tabular}{|c|c|c|c|c|}
  \hline
  $\ell$ &  $\mathcal{S}_{\text{min}}$ & $\{I_j\}_{\text{min}}$ & $\mathcal{S}_{\text{max}}$ & $\{I_j\}_{\text{max}}$\\
  \hline
  2& 1.04718  & $\{-\frac{1}{2}, \frac{1}{2}\}$ & 1.43733 & $\{-\frac{5}{2}, \frac{5}{2}\}$ \\
  \hline
 3 & 1.22805  & $\{-\frac{1}{2},\frac{1}{2}\}$ & 1.74647 & $\{-\frac{3}{2},\frac{3}{2}\}$\\
 \hline
 4 & 1.32192  & $\{-\frac{1}{2}, \frac{1}{2}\}$ & 1.9366 & $\{-\frac{5}{2}, \frac{5}{2}\}$ \\
 \hline
\end{tabular}
\end{adjustbox}
\caption{Subsystem-size dependence for $L=10$, $M=2$.}
\label{tab:L10M2}
\end{table}

\begin{table}[!htbp]
\centering
\small
\begin{adjustbox}{max width=\textwidth}
\begin{tabular}{|c|c|c|c|c|}
  \hline
  $\ell$ &  $\mathcal{S}_{\text{min}}$ & $\{I_j\}_{\text{min}}$ & $\mathcal{S}_{\text{max}}$ & $\{I_j\}_{\text{max}}$\\
  \hline
  2& 1.2314  & $\{-1, 0, 1\}$ & 1.75464 & $\{-3, -3, 2\}_{p}$ \\
  \hline
 3 & 1.40706  & $\{-1, 0, 1\}$ & 2.54621 & $\{-3, 0, 3\}$\\
 \hline
 4 & 1.4934  & $\{-1, 0, 1\}$ & 2.8093 & $\{-3, 0, 3\}$ \\
 \hline
\end{tabular}
\end{adjustbox}
\caption{Subsystem-size dependence for $L=10$, $M=3$.}
\label{tab:L10M3}
\end{table}

\begin{table}[!htbp]
\centering
\small
\begin{adjustbox}{max width=\textwidth}
\begin{tabular}{|c|c|c|c|c|}
  \hline
  $\ell$ &  $\mathcal{S}_{\text{min}}$ & $\{I_j\}_{\text{min}}$ & $\mathcal{S}_{\text{max}}$ & $\{I_j\}_{\text{max}}$\\
  \hline
  2& 1.32707  & $\{-\frac{3}{2}, -\frac{1}{2}, \frac{1}{2}, \frac{3}{2}\}$ & 1.93592 & $\{-\frac{7}{2}, -\frac{5}{2}, \frac{3}{2}, \frac{5}{2}\}_{p}$ \\
  \hline
 3 & 1.49551  & $\{-\frac{3}{2}, -\frac{1}{2}, \frac{1}{2}, \frac{3}{2}\}$ & 2.83397 & $\{-\frac{9}{2}, -\frac{1}{2}, \frac{1}{2}, \frac{9}{2}\}$\\
 \hline
 4 & 1.57754  & $\{-\frac{3}{2}, -\frac{1}{2}, \frac{1}{2}, \frac{3}{2}\}$ & 3.3082 & $\{-\frac{7}{2}, -\frac{5}{2}, \frac{1}{2}, \frac{5}{2}\}_{p}$ \\
 \hline
\end{tabular}
\end{adjustbox}
\caption{Subsystem-size dependence for $L=10$, $M=4$.}
\label{tab:L10M4}
\end{table}

\begin{table}[!htbp]
\centering
\small
\begin{adjustbox}{max width=\textwidth}
\begin{tabular}{|c|c|c|c|c|}
  \hline
  $\ell$ &  $\mathcal{S}_{\text{min}}$ & $\{I_j\}_{\text{min}}$ & $\mathcal{S}_{\text{max}}$ & $\{I_j\}_{\text{max}}$\\
  \hline
  2& 1.35244  & $\{-2, -1, 0, 1, 2\}$ & 1.99992 & $\{-3, -2, -2, 2, 3\}_{p}$ \\
  \hline
 3 & 1.51652  & $\{-2, -1, 0, 1, 2\}$ & 2.99737 & $\{-1.57067\ri, -0.5\ri, 0, 0.5\ri, 1.57067\ri\}$\\
 \hline
 4 & 1.59701  & $\{-2, -1, 0, 1, 2\}$ & 3.77094 & $\{-4, -3, -2, -1, 2\}_{p}$ \\
 \hline
\end{tabular}
\end{adjustbox}
\caption{Subsystem-size dependence for $L=10$, $M=5$; singular extrema are listed by rapidities.}
\label{tab:L10M5}
\end{table}
\begin{table}[!htbp]
\centering
\small
\begin{adjustbox}{max width=\textwidth}
\begin{tabular}{|c|c|c|c|c|}
  \hline
  $\ell$ &  $\mathcal{S}_{\text{min}}$ & $\{I_j\}_{\text{min}}$ & $\mathcal{S}_{\text{max}}$ & $\{I_j\}_{\text{max}}$\\
  \hline
  2& 1.00457  & $\{-1, 0\}_{p}$ & 1.35654 & $\{-3, 2\}_{p}$ \\
  \hline
 3 & 1.19209  & $\{-1, 0\}_{p}$ & 1.68473 & $\{-4, 3\}_{p}$\\
 \hline
 4 & 1.2994  & $\{-1, 0\}_{p}$ & 1.89022 & $\{-4, 4\}$ \\
 \hline
\end{tabular}
\end{adjustbox}
\caption{Subsystem-size dependence for $L=11$, $M=2$.}
\label{tab:L11M2}
\end{table}

\begin{table}[!htbp]
\centering
\small
\begin{adjustbox}{max width=\textwidth}
\begin{tabular}{|c|c|c|c|c|}
  \hline
  $\ell$ &  $\mathcal{S}_{\text{min}}$ & $\{I_j\}_{\text{min}}$ & $\mathcal{S}_{\text{max}}$ & $\{I_j\}_{\text{max}}$\\
  \hline
  2& 1.19523  & $\{-\frac{3}{2}, -\frac{1}{2}, \frac{1}{2}\}_{p}$ & 1.68518 & $\{-\frac{7}{2}, -\frac{7}{2}, \frac{3}{2}\}_{p}$ \\
  \hline
 3 & 1.38071  & $\{-\frac{3}{2}, -\frac{1}{2}, \frac{1}{2}\}_{p}$ & 2.43218 & $\{-\frac{7}{2}, -\frac{1}{2}, \frac{7}{2}\}_{p}$\\
 \hline
 4 & 1.48089  & $\{-\frac{3}{2}, -\frac{1}{2}, \frac{1}{2}\}_{p}$ & 2.78147 & $\{-\frac{7}{2}, -\frac{1}{2}, \frac{5}{2}\}_{p}$ \\
 \hline
\end{tabular}
\end{adjustbox}
\caption{Subsystem-size dependence for $L=11$, $M=3$.}
\label{tab:L11M3}
\end{table}

\begin{table}[!htbp]
\centering
\small
\begin{adjustbox}{max width=\textwidth}
\begin{tabular}{|c|c|c|c|c|}
  \hline
  $\ell$ &  $\mathcal{S}_{\text{min}}$ & $\{I_j\}_{\text{min}}$ & $\mathcal{S}_{\text{max}}$ & $\{I_j\}_{\text{max}}$\\
  \hline
  2& 1.30604  & $\{-2, -1, 0, 1\}_{p}$ & 1.88888 & $\{-3, -3, -2, 3\}_{p}$ \\
  \hline
 3 & 1.48488  & $\{-2, -1, 0, 1\}_{p}$ & 2.76264 & $\{-3, -3, 0, 3\}_{p}$\\
 \hline
 4 & 1.57985  & $\{-2, -1, 0, 1\}_{p}$ & 3.32262 & $\{-3, -3, 1, 3\}_{p}$ \\
 \hline
\end{tabular}
\end{adjustbox}
\caption{Subsystem-size dependence for $L=11$, $M=4$.}
\label{tab:L11M4}
\end{table}

\begin{table}[!htbp]
\centering
\small
\begin{adjustbox}{max width=\textwidth}
\begin{tabular}{|c|c|c|c|c|}
  \hline
  $\ell$ &  $\mathcal{S}_{\text{min}}$ & $\{I_j\}_{\text{min}}$ & $\mathcal{S}_{\text{max}}$ & $\{I_j\}_{\text{max}}$\\
  \hline
  2& 1.35401  & $\{-\frac{5}{2}, -\frac{3}{2}, -\frac{1}{2}, \frac{1}{2}, \frac{3}{2}\}_{p}$ & 1.98498 & $\{-\frac{9}{2}, -\frac{7}{2}, -\frac{5}{2}, \frac{1}{2}, \frac{5}{2}\}_{p}$ \\
  \hline
 3 & 1.52974  & $\{-\frac{5}{2}, -\frac{3}{2}, -\frac{1}{2}, \frac{1}{2}, \frac{3}{2}\}_{p}$ & 2.93417 &  $\{-\frac{9}{2}, -\frac{7}{2}, -\frac{5}{2}, \frac{1}{2}, \frac{5}{2}\}_{p}$\\
 \hline
 4 & 1.62391  & $\{-\frac{5}{2}, -\frac{3}{2}, -\frac{1}{2}, \frac{1}{2}, \frac{3}{2}\}_{p}$ & 3.65874 &  $\{-\frac{9}{2}, -\frac{7}{2}, -\frac{5}{2}, -\frac{3}{2}, \frac{3}{2}\}_{p}$ \\
 \hline
\end{tabular}
\end{adjustbox}
\caption{Subsystem-size dependence for $L=11$, $M=5$.}
\label{tab:L11M5}
\end{table}

\begin{table}[!htbp]
\centering
\small
\begin{adjustbox}{max width=\textwidth}
\begin{tabular}{|c|c|c|c|c|}
  \hline
  $\ell$ &  $\mathcal{S}_{\text{min}}$ & $\{I_j\}_{\text{min}}$ & $\mathcal{S}_{\text{max}}$ & $\{I_j\}_{\text{max}}$\\
  \hline
 2 & 0.964686 & $\{-\frac{1}{2}, \frac{1}{2}\}$ & 1.29461 & $\{-\frac{7}{2}, \frac{5}{2}\}_{p}$ \\
  \hline
 3 & 1.15606  & $\{-\frac{1}{2}, \frac{1}{2}\}$ & 1.61738 & $\{-\frac{5}{2}, \frac{3}{2}\}_{p}$\\
 \hline
 4 & 1.27286  & $\{-\frac{1}{2}, \frac{1}{2}\}$ & 1.83204 & $\{-\frac{7}{2}, \frac{5}{2}\}_{p}$ \\
 \hline
 5 & 1.33641  & $\{-\frac{1}{2}, \frac{1}{2}\}$ & 1.95559 & $\{-\frac{7}{2}, \frac{7}{2}\}$ \\
 \hline
\end{tabular}
\end{adjustbox}
\caption{Subsystem-size dependence for $L=12$, $M=2$.}
\label{tab:L12M2}
\end{table}

\begin{table}[!htbp]
\centering
\small
\begin{adjustbox}{max width=\textwidth}
\begin{tabular}{|c|c|c|c|c|}
  \hline
  $\ell$ &  $\mathcal{S}_{\text{min}}$ & $\{I_j\}_{\text{min}}$ & $\mathcal{S}_{\text{max}}$ & $\{I_j\}_{\text{max}}$\\
  \hline
  2& 1.15853  & $\{-1, 0, 1\}$ & 1.61286 & $\{-4, -4, 2\}_{p}$ \\
  \hline
 3 & 1.35066  & $\{-1, 0, 1\}$ & 2.35228 & $\{-4, 0, 4\}$\\
 \hline
 4 & 1.46121  & $\{-1, 0, 1\}$ & 2.7255 & $\{-3, 0, 3\}$ \\
 \hline
 5 & 1.51941  & $\{-1, 0, 1\}$ & 2.87079 & $\{-4, -1, 4\}_{p}$\\
 \hline
\end{tabular}
\end{adjustbox}
\caption{Subsystem-size dependence for $L=12$, $M=3$.}
\label{tab:L12M3}
\end{table}

\begin{table}[!htbp]
\centering
\small
\begin{adjustbox}{max width=\textwidth}
\begin{tabular}{|c|c|c|c|c|}
  \hline
  $\ell$ &  $\mathcal{S}_{\text{min}}$ & $\{I_j\}_{\text{min}}$ & $\mathcal{S}_{\text{max}}$ & $\{I_j\}_{\text{max}}$\\
  \hline
  2& 1.27875  & $\{-\frac{3}{2}, -\frac{1}{2}, \frac{1}{2}, \frac{3}{2}\}$ & 1.83375 & $\{-\frac{7}{2}, -\frac{7}{2}, \frac{1}{2}, \frac{7}{2}\}_{p}$ \\
  \hline
 3 & 1.46484  & $\{-\frac{3}{2}, -\frac{1}{2}, \frac{1}{2}, \frac{3}{2}\}$ & 2.70366 & $\{-\frac{7}{2}, -\frac{7}{2}, \frac{1}{2}, \frac{7}{2}\}_{p}$\\
 \hline
 4 & 1.56956 & $\{-\frac{3}{2}, -\frac{1}{2}, \frac{1}{2}, \frac{3}{2}\}$ & 3.32695 & $\{-\frac{7}{2}, -\frac{7}{2}, \frac{1}{2}, \frac{7}{2}\}_{p}$\\
 \hline
 5 & 1.62466 & $\{-\frac{3}{2}, -\frac{1}{2}, \frac{1}{2}, \frac{3}{2}\}$ & 3.71052 & $\{-\frac{7}{2}, -\frac{3}{2}, \frac{3}{2}, \frac{7}{2}\}$ \\
 \hline
\end{tabular}
\end{adjustbox}
\caption{Subsystem-size dependence for $L=12$, $M=4$.}
\label{tab:L12M4}
\end{table}

\begin{table}[!htbp]
\centering
\small
\begin{adjustbox}{max width=\textwidth}
\begin{tabular}{|c|c|c|c|c|}
  \hline
  $\ell$ &  $\mathcal{S}_{\text{min}}$ & $\{I_j\}_{\text{min}}$ & $\mathcal{S}_{\text{max}}$ & $\{I_j\}_{\text{max}}$\\
  \hline
  2& 1.34278  & $\{-2, -1, 0, 1, 2\}$ & 1.95887 & $\{-4, -3, -3, 2, 3\}_{p}$ \\
  \hline
 3 & 1.5238  & $\{-2, -1, 0, 1, 2\}$ & 2.92401 &  $\{-4, -3, 0, 3, 4\}$\\
 \hline
 4 & 1.6258  & $\{-2, -1, 0, 1, 2\}$ & 3.68623 &  $\{-6, -5, -1, 0, 2\}_{p}$ \\
 \hline
 5 & 1.67981 & $\{-2, -1, 0, 1, 2\}$ & 4.11333 &  $\{-4, -3, -1, 1, 3\}_{p}$ \\
 \hline
\end{tabular}
\end{adjustbox}
\caption{Subsystem-size dependence for $L=12$, $M=5$.}
\label{tab:L12M5}
\end{table}

\begin{table}[!htbp]
\centering
\small
\begin{adjustbox}{max width=\textwidth}
\begin{tabular}{|c|c|c|c|c|}
  \hline
  $\ell$ &  $\mathcal{S}_{\text{min}}$ & $\{I_j\}_{\text{min}}$ & $\mathcal{S}_{\text{max}}$ & $\{I_j\}_{\text{max}}$\\
  \hline
  2& 1.35973  & $\{-\frac{5}{2}, -\frac{3}{2}, -\frac{1}{2}, \frac{1}{2}, \frac{3}{2}, \frac{5}{2}\}$ & 2 & $\{-\frac{9}{2}, -\frac{7}{2}, -\frac{5}{2}, -\frac{3}{2}, \frac{3}{2}, \frac{5}{2}\}_{p}$ \\
  \hline
 3 & 1.53727  & $\{-\frac{5}{2}, -\frac{3}{2}, -\frac{1}{2}, \frac{1}{2}, \frac{3}{2}, \frac{5}{2}\}$ & 2.99228 &  $\{-\frac{9}{2}, -\frac{9}{2}, -\frac{7}{2}, -\frac{7}{2}, -\frac{3}{2}, \frac{5}{2}\}_{p}$\\
 \hline
 4 & 1.63792  & $\{-\frac{5}{2}, -\frac{3}{2}, -\frac{1}{2}, \frac{1}{2}, \frac{3}{2}, \frac{5}{2}\}$ & 3.87445 &  $\{-\frac{13}{2}, -\frac{11}{2}, -\frac{5}{2}, -\frac{3}{2}, -\frac{3}{2}, \frac{3}{2}\}_{p}$ \\
 \hline
 5 & 1.69151  & $\{-\frac{5}{2}, -\frac{3}{2}, -\frac{1}{2}, \frac{1}{2}, \frac{3}{2}, \frac{5}{2}\}$ & 4.60681 &  $\{-\frac{11}{2}, -\frac{5}{2}, -\frac{1}{2}, \frac{1}{2}, \frac{5}{2}\}$ \\
 \hline
\end{tabular}
\end{adjustbox}
\caption{Subsystem-size dependence for $L=12$, $M=6$.}
\label{tab:L12M6}
\end{table}

\begin{table}[!htbp]
\centering
\small
\begin{adjustbox}{max width=\textwidth}
\begin{tabular}{|c|c|c|c|c|}
  \hline
  $\ell$ &  $\mathcal{S}_{\text{min}}$ & $\{I_j\}_{\text{min}}$ & $\mathcal{S}_{\text{max}}$ & $\{I_j\}_{\text{max}}$\\
  \hline
 2 & 0.927841 & $\{-1, 0\}_{p}$ & 1.23138 & $\{-3, 3\}_{p}$ \\
  \hline
 3 & 1.1214  & $\{-1, 0\}_{p}$ & 1.55064 & $\{-4, 4\}\text{ or }\{-2, 2\}$\\
 \hline
 4 & 1.24513  & $\{-1, 0\}_{p}$ & 1.77174 & $\{-3, 3\}$ \\
 \hline
 5 & 1.31935  & $\{-1, 0\}_{p}$ & 1.92141 & $\{-5, 5\}$ \\
 \hline
\end{tabular}
\end{adjustbox}
\caption{Subsystem-size dependence for $L=13$, $M=2$.}
\label{tab:L13M2}
\end{table}

\begin{table}[!htbp]
\centering
\small
\begin{adjustbox}{max width=\textwidth}
\begin{tabular}{|c|c|c|c|c|}
  \hline
  $\ell$ &  $\mathcal{S}_{\text{min}}$ & $\{I_j\}_{\text{min}}$ & $\mathcal{S}_{\text{max}}$ & $\{I_j\}_{\text{max}}$\\
  \hline
  2& 1.12356  & $\{-\frac{3}{2}, -\frac{1}{2}, \frac{1}{2}\}$ & 1.55059 & $\{-\frac{9}{2}, -\frac{9}{2}, \frac{3}{2}\}_{p}$ \\
  \hline
 3 & 1.32078 & $\{-\frac{3}{2}, -\frac{1}{2}, \frac{1}{2}\}$ & 2.27559 & $\{-\frac{9}{2}, -\frac{1}{2}, \frac{7}{2}\}$\\
 \hline
 4 & 1.43952  & $\{-\frac{3}{2}, -\frac{1}{2}, \frac{1}{2}\}$ & 2.62989 & $\{-\frac{7}{2}, -\frac{1}{2}, \frac{5}{2}\}_{p}$ \\
 \hline
 5 & 1.50819  & $\{-\frac{3}{2}, -\frac{1}{2}, \frac{1}{2}\}$ & 2.83822 & $\{-\frac{9}{2}, \frac{1}{2}, \frac{7}{2}\}_{p}$\\
 \hline
\end{tabular}
\end{adjustbox}
\caption{Subsystem-size dependence for $L=13$, $M=3$.}
\label{tab:L13M3}
\end{table}

\begin{table}[!htbp]
\centering
\small
\begin{adjustbox}{max width=\textwidth}
\begin{tabular}{|c|c|c|c|c|}
  \hline
  $\ell$ &  $\mathcal{S}_{\text{min}}$ & $\{I_j\}_{\text{min}}$ & $\mathcal{S}_{\text{max}}$ & $\{I_j\}_{\text{max}}$\\
  \hline
  2& 1.2509  & $\{-2, -1, 0, 1\}_{p}$ & 1.77985 & $\{-4, -4, 1, 4\}_{p}$ \\
  \hline
 3 & 1.44352  & $\{-2, -1, 0, 1\}_{p}$ & 2.63971 & $\{-4, -4, 0, 4\}_{p}$\\
 \hline
 4 & 1.55635 & $\{-2, -1, 0, 1\}_{p}$ & 3.24682 & $\{-4, -2, 1, 4\}_{p}$\\
 \hline
 5 & 1.6213 & $\{-2, -1, 0, 1\}_{p}$ & 3.66031 & $\{-4, -1, 1, 4\}$ \\
 \hline
\end{tabular}
\end{adjustbox}
\caption{Subsystem-size dependence for $L=13$, $M=4$.}
\label{tab:L13M4}
\end{table}

\begin{table}[!htbp]
\centering
\small
\begin{adjustbox}{max width=\textwidth}
\begin{tabular}{|c|c|c|c|c|}
  \hline
  $\ell$ &  $\mathcal{S}_{\text{min}}$ & $\{I_j\}_{\text{min}}$ & $\mathcal{S}_{\text{max}}$ & $\{I_j\}_{\text{max}}$\\
  \hline
  2& 1.32733  & $\{-\frac{5}{2}, -\frac{3}{2}, -\frac{1}{2}, \frac{1}{2}, \frac{3}{2}\}_{p}$ & 1.92133 & $\{-\frac{7}{2}, -\frac{7}{2}, -\frac{1}{2}, \frac{7}{2}, \frac{9}{2}\}_{p}$ \\
  \hline
 3 & 1.51496  & $\{-\frac{5}{2}, -\frac{3}{2}, -\frac{1}{2}, \frac{1}{2}, \frac{3}{2}\}_{p}$ & 2.8645 &  $\{-\frac{11}{2}, -\frac{9}{2}, -\frac{5}{2}, -\frac{1}{2}, -\frac{7}{2}\}_{p}$\\
 \hline
 4 & 1.62438  & $\{-\frac{5}{2}, -\frac{3}{2}, -\frac{1}{2}, \frac{1}{2}, \frac{3}{2}\}_{p}$ & 3.64567 &  $\{-\frac{11}{2}, -\frac{9}{2}, -\frac{5}{2}, -\frac{3}{2}, \frac{5}{2}\}_{p}$ \\
 \hline
 5 & 1.68775 & $\{-\frac{5}{2}, -\frac{3}{2}, -\frac{1}{2}, \frac{1}{2}, \frac{3}{2}\}_{p}$ & 4.09207 &  $\{-\frac{9}{2}, -\frac{7}{2}, -\frac{3}{2}, \frac{3}{2}, \frac{7}{2}\}_{p}$ \\
 \hline
\end{tabular}
\end{adjustbox}
\caption{Subsystem-size dependence for $L=13$, $M=5$.}
\label{tab:L13M5}
\end{table}

\begin{table}[!htbp]
\centering
\small
\begin{adjustbox}{max width=\textwidth}
\begin{tabular}{|c|c|c|c|c|}
  \hline
  $\ell$ &  $\mathcal{S}_{\text{min}}$ & $\{I_j\}_{\text{min}}$ & $\mathcal{S}_{\text{max}}$ & $\{I_j\}_{\text{max}}$\\
  \hline
  2& 1.36046 & $\{-3, -2, -1, 0, 1, 2\}_{p}$ & 1.99095 & $\{-4, -4, -3, -3, 2, 3\}_{p}$ \\
  \hline
 3 & 1.54562  & $\{-3, -2, -1, 0, 1, 2\}_{p}$ & 2.96847 &  $\{-5, -4, -3, 0, 3, 4\}_{p}$\\
 \hline
 4 & 1.65469  & $\{-3, -2, -1, 0, 1, 2\}_{p}$ & 3.81839 &  $\{-6, -5, -3, -2, 0, 3\}_{p}$ \\
 \hline
 5 & 1.71864  & $\{-3, -2, -1, 0, 1, 2\}_{p}$ & 4.51717 &  $\{-5, -4, -3, -2, 1, 3\}_{p}$ \\
 \hline
\end{tabular}
\end{adjustbox}
\caption{Subsystem-size dependence for $L=13$, $M=6$.}
\label{tab:L13M6}
\end{table}

\FloatBarrier
\subsection[Spin 1]{Spin $1$}
\subsubsection*{Length-by-length spin-$1$ summary tables}
The representative $L=8$ spin-$1$ table remains in Section~\ref{subsec:onshell_xxxs}. The tables below collect the remaining on-shell spin-$1$ summary data. The notation follows the main-text table; when a singular extremum is involved, the rapidities are shown directly.

\begin{table}[!htbp]
\centering
\small
\begin{adjustbox}{max width=\textwidth}
\begin{tabular}{|c|c|c|c|c|c|c|}
  \hline
  $M$ & $\mathcal{N}_R$ & $\mathcal{N}_S$ & $\mathcal{S}_{\text{min}}$ & $\{I_j\}_{\text{min}}$ & $\mathcal{S}_{\text{max}}$ & $\{I_j\}_{\text{max}}$\\
  \hline
  1 & 3 & 0 & 1  & / & 1 & / \\
  \hline
 2 & 6 & 0 & 1.59715& $\{-\frac{1}{2}, -\frac{3}{2}\}_{p}$ & 1.70093 & $\{\frac{1}{2}, \frac{1}{2}\}$\\
 \hline
 3 & 5 & 1 & 1.74758  & $\{0, 0, 0\}$ & 2.5 & $\{-\ri, 0, \ri\}$ \\
 \hline
 4 & 2 & 1 & 1.90811 & $\{-\frac{1}{2}, -\frac{1}{2}, \frac{1}{2}, \frac{1}{2}\}$ & 3.02064 & {$\{-\frac{1}{2}, -\frac{1}{2}, \frac{1}{2}, \frac{1}{2}\}$} \\
   \hline
\end{tabular}
\end{adjustbox}
\caption{Summary of the on-shell spin-$1$ scan at $L=4$.}
\label{tab:S1L4}
\end{table}

\begin{table}[!htbp]
\centering
\small
\begin{adjustbox}{max width=\textwidth}
\begin{tabular}{|c|c|c|c|c|c|c|}
  \hline
  $M$ & $\mathcal{N}_R$ & $\mathcal{N}_S$ & $\mathcal{S}_{\text{min}}$ & $\{I_j\}_{\text{min}}$ & $\mathcal{S}_{\text{max}}$ & $\{I_j\}_{\text{max}}$\\
  \hline
  1 & 4 & 0 & 0.970951  & / & 0.970951 & / \\
  \hline
 2 & 10 & 0 & 1.51062 & $\{0, 0\}$ & 1.92194 & $\{-1, 1\}$\\
 \hline
 3 & 14 & 1 & 1.80944  & $\{-\frac{1}{2}, -\frac{1}{2}, -\frac{1}{2}\}_{p}$ & 2.43401 & $\{-\frac{3}{2}, -\frac{1}{2}, \frac{1}{2}\}_{p}$ \\
 \hline
 4 & 15 & 0 & 1.96038 & $\{0, 0, 0, 0\}$ & 2.92077 & $\{-1, 0, 0, 1\}$ \\
   \hline
5 & 6 & 0 & 2.89294 & $\{-\frac{3}{2}, -\frac{1}{2}, -\frac{1}{2}, \frac{1}{2}, \frac{1}{2}\}_{p}$ & 3.08418 & $\{-\frac{3}{2}, -\frac{3}{2}, -\frac{1}{2}, -\frac{1}{2}, -\frac{1}{2}\}_{p}$ \\
   \hline
\end{tabular}
\end{adjustbox}
\caption{Summary of the on-shell spin-$1$ scan at $L=5$.}
\label{tab:S1L5}
\end{table}

\begin{table}[!htbp]
\centering
\small
\begin{adjustbox}{max width=\textwidth}
\begin{tabular}{|c|c|c|c|c|c|c|}
  \hline
  $M$ & $\mathcal{N}_R$ & $\mathcal{N}_S$ & $\mathcal{S}_{\text{min}}$ & $\{I_j\}_{\text{min}}$ & $\mathcal{S}_{\text{max}}$ & $\{I_j\}_{\text{max}}$\\
  \hline
  1 & 5 & 0 & 1  & / & 1 & / \\
  \hline
 2 & 15 & 0 & 1.53181 & $\{-\frac{1}{2}, \frac{1}{2}\}$ & 1.99595 & $\{-\frac{3}{2}, \frac{1}{2}\}_{p}$\\
 \hline
 3 & 28 & 1 & 1.92872  & $\{0, 0, 0\}$ & 2.69966 & $\{-1, -1, 1\}_{p}$ \\
 \hline
 4 & 39 & 1 & 2.07 & $\{-\frac{1}{2}, -\frac{1}{2}, \frac{1}{2}, \frac{1}{2}\}$ & 3.24656 & $\{-\frac{3}{2}, -\frac{3}{2}, -\frac{1}{2}, \frac{1}{2}\}_{p}$ \\
   \hline
5 & 33 & 3 & 2.1346 & $\{0, 0, 0, 0, 0\}$ & 3.80168 & $\{-1, 0, 0, 0, 1\}$ \\
   \hline
6 & 18 & 4 & 2.17781 & $\{-\frac{1}{2}, -\frac{1}{2}, -\frac{1}{2}, \frac{1}{2}, \frac{1}{2}, \frac{1}{2}\}$ & 4.23704 & $\{-\frac{3}{2}, -\frac{3}{2}, -\frac{1}{2}, -\frac{1}{2}, -\frac{1}{2}, \frac{1}{2}\}$ \\
   \hline
\end{tabular}
\end{adjustbox}
\caption{Summary of the on-shell spin-$1$ scan at $L=6$.}
\label{tab:S1L6}
\end{table}

\begin{table}[!htbp]
\centering
\small
\begin{adjustbox}{max width=\textwidth}
\begin{tabular}{|c|c|c|c|c|c|c|}
  \hline
  $M$ & $\mathcal{N}_R$ & $\mathcal{N}_S$ & $\mathcal{S}_{\text{min}}$ & $\{I_j\}_{\text{min}}$ & $\mathcal{S}_{\text{max}}$ & $\{I_j\}_{\text{max}}$\\
  \hline
  1 & 6 & 0 & 0.985228  & / & 0.985228 & / \\
  \hline
 2 & 21 & 0 & 1.46228 & $\{0, 0\}$ & 1.96149 & $\{-2, 2\}$\\
 \hline
 3 & 48 & 1 & 1.9242  & $\{-\frac{7}{2}, -\frac{5}{2}, -\frac{3}{2}\}_{p}$ & 2.70419 & $\{-\frac{3}{2}, -\frac{3}{2}, \frac{3}{2}\}_{p}$ \\
 \hline
 4 & 84 & 0 & 2.03266 & $\{0, 0, 0, 0\}$ & 3.26689 & $\{-1, -1, 1, 1\}$ \\
   \hline
5 & 102 & 3 & 2.2118 & $\{-\frac{1}{2}, -\frac{1}{2}, -\frac{1}{2}, -\frac{1}{2}, -\frac{1}{2}\}_{p}$ & 3.72105 & $\{-\frac{3}{2}, -\frac{3}{2}, -\frac{1}{2}, \frac{1}{2}, \frac{1}{2}\}$ \\
   \hline
6 & 91 & 0 & 2.2448 & $\{0, 0, 0, 0, 0,0\}$ & 4.2932 & $\{-2, 0, 0, 0, 0, 2\}$ \\
   \hline
7 & 30 & 12 & 2.51261 & $\begin{aligned}\{&-\ri, -0.15-0.47\ri, -0.15+0.47\ri, \\&0, 0.15-0.47\ri,0.15+0.47\ri, \ri\}_{p}\end{aligned}$ & 4.49886 & $\begin{aligned}\{&-2.06\ri, -\ri, -0.83, \\ &0,0.83, \ri, 2.06\ri\}_{p}\end{aligned}$ \\
   \hline
\end{tabular}
\end{adjustbox}
\caption{Summary of the on-shell spin-$1$ scan at $L=7$; singular extrema are listed by rapidities.}
\label{tab:S1L7}
\end{table}

\FloatBarrier
\subsubsection*{Additional subsystem-size spin-$1$ data}
The tables below record supplementary spin-$1$ extrema data for selected sectors displayed in the scan.

\begin{table}[!htbp]
\centering
\small
\begin{adjustbox}{max width=\textwidth}
\begin{tabular}{|c|c|c|c|c|}
  \hline
  $M$ &  $\mathcal{S}_{\text{min}}$ & $\{I_j\}_{\text{min}}$ & $\mathcal{S}_{\text{max}}$ & $\{I_j\}_{\text{max}}$\\
  \hline
 2 & 1.41532 & $\{-\frac{1}{2}, \frac{1}{2}\}$ & 1.82923 & $\{-\frac{3}{2}, \frac{3}{2}\}_{p}$ \\
  \hline
 3 & 1.77693  & $\{0, 0, 0\}$ & 2.34573 &  $\{-1, -1, 1\}_{p}$\\
 \hline
 4 & 1.9473  & $\{-\frac{1}{2}, -\frac{1}{2}, \frac{1}{2}, \frac{1}{2}\}$ & 2.78642 &  $\{-\frac{3}{2}, -\frac{1}{2}, \frac{1}{2}, \frac{3}{2}\}$ \\
 \hline
 5 & 1.99329  & $\{0, 0, 0, 0,0\}$ & 3.0231 &  $\{-2, -1, 0, 0, 1\}$ \\
 \hline
 6 & 2.06149  & $\{-\frac{1}{2}, -\frac{1}{2}, -\frac{1}{2}, \frac{1}{2}, \frac{1}{2}, \frac{1}{2}\}$ & 3.11976 &  $\{-\frac{3}{2}, -\frac{3}{2}, -\frac{1}{2}, -\frac{1}{2}, -\frac{1}{2}, \frac{1}{2}\}_{p}$ \\
 \hline
\end{tabular}
\end{adjustbox}
\caption{Spin-$1$ extrema at $L=6$ and $\ell=2$.}
\label{tab:suppS1L6}
\end{table}

\begin{table}[!htbp]
\centering
\small
\begin{adjustbox}{max width=\textwidth}
\begin{tabular}{|c|c|c|c|c|}
  \hline
  $M$ &  $\mathcal{S}_{\text{min}}$ & $\{I_j\}_{\text{min}}$ & $\mathcal{S}_{\text{max}}$ & $\{I_j\}_{\text{max}}$\\
  \hline
 2 & 1.30038 & $\{0, 0\}$ & 1.71721 & $\{-2, 1\}_{p}$ \\
  \hline
 3 & 1.68713  & $\{-\frac{7}{2}, -\frac{5}{2}, -\frac{3}{2}\}_{p}$ & 2.24027 &  $\{-\frac{3}{2}, -\frac{3}{2}, \frac{3}{2}\}_{p}$\\
 \hline
 4 & 1.85687  & $\{0, 0, 0, 0\}$ & 2.65242 &  $\{-1, -1, 1, 1\}$ \\
 \hline
 5 & 1.99766  & $\{-\frac{1}{2}, -\frac{1}{2}, -\frac{1}{2}, -\frac{1}{2}, -\frac{1}{2}\}$ & 2.91311 &  $\{-\frac{3}{2}, -\frac{3}{2}, -\frac{1}{2}, \frac{1}{2}, \frac{3}{2}\}$ \\
 \hline
 6 & 2.06675  & $\{0, 0, 0, 0, 0, 0\}$ & 3.07029 &  $\{-2, 0, 0, 0, 0, 2\}$ \\
 \hline
 7 & 2.1797  & $\begin{aligned}\{&-\ri, -0.15-0.475\ri, -0.15+0.475\ri, \\&0, 0.15-0.475\ri, 0.15+0.475\ri, \ri\}\end{aligned}$ & 3.16067 &  $\{-\frac{9}{2}, -\frac{7}{2}, -\frac{1}{2}, -\frac{1}{2}, -\frac{1}{2}, \frac{1}{2}, \frac{1}{2}\}_{p}$ \\
 \hline
\end{tabular}
\end{adjustbox}
\caption{Spin-$1$ extrema at $L=7$ and $\ell=2$; singular extrema are listed by rapidities.}
\label{tab:suppS1L7}
\end{table}

\begin{table}[!htbp]
\centering
\small
\begin{adjustbox}{max width=\textwidth}
\begin{tabular}{|c|c|c|c|c|}
  \hline
  $\ell$ &  $\mathcal{S}_{\text{min}}$ & $\{I_j\}_{\text{min}}$ & $\mathcal{S}_{\text{max}}$ & $\{I_j\}_{\text{max}}$\\
  \hline
 2 & 1.21014 & $\{-\frac{1}{2}, \frac{1}{2}\}$ & 1.61487 & $\{-\frac{5}{2}, \frac{3}{2}\}_{p}$ \\
  \hline
 3 & 1.39345  & $\{-\frac{1}{2}, \frac{1}{2}\}$ & 1.90851 &  $\{-\frac{5}{2}, \frac{5}{2}\}$\\
 \hline
\end{tabular}
\end{adjustbox}
\caption{Spin-$1$ subsystem-size dependence at $L=8$, $M=2$.}
\label{tab:S1L8M2}
\end{table}

\begin{table}[!htbp]
\centering
\small
\begin{adjustbox}{max width=\textwidth}
\begin{tabular}{|c|c|c|c|c|}
  \hline
  $\ell$ &  $\mathcal{S}_{\text{min}}$ & $\{I_j\}_{\text{min}}$ & $\mathcal{S}_{\text{max}}$ & $\{I_j\}_{\text{max}}$\\
  \hline
 2 & 1.58 & $\{-4, -3, -2\}_{p}$ & 2.10798 & $\{-2, -2, 1\}_{p}$ \\
  \hline
 3 & 1.84076  & $\{-4, -3, 2\}_{p}$ & 2.76272 &  $\{-2, 0, 2\}$\\
 \hline
\end{tabular}
\end{adjustbox}
\caption{Spin-$1$ subsystem-size dependence at $L=8$, $M=3$.}
\label{tab:S1L8M3}
\end{table}

\begin{table}[!htbp]
\centering
\small
\begin{adjustbox}{max width=\textwidth}
\begin{tabular}{|c|c|c|c|c|}
  \hline
  $\ell$ &  $\mathcal{S}_{\text{min}}$ & $\{I_j\}_{\text{min}}$ & $\mathcal{S}_{\text{max}}$ & $\{I_j\}_{\text{max}}$\\
  \hline
 2 & 1.77933 & $\{-\frac{1}{2}, -\frac{1}{2}, \frac{1}{2}, \frac{1}{2}\}$ & 2.53838 & $\{-\frac{3}{2}, -\frac{3}{2}, \frac{3}{2}, \frac{3}{2}\}$ \\
  \hline
 3 & 1.99034  & $\{-\frac{1}{2}, -\frac{1}{2}, \frac{1}{2}, \frac{1}{2}\}$ & 3.25645 &  $\{-\frac{3}{2}, -\frac{3}{2}, \frac{3}{2}, \frac{3}{2}\}$\\
 \hline
\end{tabular}
\end{adjustbox}
\caption{Spin-$1$ subsystem-size dependence at $L=8$, $M=4$.}
\label{tab:S1L8M4}
\end{table}

\begin{table}[!htbp]
\centering
\small
\begin{adjustbox}{max width=\textwidth}
\begin{tabular}{|c|c|c|c|c|}
  \hline
  $\ell$ &  $\mathcal{S}_{\text{min}}$ & $\{I_j\}_{\text{min}}$ & $\mathcal{S}_{\text{max}}$ & $\{I_j\}_{\text{max}}$\\
  \hline
 2 & 1.95883 & $\{0, 0, 0, 0, 0\}$ & 2.79889 & $\{-2, -2, -1, 1, 1\}_{p}$ \\
  \hline
 3 & 2.22077  & $\{0, 0, 0, 0, 0\}$ & 3.70576 &  $\{-2, -1, 0, 1, 2\}$\\
 \hline
\end{tabular}
\end{adjustbox}
\caption{Spin-$1$ subsystem-size dependence at $L=8$, $M=5$.}
\label{tab:S1L8M5}
\end{table}

\begin{table}[!htbp]
\centering
\small
\begin{adjustbox}{max width=\textwidth}
\begin{tabular}{|c|c|c|c|c|}
  \hline
  $\ell$ &  $\mathcal{S}_{\text{min}}$ & $\{I_j\}_{\text{min}}$ & $\mathcal{S}_{\text{max}}$ & $\{I_j\}_{\text{max}}$\\
  \hline
 2 & 2.05125 & $\{-\frac{1}{2}, -\frac{1}{2}, -\frac{1}{2}, \frac{1}{2}, \frac{1}{2}, \frac{1}{2}\}$ & 2.98061 & $\{-\frac{5}{2}, -\frac{3}{2}, -\frac{1}{2}, \frac{1}{2}, \frac{1}{2}, \frac{3}{2}\}_{p}$ \\
  \hline
 3 & 2.26971  & $\{-\frac{1}{2}, -\frac{1}{2}, -\frac{1}{2}, \frac{1}{2}, \frac{1}{2}, \frac{1}{2}\}$ & 4.14968 &  $\{-\frac{7}{2}, -\frac{5}{2}, -\frac{3}{2}, -\frac{1}{2}, \frac{1}{2}, \frac{1}{2}\}_{p}$\\
 \hline
\end{tabular}
\end{adjustbox}
\caption{Spin-$1$ subsystem-size dependence at $L=8$, $M=6$.}
\label{tab:S1L8M6}
\end{table}

\begin{table}[!htbp]
\centering
\small
\begin{adjustbox}{max width=\textwidth}
\begin{tabular}{|c|c|c|c|c|}
  \hline
  $\ell$ &  $\mathcal{S}_{\text{min}}$ & $\{I_j\}_{\text{min}}$ & $\mathcal{S}_{\text{max}}$ & $\{I_j\}_{\text{max}}$\\
  \hline
 2 & 2.06833 & $\{0, 0, 0, 0, 0, 0, 0\}$ & 3.1049 & $\{-5, -4, -1, -1, 0, 1\}_{p}$ \\
  \hline
 3 & 2.29695  & $\{0, 0, 0, 0, 0, 0, 0\}$ & 4.41397 &  $\{-3, -3, -2, -2, -2, 0, 1\}_{p}$\\
 \hline
\end{tabular}
\end{adjustbox}
\caption{Spin-$1$ subsystem-size dependence at $L=8$, $M=7$.}
\label{tab:S1L8M7}
\end{table}

\begin{table}[!htbp]
\centering
\small
\begin{adjustbox}{max width=\textwidth}
\begin{tabular}{|c|c|c|c|c|}
  \hline
  $\ell$ &  $\mathcal{S}_{\text{min}}$ & $\{I_j\}_{\text{min}}$ & $\mathcal{S}_{\text{max}}$ & $\{I_j\}_{\text{max}}$\\
  \hline
 2 & 2.10865 & $\{-\frac{1}{2}, -\frac{1}{2}, -\frac{1}{2}, -\frac{1}{2}, \frac{1}{2}, \frac{1}{2}, \frac{1}{2}, \frac{1}{2}\}$ & 3.16968 & $\{-\frac{11}{2}, -\frac{9}{2}, -\frac{7}{2}, -\frac{5}{2}, -\frac{3}{2}, -\frac{1}{2}, \frac{3}{2}, \frac{5}{2}\}_{p}$ \\
  \hline
 3 & 2.31972 & $\{-\frac{1}{2}, -\frac{1}{2}, -\frac{1}{2}, -\frac{1}{2}, \frac{1}{2}, \frac{1}{2}, \frac{1}{2}, \frac{1}{2}\}$ & 4.5242 &  $\{-\frac{7}{2}, -\frac{5}{2}, -\frac{5}{2}, -\frac{3}{2}, -\frac{1}{2}, -\frac{1}{2}, \frac{1}{2}, \frac{1}{2}\}_{p}$\\
 \hline
\end{tabular}
\end{adjustbox}
\caption{Spin-$1$ subsystem-size dependence at $L=8$, $M=8$.}
\label{tab:S1L8M8}
\end{table}

\FloatBarrier
\subsection[Spin 3/2]{Spin $\frac{3}{2}$}
\subsubsection*{Length-by-length spin-$\frac{3}{2}$ summary tables}

The representative $L=6$ spin-$\frac{3}{2}$ table remains in Section~\ref{subsec:onshell_xxxs}. The tables below collect the remaining on-shell summary data. The notation follows the preceding supplementary tables; singular extrema are listed directly by rapidities.

\begin{table}[!htbp]
\centering
\small
\begin{adjustbox}{max width=\textwidth}
\begin{tabular}{|c|c|c|c|c|c|c|}
  \hline
  $M$ & $\mathcal{N}_R$ & $\mathcal{N}_S$ & $\mathcal{S}_{\text{min}}$ & $\{I_j\}_{\text{min}}$ & $\mathcal{S}_{\text{max}}$ & $\{I_j\}_{\text{max}}$\\
  \hline
  1 & 3 & 0 & 1  & / & 1 & / \\
  \hline
 2 & 6 & 0 & 1.55639 & $\{-\frac{3}{2}, -\frac{1}{2}\}_{p}$ & 1.88771 & $\{-\frac{1}{2}, \frac{1}{2}\}$\\
 \hline
 3 & 10 & 0 & 1.98244  & $\{-2, -1, 0\}_{p}$ & 2.36801 & $\{-1, 0, 0\}_{p}$ \\
 \hline
 4 & 10 & 1 & 2.16731 & $\{-\frac{1}{2}, -\frac{1}{2}, \frac{1}{2}, \frac{1}{2}\}$ & 2.78631 & $\{-1.5\ri, -0.5\ri, 0.5\ri, 1.5\ri\}_{p}$ \\
   \hline
5 & 8 & 1 & 2.23256 & $\{-1, 0, 0, 0, 1\}$ & 3.39047 & $\{-1.5\ri, -0.5\ri, 0, 0.5\ri, 1.5\ri\}$ \\
   \hline
6 & 2 & 4 & 2.40071 & $\{-\frac{3}{2}, -\frac{1}{2}, -\frac{1}{2}, \frac{1}{2}, \frac{1}{2}, \frac{3}{2}\}$ & 3.66087 & $\{-2.665\ri, -1.5\ri, 0.5\ri, 0.5\ri, 1.5\ri, 2.665\ri\}_{p}$ \\
   \hline
\end{tabular}
\end{adjustbox}
\caption{Summary of the on-shell spin-$\frac{3}{2}$ scan at $L=4$; singular extrema are listed by rapidities.}
\label{tab:S32L4}
\end{table}

\begin{table}[!htbp]
\centering
\small
\begin{adjustbox}{max width=\textwidth}
\begin{tabular}{|c|c|c|c|c|c|c|}
  \hline
  $M$ & $\mathcal{N}_R$ & $\mathcal{N}_S$ & $\mathcal{S}_{\text{min}}$ & $\{I_j\}_{\text{min}}$ & $\mathcal{S}_{\text{max}}$ & $\{I_j\}_{\text{max}}$\\
  \hline
  1 & 4 & 0 & 0.97095  & / & 0.97095 & / \\
  \hline
 2 & 10 & 0 & 1.50008 & $\{-2, -1\}_{p}$ & 1.93807 & $\{-1, 1\}$\\
 \hline
 3 & 20 & 0 & 1.85124  & $\{-\frac{1}{2}, -\frac{1}{2}, \frac{1}{2}\}_{p}$ & 2.4516 & $\{-\frac{3}{2}, -\frac{1}{2}, \frac{1}{2}\}_{p}$ \\
 \hline
 4 & 30 & 0 & 2.21545 & $\{-3, -2, -1, 0\}_{p}$ & 2.95731 & $\{-1, 0, 0, 1\}$ \\
   \hline
5 & 36 & 1 & 2.34857 & $\{-\frac{3}{2}, -\frac{1}{2}, -\frac{1}{2}, -\frac{1}{2}, \frac{1}{2}\}_{p}$ & 3.32025 & $\{-\frac{5}{2}, -\frac{3}{2}, -\frac{1}{2}, \frac{1}{2}, \frac{1}{2}\}_{p}$ \\
   \hline
6 & 34 & 4 & 2.46692 & $\{-1, 0, 0, 0, 1, 1\}_{p}$ & 3.69839 & $\{-2, -1, -1, -1, 0, 0\}_{p}$ \\
   \hline
\end{tabular}
\end{adjustbox}
\caption{Summary of the on-shell spin-$\frac{3}{2}$ scan at $L=5$.}
\label{tab:S32L5}
\end{table}

\FloatBarrier
\subsubsection*{Additional spin-$\frac{3}{2}$ data}

The table below records supplementary subsystem-size data for the $L=6$ scan.

\begin{table}[!htbp]
\centering
\small
\begin{adjustbox}{max width=\textwidth}
\begin{tabular}{|c|c|c|c|c|}
  \hline
  $M$ &  $\mathcal{S}_{\text{min}}$ & $\{I_j\}_{\text{min}}$ & $\mathcal{S}_{\text{max}}$ & $\{I_j\}_{\text{max}}$\\
  \hline
 2 & 1.41534 & $\{-\frac{5}{2}, -\frac{3}{2}\}_{p}$ & 1.82351 & $\{-\frac{3}{2}, \frac{3}{2}\}$ \\
  \hline
 3 & 1.68692  & $\{-1, 0, 1\}$ & 2.41009 &  $\{-1, -1, 1\}_{p}$\\
 \hline
 4 & 2.04584  & $\{-\frac{7}{2}, -\frac{5}{2}, -\frac{3}{2}, -\frac{1}{2}\}_{p}$ & 2.84307 &  $\{-\frac{3}{2}, -\frac{1}{2}, \frac{1}{2}, \frac{1}{2}\}_{p}$ \\
 \hline
 5 & 2.30695  & $\{-4, -3, -2, -1, 0\}_{p}$ & 3.2362 &  $\{-1, -1, 0, 0, 1\}_{p}$ \\
 \hline
 6 & 2.43285  & $\{-\frac{3}{2}, -\frac{1}{2}, -\frac{1}{2}, \frac{1}{2}, 
 \frac{1}{2}, \frac{3}{2}\}$ & 3.55233 &  $\{-\frac{1}{2}, -\frac{1}{2}, -\frac{1}{2}, \frac{1}{2}, 
 \frac{1}{2}, \frac{1}{2}\}$ \\
 \hline
 7 & 2.50492  & $\{-1, -1, 0, 0, 0, 1, 1\}$ & 3.74283 &  $\{-1, -1, -1, 0, 1, 1, 2\}_{p}$ \\
 \hline
 8 & 2.4988  & $\{-\frac{3}{2}, -\frac{1}{2}, -\frac{1}{2}, -\frac{1}{2}, \frac{1}{2}, \frac{1}{2}, \frac{1}{2}, \frac{3}{2}\}$ & 3.88199 &  $\begin{aligned}\{&-2.535\ri, -1.5\ri, -0.898358, -0.5\ri,\\ &0.898, 0.5\ri, 1.5\ri, 2.535\ri\}\end{aligned}$ \\
 \hline
 9 & 2.58816  & $\{-2, -1, -1, 0, 0, 0, 1, 1, 2\}$ & 3.94586 &  $\{-2, -2, -1, -1, -1, 0, 0, 0,1\}_{p}$ \\
 \hline
\end{tabular}
\end{adjustbox}
\caption{Spin-$\frac{3}{2}$ extrema at $L=6$ and $\ell=2$; singular extrema are listed by rapidities.}
\label{tab:suppS32L6}
\end{table}

\FloatBarrier
\endgroup
\section{Regularization for Singular Strange Solutions}\label{app:GReg}

The transfer matrix acts on a Bethe state as
\begin{align}
    \label{eq:ABA_app}
    &\mathrm{T}(v)\mathbf{B}^{(s)}(\bar{u})|0\rangle = t(v)\mathbf{B}^{(s)}(\bar{u})|0\rangle + \sum_k \mathrm{F}^{(s)}_{k}(v,\bar{u})\mathbf{B}^{(s)}(v)\mathbf{B}^{(s)}(\bar{u}_k)|0\rangle.
\end{align}
Here, we use the compact notation ${\mathbf{B}}^{(s)}(\bar{u})=\mathbf{B}^{(s)}(u_1)\dots \mathbf{B}^{(s)}(u_M)$, and $\mathbf{B}^{(s)}(\bar{u}_k)$ means the absence of $\mathbf{B}^{(s)}(u_k)$. 
A general regularization for a singular root with multiplicity $r$ is given by
\begin{equation}\label{eq:GsingularRegulators}
u^{\epsilon}_{(k,j)}=(s-k+1)\ri+\epsilon+c_{k,j}\epsilon^{L/r}, \qquad 1 \leq k \leq 2s+1,\qquad 1\leq j \leq r.
\end{equation}

For the $\mathrm{XXX}_1$ case we discussed in the main body, as $\epsilon \to 0$, the wanted term $\mathbf{B}^{(s)}(\bar{u}^{\epsilon})|0\rangle \to \epsilon^{L}$. To ensure that $\mathbf{B}^{(s)}(\bar{u})|0\rangle$ remains an eigenstate of the transfer matrix, we need to require the unwanted terms to vanish as $\epsilon^{L+1}$. For unrepeated singular roots $\{\ri,-\ri\}$, $\mathbf{B}^{(s)}(v)\mathbf{B}^{(s)}(\bar{u}_k)|0\rangle \sim 1$, so we impose the same condition as in the nonsingular (regular) case, namely $\mathrm{F}_{k}(v,\bar{u}) \sim \epsilon^{L+1}$. 
However, the vectors $\mathbf{B}^{(s)}(v)\mathbf{B}^{(s)}(\bar{u}_{(2,i)})|0\rangle$ associated with the repeated roots are not linearly independent, and their asymptotic behavior is
\begin{equation}
   \mathbf{B}^{(s)}(v)\mathbf{B}^{(s)}(\bar{u}_{(2,i)}) |0\rangle \sim \epsilon^{L/2} \prod_{j \neq i} c_{2,j} |v\rangle,
\end{equation}
where $|v\rangle$ is universal for each distinct repeated singular root; the asymptotics differ only by the coefficient $c_{2,j}$ determined by the regulators. 
As we require this term to vanish at $\epsilon^L$,
\begin{equation}\label{repeated_roots_constraints}
\mathrm{F}^{(s)}_{(2,1)}(v,\bar{u})\mathbf{B}^{(s)}(v)\mathbf{B}^{(s)}(\bar{u}_{(2,1)}) |0\rangle + \mathrm{F}^{(s)}_{(2,2)}(v,\bar{u})\mathbf{B}^{(s)}(v)\mathbf{B}^{(s)}(\bar{u}_{(2,2)}) |0\rangle \sim \epsilon^{L+1},
\end{equation}
This will provide the algebraic equations for the $c_{k,j}$.
Such analysis generalizes to singular roots with higher multiplicity.
With general multiplicity $r_k$, the corresponding unwanted terms are as follows:
\begin{align}
    \label{eq:HigerMulti_ABA}
    \mathbf{B}^{(s)}(v)\mathbf{B}^{(s)}(\bar{u}_{(k,i)})|0\rangle \sim \epsilon^{L\left(1-\frac{1}{r_k}\right)}\prod_{j\neq i}c_{k,j}|v\rangle.
\end{align}
The condition generalizes to:
\begin{equation}
    \label{eq:higherMultipicity2}
    \sum_j \mathrm{F}^{(s)}_{(k,j)}(v,\bar{u}) \prod_{i \neq j} c_{k,i} \sim \epsilon^{\frac{L}{r_k} + 1},\qquad 1\leq i,j
    \leq r_k.
\end{equation}

\bibliographystyle{utphys}
\bibliography{refs}

\end{document}